\documentclass{aa}

\newcommand{\Wlam}   {$W_{\lambda}$~}
\newcommand{\Wlams}  {$W_{\lambda}$s~}

\newcommand{\Teff}   {$T_{\rm eff}$~}
\newcommand{\Teffs}  {$T_{\rm eff}$s~}

\newcommand{\feh} {$[$Fe/H$]$~}
\newcommand{\fehs} {$[$Fe/H$]$s~}

\usepackage{graphicx,natbib}
\usepackage{txfonts}
\bibpunct{(}{)}{;}{a}{}{,}
\usepackage{color}

\begin{document}

\title{A photometric and spectroscopic survey of solar twin stars within 50
parsecs of the Sun:}

\subtitle{I. Atmospheric parameters and color similarity to the
Sun}

\author{G. F. Porto de Mello\inst{1}, R. da Silva\inst{1}\fnmsep
\thanks{Present address: INAF, Osservatorio Astronomico di Roma, 00040, Monte Porzio Catone, Italy}, L. da Silva\inst{2} \and R. V. de
Nader\inst{1}\fnmsep\thanks{Based on spectroscopic observations
collected at the Observat\'orio do Pico dos Dias (OPD), operated
by the Laborat\'orio Nacional de Astrof\'{\i}sica, CNPq, Brazil,
and the European Southern Observatory (ESO), within the ON/ESO and
ON/IAG agreements, under FAPESP project n$^{\circ}$
1998/10138-8.}}

\offprints{G. F. Porto de Mello, gustavo@astro.ufrj.br}

\institute{Universidade Federal do Rio de Janeiro, Observat\'orio
do Valongo, Ladeira do Pedro Antonio 43, CEP: 20080-090 Rio de Janeiro, RJ, Brazil\\
\email{gustavo@astro.ufrj.br,ronaldo.dasilva@oa-roma.inaf.it,rvnader@astro.ufrj.br}
\and Observat\'orio Nacional, Rua Gen.
Jos\'e Cristino 77, CEP: 20921-400, Rio de Janeiro, Brazil\\
\email{licio@on.br}}

\date{Received; accepted}

\authorrunning{Porto de Mello et al.}

\titlerunning{Spectroscopic and photometric survey of solar twins}

\abstract
{Solar twins and analogs are fundamental in the characterization
of the Sun's place in the context of stellar measurements, as they
are in understanding how typical the solar properties are in its
neighborhood. They are also important for representing sunlight
observable in the night sky for diverse photometric and
spectroscopic tasks, besides being natural candidates for
harboring planetary systems similar to ours and possibly even
life-bearing environments.}
{We report a photometric and spectroscopic survey of solar twin
stars within 50 parsecs of the Sun. Hipparcos absolute magnitudes
and ($B-V$)$^{\rm Tycho}$ colors were used to define a 2$\sigma$
box around the solar values, where 133 stars were considered.
Additional stars resembling the solar UBV colors in a broad sense,
plus stars present in the lists of Hardorp, were also selected.
All objects were ranked by a color-similarity index with respect
to the Sun, defined by $uvby$ and $BV$ photometry.}
{Moderately high-resolution, high-S/N spectra were used for a
subsample of equatorial-southern stars to derive \Teff, $\log{g}$
(both ionization and evolutionary), and spectroscopic $[$Fe/H$]$
with average internal errors better than 50 K, 0.20 dex, and 0.08
dex,respectively. Ages and masses were estimated from theoretical
HR diagrams.}
{The color-similarity index proved very successful, since none of
the best solar-analog and twin candidates that were
photometrically and spectroscopically found to be good solar
matches differed from the Sun by more than 3$\sigma$ in their
colors. We identify and quantitatively rank many new excellent
solar analogs, which are fit to represent the Sun in the night sky
to varying degrees of accuracy and in a wide range of contexts.
Some of them are faint enough (V$^{\rm Tycho}\sim$ 8.5) to be of
interest for moderately large telescopes. We also identify two
stars with near-UV spectra indistinguishable from the Sun's,
although only \object{HD\,140690} also has atmospheric parameters
matching the Sun's, besides a high degree of photometric fidelity.
This object is proposed as a prime solar analog in both the UV and
visible ranges, a rare object. We present five new ``probable''
solar twin stars, besides five new ``possible'' twins, the best
candidates being \object{HD\,98649}, \object{HD\,118598},
\object{HD\,150248}, and \object{HD\,164595}. Masses and
isochronal ages for the best solar twin candidates lie very close
to the solar values within uncertainties, but chromospheric
activity levels range somewhat. We propose that the solar twins be
emphasized in the ongoing searches for extra-solar planets and
SETI searches.}
{}


\maketitle

\section{Introduction}

The Sun occupies a very special place in stellar studies, being
still the most fundamental and dependable reference object in
stellar astrophysics. It remains the one star for which extremely
important parameters can be determined from first principles, such
as the effective temperature \Teff from directly observed
irradiance \citep{neckel1986}, age from nucleochronology and/or
undisturbed meteorite differentiates
\citep{guenther1989,gancarzwasserburg77}, and mass from planetary
motion and asteroseismology. Nevertheless, to fully exploit the
potential reference position of the Sun as a star, we need to know
its place in the parameter space of stellar observations. This is
no easy task: the extremely detailed observations available for
the Sun are only tied with difficulty to the woefully less
detailed stellar measurements. The main reasons for this in
photometry and spectrophotometry are the difficulties in
geometrically treating the solar image and scattered light; the
immense flux difference between the Sun and stars, a factor of
$\sim$10$^{\rm 11}$; and the impossibility of observing the Sun at
night, which taxes the stability of instrumentation. The
photometric properties of the Sun in the various photometric
systems in use, thus, remain uncertain despite protracted efforts
employing a variety of direct and indirect methods \citep[e.g.,][
and many others]{tugkaler1982, hayes1985, saxnerhammarback1985,
neckel1986, frieletal93, gray1992, gray1994,
portodemellodasilvatwin1997, ramirezmelendez2005b,
holmbergetal2006, casagrandeetal2006, melendezetal2010}.

The identification of stars closely resembling the Sun plays an
extremely interesting role in this task
\citep{cayreldestrobel1996} and raises considerable interest on
diverse astrophysical fronts. Solar twin stars were defined  by
\citet{cayrelbentolila1989} as (non binary) stars identical to the
Sun, within the observational uncertainties, in all fundamental
astrophysical parameters, such as mass, chemical composition,
\Teff, surface gravity, photospheric velocity fields, magnetic
activity, age, luminosity, and asteroseismological properties.
Such stars should have spectra that are indistinguishable from the
solar one. It is debatable whether such stars should be
detectable, or even actually exist
\citep{cayreldestrobeletal1981}. Even though uncertainties in
determining fundamental stellar parameters have been decreasing
steadily, minute differences from star to star may simply be too
small to be distinguished. For instance, very slight variances in
chemical composition or details of internal structure between two
stars can lead to sizable disparities of observable spectral
properties and evolutionary states, and turn them into very
dissimilar objects indeed.

Solar analogs, by contrast, are unevolved, or hardly evolved,
solar-type stars that merely share the solar atmospheric
parameters and are thus expected to have very similar colors and
spectral flux distributions to the Sun. We feel the distinction
between solar twins and analogs has not been sufficiently stressed
in the literature, and we thus take some time to point out some
key issues. Solar twins, on the one hand, are expected to match
every conceivable solar physical property, and therefore to
materialize in a star all the photometric and spectroscopic solar
properties, under the reasonable assumption that a perfect
physical match would automatically lead to the same observables.
Solar analogs, on the other hand, merely have the atmospheric
parameters loosely similar to the solar ones, to degrees of
similarity that have been taken at different values by different
authors (alas, adding to the confusion). Such stars are expected
to possess spectrophotometric quantities, including colors,
similar to the solar ones, but we note that, due to the various
degeneracies of the problem, which we discuss below, stars with
colors resembling the Sun may not turn out to be solar analogs.

Solar analogs, and of course also solar twins, may be very useful
in providing a proxy for sunlight in the night sky specifically
for spectrophotometry of solar system bodies and other calibration
purposes. These solar surrogates are very important for those
cases when techniques that can be applied in daytime, such as
observing the clear blue sky or solar radiation reflected off
telescope domes, are not an option. Ideally these solar analogs
should be faint enough for adequate use by large telescopes, and
be observable with the same instrumentation as used for working
with very faint targets, such as small and distant asteroids,
besides being observable without the need of stoppers or neutral
density filters, which always add some measure of uncertainty.
Additionally, both solar twins and solar analogs are expected to
help pin down the solar color indices better.

Moreover, solar twins may be expected to have followed an
evolutionary history similar to that of the Sun. There is some
evidence that the Sun may be metal-richer than the average G-type
thin disk star in its neighborhood \citep{rochapintomaciel96},
though we note that recent data has cast doubt upon this claim, as
judged by a revised solar \citep{asplundetal2009} and interstellar
medium \citep{nievaprzybilla2012} composition, as well as results
from nearby solar-type stars \citep{adibekyanetal2013}. It also
seems to be part of a stellar population that is heavily depleted
in lithium \citep[e.g.,][]{pasquinietal94,takedaetal2007}, and it
may possess lower-than-average chromospheric activity for its age
\citep{halllockwood2000,halletal2007}, have more subdued
photometric variability than stars with similar properties
\citep{lockwoodetal2007,radicketal1998} \citep[but
see][]{halletal2009}, and have a slightly longer rotational period
than stars of the same age \citep{pacepasquini2004}. In addition,
the Sun seems to lead most of the local stars of similar age and
metallicity in the velocity component towards the galactic
rotation \citep{cayreldestrobel1996,portodemelloetal2006}. Adding
to these putative peculiarities \citep[for an interesting review
of this topic, see][]{gustafsson1998}, the Sun occupies a position
very close to the Galactic corotation \citep{lepineetal2001},
whereby the Sun shares the rotational velocity of the spiral arms
and the number of passages through them is presumably minimized.
These characteristics may have a bearing on the Sun's ability to
maintain Earth's biosphere on long timescales
\citep{leitchvasisht98,gonzalezetal2001,portodemelloetal2009}.

Is the Sun an atypical star for its age and galactic position? A
sample of nearby solar twins may help gauge the solar status in
the local population of middle-aged G-type stars better. And, last
but not least, solar twin stars would be natural choices when
searching for planetary systems similar to our own, as well as
presenting interesting targets to the ongoing SETI programs
\citep{tarter2001} and the planned interferometric probes aimed at
detecting life, remotely, in extra
solar Earth-like planets by way
of biomarkers \citep{seguraetal2003}.

The search for solar analogs was initially stimulated by
\citet[and references therein]{hardorp1982} when attempting to
identify stars with UV spectra matching the solar one, as judged
mainly by the CN feature around $\lambda$3870. Hardorp classed
stars by magnitude differences of their spectral features to the
Sun's (represented by Galilean satellites), and his solar analog
lists are still widely referred to nowadays
\citep[e.g.,][]{alvarezcandaletal2007, milanietal2006}. This
prompted an effort by \citet{cayreldestrobeletal1981},
\citet{cayrelbentolila1989}, and \citet{frieletal93} to check that
Hardorp's best candidates stood up to detailed spectroscopic
analysis: this subject received a thorough review by
\cite{cayreldestrobel1996}. Subsequently,
\citet{portodemellodasilvatwin1997} used a detailed spectroscopic
and evolutionary state analysis to show that 18 Sco (\object{HR\,
6060}, \object{HD\, 146233}) was a nearly perfect match for the
Sun as judged by colors, chemical composition, \Teff, surface
gravity, luminosity, mass, and age, thereby confirming that the 16
Cyg A and B pair (\object{HD\,186408} and \object{HD\,186427}),
previously pointed to by the Cayrel de Strobel group as the best
solar twins, were older, less active, and more luminous than the
Sun, though possessing \Teff and metallicity very near the Sun's.
\citet{glushnevaetal2000} analyzed the spectral energy
distributions of solar analogs from Hardorp's lists, concluding
that 16 Cyg A and B are the closest matches to the solar
distribution, followed closely by 18 Sco, but, as did
\citet{portodemellodasilvatwin1997}, they found the two former
objects to be more luminous than the Sun, concluding that they are
not true solar twins. \citet{soubirantriaud2004} have analyzed
moderately high-resolution, homogeneous ELODIE spectra by
comparing the stars with spectra of Moon and Ceres in an automated
$\chi^2$ method measuring over 30 000 resolution elements. They
confirm that \object{HD\, 146233} is the best match for the Sun
and conclude that both photometric and spectroscopic data must be
assembled to find real solar twins. These authors also found a
very large dispersion in the published atmospheric parameters of
solar analog candidates. \citet{galeevetal2004} have
spectroscopically analyzed 15 photometric analogs of the Sun,
presenting \object{HD\, 146233} and \object{HD\, 186427} as the
best analogs, along with \object{HD\, 10307} and \object{HD\,
34411}, also concurring that photometric and spectroscopic data
must be merged for a precise determination of similarity to the
Sun.

\citet{kingetal2005} suggest that \object{HD\, 143436} is as good
a solar twin as \object{HD\, 146233}. \citet{melendezetal2006}
present \object{HD\, 98618} as another star as close to the solar
properties as \object{HD\, 146233}, and \citet{melendezetal2007}
claim that the best solar twins ever are \object{HD\, 101364} and
\object{HD\, 133600}, since they not only reproduce all the solar
fundamental parameters but also have a similar lithium abundance
\citep[see also ][]{melendezetal2012}. \citet{takedaetal2007} draw
attention to the importance of the lithium abundance as a record
of the stellar history of mixing and rotational evolution,
concluding that slow rotation induces greater depletion. Finally,
\cite{donascimentojretal2009} show that the lithium depletion
history of solar analogs is critically mass-dependent and suggest
that, among the proposed solar twins, the best match for the solar
convective properties, including the Li abundance, is \object{HD\,
98618}. This star also seems to fit the solar mass and age very
closely. Israelian et al.(2004, 2009) suggest that an enhanced
depletion of lithium is linked to the presence of planetary
companions; however, this claim has been questioned by
\cite{luckheiter2006}, \cite{ghezzietal2010}, and
\cite{baumannetal2010}. It is possible that the very low lithium
abundance of the Sun and other stars may be yet another piece of
the major observational and theoretical puzzle of planetary
formation.

As part of an ongoing effort at a complete survey of solar analog
stars nearer than 50 pc, this paper reports a volume-limited,
homogeneous, and systematic photometric and spectroscopic survey
of solar twin stars, approximately restricted to $\delta$ $\leq$
$+$30$^\circ$ in what pertains to spectroscopic observations. It
is, however, photometrically complete and all-sky within d $\leq$
40 pc and V$^{\rm Tycho}$ $\leq$ 8, and partially complete (owing
to lack of photometry) within 40 pc $<$ d $<$ 50 pc and 8 $<$
V$^{\rm Tycho}$ $<$ 9. The best candidates will be subjected to
detailed spectroscopic analysis that employs higher resolution
spectra in a forthcoming paper. In section 2 the selection of the
sample is described. In section 3 we describe the results of the
photometric similarity analysis. In section 4 the observational
data are presented, and the spectroscopic analysis is described in
section 5. In section 6 we discuss the spectroscopic results and
obtain masses and ages in an evolutionary analysis, presenting the
best candidates, and we draw the conclusions in section 7. A new
photometric calibration of \Teff on colors and metallicity, based
on IRFM data and tailored specifically to solar analog stars and
MARCS model atmosphere analysis, is presented in the Appendix.

\begin{figure}
\centering
\includegraphics[width=8.5cm]{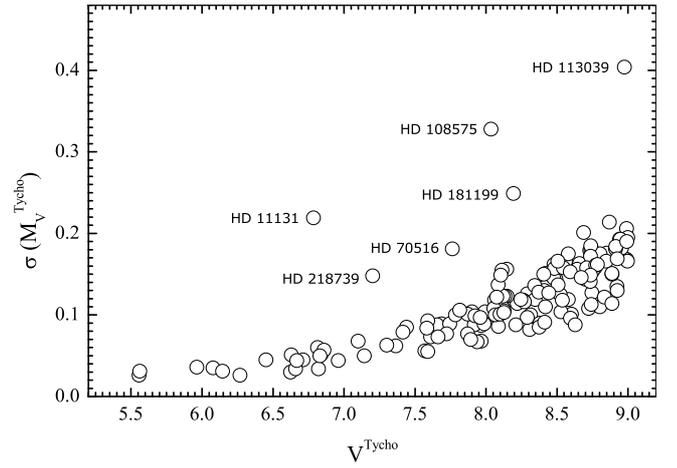}
\caption{Uncertainties in the M$_{\rm V}^{\rm Tycho}$ absolute
magnitudes for stars of the final sample of 158 non-binary stars
selected from the Hipparcos catalog. Outliers with large errors in
absolute magnitude are identified by HD numbers.}
\label{vtsigmamvt}
\end{figure}

\section{Sample selection}

In a solar-analog hunt, by the very nature of the objects, the
selection of candidates must be initiated photometrically by
colors and absolute magnitudes. The next step in the selection
process must be spectroscopic, since atmospheric parameters and
luminosities of the candidate objects will be compared to those of
the Sun. An important question, therefore, is the availability of
consistent \Teff scales where the Sun may be accurately placed
both photometrically and spectroscopically.

\cite{portodemelloetal2008} discussed, in their analysis of the
atmospheric parameters of the $\alpha$ Centauri binary system,
possible offsets between the various published photometric and
spectroscopic \Teff scales. While one conclusion was that there is
no consensus as yet on the existence of inconsistencies between
the two scales, there is evidence that for stars that are
substantially cooler than the Sun, \Teff $\leq$ 5300\,K, NLTE, and
other effects may be precluding strict consistency between them
\citep[e.g.,][]{yongetal2004}. For stars with \Teff that are not
too dissimilar from the Sun, good agreement can be expected
between photometric and spectroscopic \Teffs. This is an important
issue, since the properties of solar twins and analogs must be
equal to those of the Sun in a variety of contexts, such as in
narrow and broad-band photometry and in low and high-resolution
spectroscopy. On the other hand, a differential spectroscopic
approach allows direct comparison between Sun and stars.

In the appendix, we present a new photometric calibration for
solar-type stars for many colors in regular use, including the
($B-V$)$^{\rm Johnson}$, ($B-V$)$^{\rm Tycho}$, and Str\"omgren
($b-y$) indices, based on published \Teffs employing the infrared
flux method. The Paschen continuum colors have been
metallicity-calibrated using only spectroscopic metallicities from
detailed analyses. Our solar twin selection process starts from
the Hipparcos catalog \citep{hipparcos1997} photometry and is
subsequently refined with ($B-V$) and Str\"omgren color indices.
From the calibrations described in the Appendix, solar color
indices were derived and have been the basis for our selection of
solar twin candidates.

From our photometric calibrations, adopting for the Sun \Teff $=$
5777\,K \citep{neckel1986}, we obtain\\

($B-V$)$_\odot$$^{\rm Johnson}$ $=$ 0.653\\

($B-V$)$_\odot$$^{\rm Tycho}$ $=$ 0.737\\

($b-y$)$_\odot$ $=$ 0.409.\\

These values are in good agreement, within quoted errors, with the
determinations of \cite{holmbergetal2006} and
\cite{casagrandeetal2010}, for all three colors, and with
($b-y$)$_\odot$ as given by \cite{melendezetal2010}.

For the $m_{\rm 1}$ index, we adopted the same procedure as
employed by \cite{portodemellodasilvatwin1997} to derive the solar
($B-V$) and ($U-B$) colors. A sample of nine stars,
spectroscopically analyzed with homogeneous methods, with solar
metallicity, and a narrow range of \Teffs around the Sun leads to,
interpolating the solar \Teff in an $m_{\rm 1}$ $\it versus$ \Teff regression:\\

$m_{\rm 1}^\odot$ $=$ 0.217 (F catalog: see below).\\

This same procedure was applied to the Paschen continuum colors
and led to ($B-V$)$_\odot$$^{\rm Johnson}$ $=$ 0.648,
($B-V$)$_\odot$$^{\rm Tycho}$ $=$ 0.730, and ($b-y$)$_\odot$ $=$
0.406, in excellent agreement with the solar colors derived
directly from the photometric calibrations. Our $m_{\rm 1}^\odot$
also agrees very well with the recent derivation of
\cite{melendezetal2010}.

The initial selection process sets up a 2$\sigma$ box around the
($B-V$)$_\odot^{\rm Tycho}$ and the solar absolute magnitude in
the Tycho system, M$_{\rm V_{\odot}}^{\rm Tycho}$. To obtain the
latter figure, we compared the Sun to the solar twin \object{HD\,
146233} (18
Sco) \citep{portodemellodasilvatwin1997}, and set\\

M$_{\rm V_{18~Sco}}^{\rm Tycho}$ $-$ M$_{\rm V_{\odot}}^{\rm
Tycho}$ $=$ M$_{\rm V_{18~Sco}}^{\rm Johnson}$ $-$ M$_{\rm V_{\odot}}^{\rm Johnson}$.\\

Regarding the similarity between the V and V$^{\rm Tycho}$ bands
and the very slight absolute magnitude difference between the Sun
and \object{HD\, 146233} in the Johnson V band, this procedure
should not introduce any systematic error. We take M$_{\rm
V_{\odot}}$ $=$ 4.82 \citep{neckel1986}, and from the Hipparcos
parallax obtain M$_{\rm V_{18 Sco}}^{\rm Johnson}$ $=$ 4.77 and
M$_{\rm V_{18~Sco}}^{\rm Tycho}$ $=$ 4.83, to derive M$_{\rm
V_{\odot}}^{\rm Tycho}$ $=$ 4.88 $\pm$ 0.03, to which we formally
attach the uncertainty of M$_{\rm V_{18~Sco}}^{\rm Tycho}$.

The widths of the 2$\sigma$ ($B-V$)$^{\rm Tycho}$ vs. M$_{\rm
V}^{\rm Tycho}$ box were arrived at by an iterative procedure. The
uncertainties in ($B-V$)$^{\rm Tycho}$ and M$_{\rm V}^{\rm
Tycho}$, are obtained from the uncertainties of the B$^{\rm
Tycho}$ and V$^{\rm Tycho}$ bands respectively, and the
uncertainty in the parallax and the V$^{\rm Tycho}$ band.
Experimentation with arbitrary widths revealed that the average
uncertainties were a function of the magnitude limit, being
independent of the absolute magnitude and color indices of the
selected stars. The Hipparcos catalog is formally complete down to
V$^{\rm Tycho}$ $\sim$ 9.0, an apparent magnitude that translates
to a distance of 67 parsecs for a star with the same luminosity as
the Sun. The uncertainty in M$_{\rm V}^{\rm Tycho}$ increases
smoothly as magnitude increases, and there is a small
discontinuity at V$^{\rm Tycho}$ $\sim$ 8.0
(Fig.~\ref{vtsigmamvt}). At this magnitude, approximately, the
completeness limit of the $uvby$ catalogs also lies
\citep{olsen1983, olsen1993, olsen1994a, olsen1994b}; indeed, the
completeness of these catalogs was lost at V$^{\rm Tycho}$ $\sim$
8.1, for the samples selected in the first iterative runs. Our
sample was therefore divided at V$^{\rm Tycho}$ $=$ 8.0. The
2$\sigma$ limits of the box were chosen so that the box widths
corresponded to the average uncertainties of the ($B-V$)$^{\rm
Tycho}$ and M$_{\rm V}^{\rm Tycho}$ for the stars inside the box.
This was satisfied by $<$$\sigma$$>$ (M$_{\rm V}^{\rm Tycho}$) $=$
0.07 and $<$$\sigma$$>$ (($B-V$)$^{\rm Tycho}$) $=$ 0.013, for
V$^{\rm Tycho}$ $\leq$ 8.0 stars. The corresponding values for the
8.0 $<$ V$^{\rm Tycho}$ $<$ 9.0 stars are 0.013 and 0.020,
respectively, but the figures for V$^{\rm Tycho}$ $\leq$ 8.0 stars
were used to define both boxes. We chose to enforce strict
consistency for the brighter sample, for which $uvby$ photometry
is complete and for which better spectroscopic data could be
secured. After binary or suspected binary stars were removed from
the list, 158 stars were retained, 52 having V$^{\rm Tycho}$
$\leq$ 8.0. The completeness of the availability of the
($B-V$)$^{\rm Tycho}$ color in the Hipparcos catalog for V$^{\rm
Tycho}$ $\leq$ 8.0 stars of all spectral types is 92\%. This
figure increases to 95\% for G-type stars. A 2$\sigma$ box is thus
seen to be a practical limit that allows the working sample to be
observed spectroscopically in a reasonable amount of time.

To this sample we added some stars selected in the Bright Star
Catalog \citep{hoffleitjaschek1991, hoffleit1991} solely for
having both ($B-V$) and ($U-B$) colors similar to the solar ones,
plus a few stars from \citet{hardorp1982}. For the solar colors,
we used ($B-V$)$^\odot$ = 0.653 $\pm$ 0.003 (as above) and
($U-B$)$^\odot$ = 0.178 $\pm$ 0.013 \citep[the latter
from][]{portodemellodasilvatwin1997}. Some of the former stars
have also been considered by \citet{hardorp1982} to have UV
spectral features very similar to the solar ones.

\begin{figure*}
\begin{center}
\resizebox{8.5cm}{!}{\includegraphics{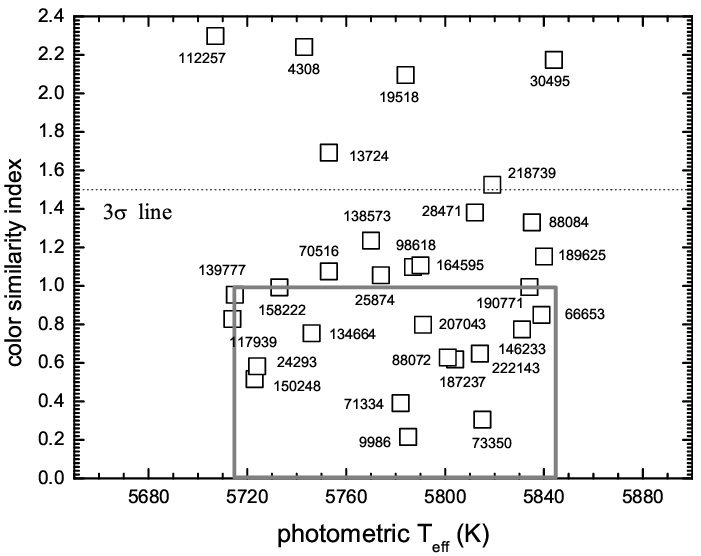}}
\resizebox{8.5cm}{!}{\includegraphics{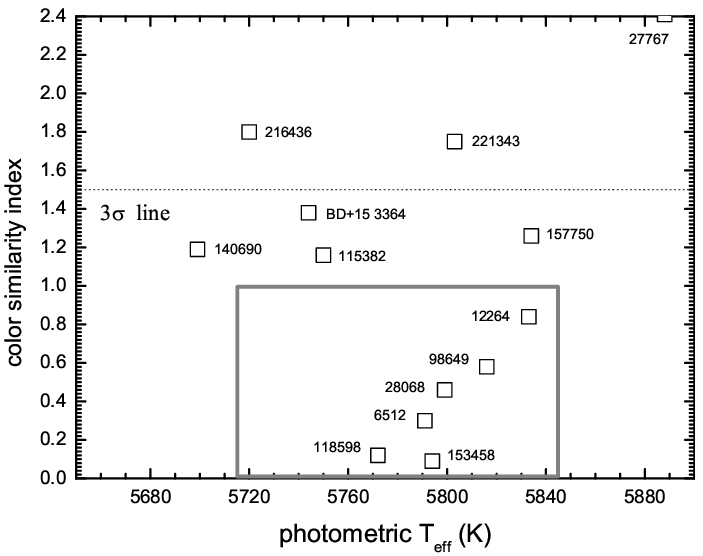}}
\resizebox{8.5cm}{!}{\includegraphics{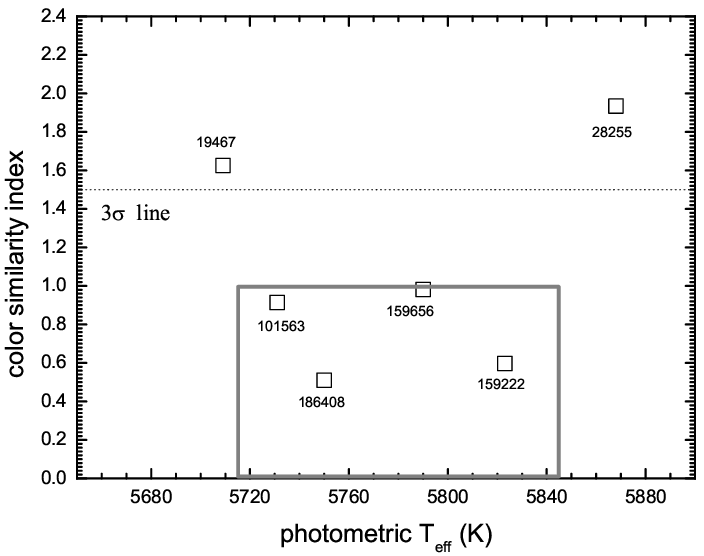}}
\end{center}
\caption[]{{\it Left}. The color similarity index S$_{\rm C}$
plotted {\it versus} the photometric \Teffs for the 52 stars with
V$^{\rm Tycho}$ $\leq$ 8.0. The box contains, in the S$_{\rm C}$
axis, stars with S$_{\rm C}$ $\leq$ 1.00 within 2$\sigma$ of the
solar one (defined as zero) in the ordinate. The width of the box
is set by the $\sigma$(\Teff$^{\rm phot}$) = 65 K uncertainty in
the photometric \Teffs. The dotted horizontal line defines the
3$\sigma$ limit in S$_{\rm C}$. Stars are identified by HD
numbers. {\it Right}. The same as the left panel, but for the 8.0
$<$ V$^{\rm Tycho}$ $\leq$ 9.0 stars, only those observed
spectroscopically. {\it Below}. Same as above for stars selected
by their UBV similarity to the Sun, presence in the lists of
\cite{hardorp1982}, or both.} \label{scindex}
\end{figure*}

\section{Photometric similarity index}

The two-dimensional 2$\sigma$ ((B-V)$^{\rm Tycho}$, M$_{\rm
V}^{\rm Tycho}$) box would automatically select solar twins were
it not for the metallicity dimension, since surface gravity
effects are negligible in the narrow color-magnitude interval
involved here. Stars selected by means of color have a dispersion
in \Teff corresponding to a dispersion in [Fe/H]. We use
throughout the notation $[$A/B$]$ = log N(A)/N(B)$_{\rm star}$ -
log N(A)/N(B)$_{\rm Sun}$, where N denotes the number abundance of
a given element. Thus, a metal-rich star cooler than the Sun may
mimic the solar colors, as may a star hotter than the Sun but
metal-richer. To narrow down further the candidate list to be
observed spectroscopically, we defined a color-similarity index
S$_{\rm C}$ with respect to the Sun:

\begin{equation}
\label{simil_fot} S_{\rm C} = \alpha \sum_{C_i} \left\{
\frac{\big({C_i}^\bigstar -
{C_i}^\odot\big)^2}{\big(\overline{\sigma}_{C_i}\big)^2} \right\}
\end{equation}

\noindent where $C_i$ represents the color indices $(B-V)$,
($B-V$)$^{\rm Tycho}$, $(b-y)$, and $m_{\rm 1}$, and $\alpha$ is
an arbitrary normalization constant. The last color is essentially
a photometric metallicity dimension, while the three previous
colors are independent measurements of the stellar Paschen
continuum. This index then simultaneously reflects the gradient of
the Paschen continuum and the strength of metal lines. Attempts to
employ the $\beta$ and ($V-K$) colors in the definition of the
index had to be abandoned owing to the large incompleteness of
such data for the program stars. The index S$_{\rm C}$ expresses a
simple sum of quadratic differences with respect to the adopted
solar colors, weighted by the average error
of each color. The average color errors for the V$^{\rm Tycho}$ $\leq$ 8.0 stars are\\

$<$$\sigma$$>$($B-V$)$^{\rm Johnson}$ $=$ 0.009\\

$<$$\sigma$$>$ ($B-V$)$^{\rm Tycho}$ $=$ 0.013\\

$<$$\sigma$$>$ ($b-y$) $=$ 0.003\\

$<$$\sigma$$>$ ($m_{\rm 1}$) $=$ 0.005.\\

The ($B-V$)$^{\rm Johnson}$ and ($B-V$)$^{\rm Tycho}$ errors were
directly obtained from the Hipparcos catalog, and the ($b-y$) and
$m_{\rm 1}$ errors are given by  Olsen (1983, 1993, 1994a, 1994b)
for each object. For the 106 stars within 8.0 $<$ V$^{\rm Tycho}$
$<$ 9.0, $uvby$ photometry is only available for
68 stars. The corresponding average color errors for the fainter targets are\\

$<$$\sigma$$>$ ($B-V$)$^{\rm Johnson}$ $=$ 0.013\\

$<$$\sigma$$>$ ($B-V$)$^{\rm Tycho}$ $=$ 0.020\\

$<$$\sigma$$>$ ($b-y$) $=$ 0.003\\

$<$$\sigma$$>$ ($m_{\rm 1}$) $=$ 0.005.\\

\citet{olsen1993} discusses systematic discrepancies between the
photometry of southern stars (which he calls the "F" catalog of
Olsen 1983) and northern stars \citep[the "G"
catalog]{olsen1993,olsen1994a,olsen1994b}, providing
transformations for the homogenization of the photometry. We
employed these transformations to convert all the ($b-y$) and
$m_{\rm 1}$ indices to the "F" catalog of \citet{olsen1983}, since
more than half of our prime targets, the V$^{\rm Tycho}$ $\leq$
8.0 stars, have their photometry in this catalog. Since the color
similarity index will be an important tool in the forthcoming
discussion, it is very important that the reader keeps in mind
that all the Str\"omgren photometry discussed here is compatible
with the catalog of \citet{olsen1983}.

The color-similarity index was computed for the sample taking the
different color errors for the V$^{\rm Tycho}$ $\leq$ 8.0 and 8.0
$<$ V$^{\rm Tycho}$ $\leq$ 9.0 stars into account. The solar
colors obviously correspond to an index S$_{\rm C}$ $=$ 0.00, and
the $\alpha$ constant was adjusted in each case so that a
2$\sigma$ uncertainty in S$_{\rm C}$ was equal to unity. This was
obtained by inserting the solar color themselves into the index
equation, added by twice their corresponding errors.

\begin{figure}
\centering
\includegraphics[width=8.5cm]{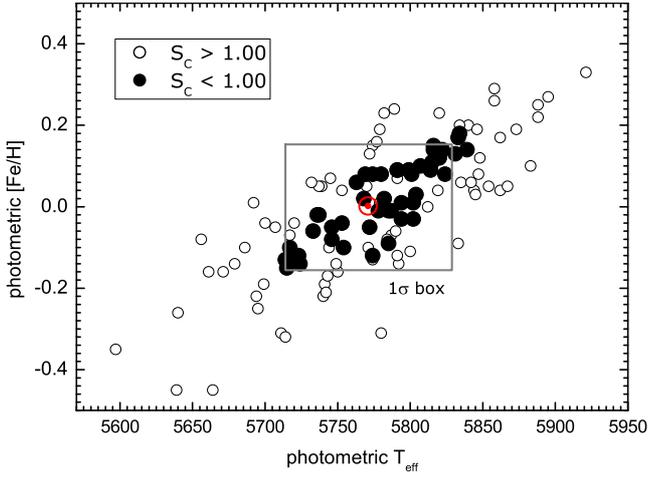}
\caption{Photometric \Teffs and [Fe/H]s, separated by the color
similarity index S$_{\rm C}$. The gray box limits the 1$\sigma$
errors in the photometric \Teffs and [Fe/H]s, and is centered on
the solar parameters (the Sun is identified by its usual symbol).
Black circles are those stars within a 2$\sigma$ color similarity
with the Sun.} \label{teff-feh-phot}
\end{figure}

In Fig~\ref{scindex} the color similarity index for the V$^{\rm
Tycho}$ $\leq$ 8.0 stars with S$_{\rm C}$ $\leq$ 2.4 is shown. The
stars located inside the 2$\sigma$ box are the best candidates, in
principle, since they have the Tycho absolute magnitude and color
compatible with the solar ones within a formal 2$\sigma$ limit. A
similar diagram was obtained for the 8.0 $<$ V$^{\rm Tycho}$
$\leq$ 9.0 stars. The number of stars within the 2$\sigma$ boxes
for each case was 16 for the brighter and 28 for the fainter
candidates. As an initial estimation of the atmospheric
parameters, the $m_{\rm 1}$-[Fe/H] calibration of
\citet{mcnamarapowell1985}, along with the photometric \Teff
calibrations for ($B-V$), ($B-V)^{\rm Tycho}$ and ($b-y$) detailed
in the Appendix, were used to obtain photometric \Teff and [Fe/H]
parameters. Though superseded by recent works, the relation of
\citet{mcnamarapowell1985} provides a simple linear $m_{\rm
1}$-[Fe/H] relationship for solar-type stars, which is very
accurate in a narrow interval around the Sun and well linked to
the Hyades iron abundance [Fe/H] $= +$0.12 \citep{paulsonetal2003,
cayreletal1985}. These photometrically derived atmospheric
parameters are plotted in Fig~\ref{teff-feh-phot} for all the
sample stars for which $uvby$ photometry is available and the
S$_{\rm C}$ defined. The solar colors, once entered into this set
of calibrations, yields \Teff$_\odot$ $=$ 5774\,K and [Fe/H] $=$
-0.03, assuring us that no significant systematic error is
incurred by the procedure. The photometrically derived [Fe/H]s,
when compared with the final spectroscopic ones (see section 4),
define a linear regression with a low correlation coefficient of R
$=$ 0.32, but no systematic deviation from the identity line.
Under the reasonable hypothesis of statistical independence, the
total dispersion of the regression, $\sigma$ $=$ 0.16 dex, yields
$\sigma$([Fe/H]$^{\rm phot}$) $=$ 0.14 dex, when the errors of the
spectroscopically determined [Fe/H] are taken into account
(section 4). When entered into the \Teff calibrations, this
uncertainty in [Fe/H]$^{\rm phot}$, along with the already
determined errors in the colors, fixes the uncertainty of the
photometrically derived \Teffs. We obtained $\sigma$(\Teff$^{\rm
phot}$) $=$ 65 K, a value that surpasses the expected internal
dispersion of \Teffs derived from the ($B-V$), ($B-V)^{\rm Tycho}$
and ($b-y$) calibrations, as discussed in the Appendix. This can
be explained by the larger errors in colors and [Fe/H] of the
sample stars, as compared with the much brighter stars used to
obtain the photometric calibrations. For stars not observed
spectroscopically (section 4), these are our final determination
of the atmospheric parameters and absolute magnitudes. For all the
sample, the colors, color similarity indices, along with the
photometrically derived \Teff, [Fe/H], and absolute magnitudes (in
the V$^{\rm Tycho}$ band) are shown in Tables 1, 2, and 3.

In Fig~\ref{teff-feh-phot}, those stars with a 2$\sigma$ color
similarity with the Sun are clearly contained within the box that
defines atmospheric parameters \Teff and [Fe/H] within 65 K and
0.14 dex, respectively, from the solar values. It will be shown
below that the color similarity index is indeed very efficient in
selecting stars which not only are photometrically similar to the
Sun, but also possess atmospheric parameters contained within a
narrow interval of the solar ones, as determined from a
spectroscopic model atmospheres analysis. This approach allows a
fast and convenient photometric selection of stars resembling any
desired set of colors and atmospheric parameters, expediently
diminishing the necessity of spectroscopic follow-up, provided
that the colors are sufficiently accurate.

\begin{table*} \centering \caption[]{Photometric and spectroscopic data for the
program stars with $V_{\rm T}$ $\leq$ 8.0. The second to sixth
columns provide $V^{\rm Tycho}$, ($B-V$)$^{\rm Johnson}$,
($B-V$)$^{\rm Tycho}$, ($b-y$), and $m_{\rm 1}$, respectively. The
seventh column gives the photometric similarity index with respect
to the solar colors (see text). Columns 8 and 9 list the
photometrically derived effective temperature \Teff and
metallicity. Columns 10 and 11 list the visual absolute magnitude
from the Tycho $V_{\rm T}$ band and its corresponding uncertainty,
respectively. Columns 12, 13, and 14 list the S/N ratios of the
spectroscopic observations as follows: $\lambda$6145 and
$\lambda$6563 ranges from the OPD coud\'e spectrograph; FEROS
spectrograph. The last column merely notes if low-resolution UV
spectra from the Boller-Chivens Cassegrain spectrograph of the ESO
1.52m telescope have been obtained.}
 \label{parphot1}
{\tiny
\begin{tabular}{ l @{\hspace{1em}} c @{\hspace{1em}} c @{\hspace{1em}} c @{\hspace{1em}} c @{\hspace{1em}}
c @{\hspace{1em}} c @{\hspace{1em}} c @{\hspace{1em}} c
@{\hspace{1em}} c @{\hspace{1em}} c @{\hspace{1em}} c
@{\hspace{1em}} c @{\hspace{1em}} c @{\hspace{1em}} c
@{\hspace{1em}} c @{\hspace{1em}} c }

\noalign{\smallskip} \hline \hline \noalign{\smallskip}

HD  & $V^{\rm Tycho}$ & $(B-V)^{\rm Johnson}$ & $(B-V)^{\rm
Tycho}$ & $(b-y)$ & $m_{\rm 1}$ & $S_{\rm C}$ & \Teff$^{\rm phot}$
& [Fe/H]$_{\rm phot}$ & $M_{\rm V}^{\rm Tycho}$ & $\sigma$
($M_{\rm V}^{\rm Tycho}$) & $\lambda$6145 & $\lambda$6563 & FEROS
& UV \cr

\noalign{\smallskip} \hline \noalign{\smallskip}

1835   & 6.5 & 0.659 & 0.758 & 0.422 & 0.225 & 4.03  & 5686 &
$-$0.10 & 4.90 & 0.05 & --  & --  & --  & \cr

4308 & 6.6 & 0.655 & 0.723 & 0.402 & 0.198 & 2.24  & 5743 &
$-$0.17 & 4.93 & 0.04 & 280 & 180 & --  & $\surd$ \cr

9986 & 6.8 & 0.648 & 0.720 & 0.408 & 0.217 & 0.22  & 5785 &
$-$0.01 & 4.79 & 0.06 & --  & --  & 220 & \cr

11131 & 6.8 & 0.654 & 0.711 & 0.403 & 0.188 & 4.85  & 5711 &
$-$0.31 & 4.98 & 0.23 & --  & --  & --  & \cr

13724 & 8.0 & 0.667 & 0.746 & 0.414 & 0.228 & 1.69  & 5753 & +0.04
& 4.78 & 0.10 & 70 & 140 & --  & $\surd$ \cr

18757  & 6.7 & 0.634 & 0.729 & 0.417 & 0.191 & 5.40  & 5639 &
$-$0.45 & 4.92 & 0.05 & -- & --  & --  & \cr

19518 & 7.9 & 0.642 & 0.731 & 0.399 & 0.203 & 2.10  & 5784 &
$-$0.08 & 4.84 & 0.11 & 120 & --  & 450 & $\surd$ \cr

24293 & 7.9 & 0.658 & 0.727 & 0.412 & 0.211 & 0.58  & 5724 &
$-$0.14 & 4.79 & 0.10 & 60 & 90 & --  & $\surd$ \cr

25680 & 6.0 & 0.620 & 0.720 & 0.397 & 0.212 & 3.45  & 5854 & +0.05
& 4.85 & 0.05 & --  & --  & -- & \cr

25874 & 6.8 & 0.667 & 0.747 & 0.410 & 0.227 & 1.06 & 5774 & +0.08
& 4.77 & 0.04 & 170 & 170 & --  & \cr

28471 & 8.0 & 0.650 & 0.717 & 0.399 & 0.210 & 1.38  & 5812 & +0.00
& 4.79 & 0.08 & --  & --  & --  & \cr

28701 & 7.9 & 0.650 & 0.710 & 0.402 & 0.186 & 5.64  & 5714 &
$-$0.32 & 4.76 & 0.08 & --  & --  & --  & \cr

28821 & 7.7 & 0.683 & 0.747 & 0.422 & 0.205 & 5.31  & 5597 &
$-$0.35 & 4.85 & 0.10 & 80 & 170 & --  & $\surd$ \cr

30495 & 5.6 & 0.632 & 0.710 & 0.399 & 0.213 & 2.18  & 5844 & +0.04
& 4.94 & 0.04 & 260 & 380 & --  & \cr

32963  & 7.7 & 0.664 & 0.743 & 0.403 & 0.237 & 2.46  & 5858 &
+0.29 & 4.92 & 0.10 & -- & --  & 410 & $\surd$ \cr

35041  & 7.7 & 0.636 & 0.731 & 0.395 & 0.213 & 2.58  & 5847 &
+0.08 & 4.89 & 0.10 & -- & --  & 250 & $\surd$ \cr

37773  & 7.8 & 0.692 & 0.746 & 0.429 & 0.236 & 10.90 & 5656 &
$-$0.08 & 4.90 & 0.11 & -- & --  & 610 & $\surd$ \cr

64184 & 7.6 & 0.675 & 0.755 & 0.420 & 0.218 & 3.45  & 5661 &
$-$0.16 & 4.95 & 0.07 & --  & --  & --  & \cr

66653 & 7.6 & 0.655 & 0.726 & 0.400 & 0.222 & 0.85  & 5839 & +0.14
& 4.80 & 0.06 & 130 & --  & 560 & $\surd$ \cr

68168 & 7.4 & 0.667 & 0.729 & 0.419 & 0.224 & 2.52  & 5717 &
$-$0.07 & 4.79 & 0.09 & -- & --  & 630 & $\surd$ \cr

70516 & 7.8 & 0.652 & 0.725 & 0.415 & 0.222 & 1.08  & 5753 &
$-$0.04 & 4.93 & 0.19 & -- & --  & --  & \cr

71334  & 7.9 & 0.643 & 0.746 & 0.408 & 0.220 & 0.39  & 5782 &
+0.02 & 4.92 & 0.09 & -- & 220 & 660 & $\surd$ \cr

73350  & 6.8 & 0.655 & 0.731 & 0.405 & 0.224 & 0.31 & 5815 & +0.11
& 4.95 & 0.07 & 180 & 210 & 700 & \cr

76151 & 6.1 & 0.661 & 0.752 & 0.410 & 0.239 & 2.93 & 5820 & +0.23
& 4.91 & 0.04 & --  & --  & --  & \cr

77006 & 8.0 & 0.651 & 0.707 & 0.384 & 0.202 & 9.07 & 5867 & +0.05
& 4.82 & 0.11 & --  & --  & -- & \cr

86226 & 8.0 & 0.647 & 0.719 & 0.381 & 0.204 & 10.48 & 5883 & +0.10
& 4.85 & 0.10 & -- & --  & --  & \cr

88072 & 7.6 & 0.647 & 0.753 & 0.404 & 0.221 & 0.63  & 5801 & +0.08
& 4.74 & 0.10 & --  & --  & 250 & \cr

88084 & 7.6 & 0.649 & 0.712 & 0.399 & 0.215 & 1.33  & 5835 & +0.06
& 4.87 & 0.08 & -- & --  & -- & \cr

98618  & 7.7 & 0.642 & 0.713 & 0.404 & 0.208 & 1.10 & 5787 &
$-$0.07 & 4.78 & 0.09 & -- & --  & --  & \cr

108575 & 8.0 & 0.68  & 0.745 & 0.417 & 0.234 & 3.37  & 5745 &
+0.07 & 4.90 & 0.34 & --  & --  & -- & \cr

112257 & 7.9 & 0.665 & 0.758 & 0.417 & 0.223 & 2.30 & 5707 &
$-$0.05 & 4.76 & 0.11 & --  & --  & --  & \cr

114174 & 6.9 & 0.667 & 0.757 & 0.418 & 0.233 & 3.72  & 5737 &
+0.05 & 4.76 & 0.07 & -- & --  & --  & \cr

117939 & 7.4 & 0.669 & 0.738 & 0.409 & 0.208 & 0.83  & 5714 &
$-$0.13 & 4.96 & 0.07 & 140 & 120 & --  & \cr

134664 & 7.8 & 0.662 & 0.739 & 0.404 & 0.207 & 0.76  & 5746 &
$-$0.08 & 4.83 & 0.12 & 100 & 180 & --  & \cr

138573 & 7.3 & 0.656 & 0.745 & 0.413 & 0.228 & 1.24  & 5770 &
+0.05 & 4.85 & 0.07 & 130 & 120 & -- & \cr

139777 & 6.7 & 0.665 & 0.723 & 0.413 & 0.211 & 0.95 & 5715 &
$-$0.15 & 4.94 & 0.04 & -- & --  & --  & \cr

142072 & 7.9 & 0.670 & 0.749 & 0.420 & 0.210 & 3.21  & 5640 &
$-$0.26 & 4.83 & 0.11 & 120 & 170 & --  & \cr

145825 & 6.6 & 0.646 & 0.727 & 0.395 & 0.228 & 2.74  & 5895 &
+0.27 & 4.93 & 0.06 & 180 & 230 & --  & \cr

146233 & 5.6 & 0.652 & 0.736 & 0.400 & 0.221 & 0.78  & 5831 &
+0.13 & 4.83 & 0.04 & 340 & 230 & 380 & $\surd$ \cr

150248 & 7.1 & 0.653 & 0.740 & 0.412 & 0.212 & 0.52  & 5723 &
$-$0.12 & 4.83 & 0.08 & 180 & 200 & --  & \cr

155114 & 7.6 & 0.637 & 0.720 & 0.396 & 0.197 & 4.27  & 5791 &
$-$0.12 & 4.85 & 0.09 & 140 & 120 & --  & $\surd$ \cr

158222 & 7.9 & 0.667 & 0.727 & 0.414 & 0.219 & 0.99  & 5733 &
$-$0.06 & 4.83 & 0.08 & -- & -- & --  & \cr

164595 & 7.1 & 0.635 & 0.722 & 0.404 & 0.209 & 1.11 & 5790 &
$-$0.06 & 4.84 & 0.06 & 190 & 160 & --  & \cr

187237 & 7.0 & 0.660 & 0.718 & 0.402 & 0.215 & 0.62  & 5804 &
+0.03 & 4.88 & 0.05 & -- & -- & 100 & \cr

189567 & 6.1 & 0.648 & 0.718 & 0.399 & 0.199 & 2.76 & 5774 &
$-$0.13 & 4.90 & 0.04 & -- & --  & -- & \cr

189625 & 7.4 & 0.654 & 0.729 & 0.406 & 0.232 & 1.15 & 5840 & +0.20
& 4.74 & 0.09 & 150 & 110 & --  & $\surd$ \cr

190771 & 6.3 & 0.654 & 0.732 & 0.406 & 0.231 & 1.00  & 5834 &
+0.18 & 4.89 & 0.04 & 290 & 150 & -- & $\surd$ \cr

202628 & 6.8 & 0.637 & 0.710 & 0.396 & 0.210 & 2.83 & 5845 & +0.03
& 4.95 & 0.06 & -- & -- & --  & \cr

207043 & 7.7 & 0.660 & 0.737 & 0.410 & 0.228 & 0.80  & 5791 &
+0.09 & 5.01 & 0.08 & 130 & 160 & -- & \cr

214385 & 8.0 & 0.640 & 0.723 & 0.403 & 0.195 & 3.03  & 5740 &
$-$0.22 & 4.97 & 0.11 & 120 & -- & --  & $\surd$ \cr

218739 & 7.2 & 0.658 & 0.716 & 0.398 & 0.212 & 1.53  & 5819 &
+0.04 & 4.86 & 0.16 & -- & -- & -- & \cr

222143 & 6.7 & 0.665 & 0.724 & 0.402 & 0.220 & 0.65 & 5814 & +0.09
& 4.85 & 0.05 & --  & --  & --  & \cr

\noalign{\smallskip} \hline \hline
\end{tabular}
}
\end{table*}

\begin{table*}
\centering \caption[]{The same as Table~\ref{parphot1} for the 70
program stars with 8.0 $< V_{\rm T}$ $\leq$ 9.0 and available
$uvby$ photometry. Owing to a management error, we also observed
two stars not inside the Hipparcos 2$\sigma$ box, \object{HD\,
140690}, and \object{HD\, 216436}, plus one star inside the box,
but without $uvby$ photometry, \object{HD\, 8291}: the [Fe/H] of
the latter was arbitrarily set to solar.} \label{parphot2}

{\tiny
\begin{tabular}{ l @{\hspace{1em}} c @{\hspace{1em}} c @{\hspace{1em}} c @{\hspace{1em}} c @{\hspace{1em}}
c @{\hspace{1em}} c @{\hspace{1em}} c @{\hspace{1em}} c
@{\hspace{1em}} c @{\hspace{1em}} c @{\hspace{1em}} c
@{\hspace{1em}} c @{\hspace{1em}} c @{\hspace{1em}} c
@{\hspace{1em}} c @{\hspace{1em}} c }\\[-0.65cm]

\noalign{\smallskip} \hline \hline \noalign{\smallskip}

HD  & $V^{\rm Tycho}$ & $(B-V)^{\rm Johnson}$ & $(B-V)^{\rm
Tycho}$ & $(b-y)$ & $m_{\rm 1}$ & $S_{\rm C}$ & \Teff$^{\rm phot}$
& [Fe/H]$_{\rm phot}$ & $M_{\rm V}^{\rm Tycho}$ & $\sigma$
($M_{\rm V}^{\rm Tycho}$) & $\lambda$6145 & $\lambda$6563 & FEROS
& UV \cr

\noalign{\smallskip} \hline \noalign{\smallskip}

3810 & 8.8 & 0.638 & 0.713 & 0.396 & 0.196 & 4.19  & 5792 &
$-$0.14 & 4.96 & 0.17 & --  & --  & --  & \cr

6512 & 8.2 & 0.656 & 0.746 & 0.407 & 0.224 & 0.30  & 5791 &
$+$0.09 & 4.81 & 0.12 & --  & --  & --  & $\surd$ \cr

8291 & 8.7 & 0.638 & 0.736 & --    & --    & --    & 5794 &
$+$0.00 & 4.83 & 0.17 & 90  & 80 & --  & $\surd$ \cr

7678 & 8.3 & 0.646 & 0.718 & 0.399 & 0.223 & 1.18  & 5862 &
$+$0.17 & 4.91 & 0.11 & --  & --  & --  & \cr

12264 & 8.1 & 0.660 & 0.736 & 0.401 & 0.225 & 0.84  & 5833 &
$+$0.17 & 4.85 & 0.13 & 80  & 110 & --  & $\surd$ \cr

15507 & 8.7 & 0.670 & 0.749 & 0.423 & 0.240 & 6.46  & 5732 &
$+$0.06 & 4.87 & 0.14 & --  & --  & --  & \cr

15632 & 8.1 & 0.666 & 0.749 & 0.413 & 0.222 & 0.81  & 5736 &
$-$0.02 & 4.99 & 0.12 & --  & --  & --  & \cr

17439 & 8.7 & 0.668 & 0.747 & 0.416 & 0.237 & 3.33  & 5772 &
$+$0.13 & 5.00 & 0.12 & --  & --  & --  & \cr

19617 & 8.8 & 0.682 & 0.746 & 0.430 & 0.245 & 11.89 & 5692 &
$+$0.01 & 4.99 & 0.19 & --  & --  & --  & \cr

21543 & 8.3 & 0.619 & 0.713 & 0.388 & 0.176 & 14.44 & 5780 &
$-$0.31 & 5.00 & 0.14 & --  & --  & --  & \cr

26736 & 8.1 & 0.657 & 0.709 & 0.407 & 0.232 & 1.31  & 5846 &
$+$0.19 & 4.83 & 0.13 & --  & --  & --  &  \cr

26767 & 8.1 & 0.640 & 0.724 & 0.395 & 0.224 & 2.41  & 5888 &
$+$0.22 & 4.88 & 0.17 & --  & --  & --  & $\surd$ \cr

27857 & 8.1 & 0.657 & 0.715 & 0.402 & 0.213 & 0.53  & 5802 &
$+$0.01 & 4.81 & 0.13 & --  & --  & --  & \cr

28068 & 8.1 & 0.651 & 0.734 & 0.409 & 0.226 & 0.46  & 5799 &
$+$0.09 & 4.80 & 0.16 & --  & --  & --  & $\surd$ \cr

31130 & 8.9 & 0.655 & 0.730 & 0.407 & 0.225 & 0.32  & 5807 &
$+$0.10 & 4.97 & 0.14 & --  & --  & --  & \cr

34599 & 8.4 & 0.660 & 0.737 & 0.405 & 0.222 & 0.21  & 5799 &
$+$0.09 & 5.00 & 0.10 & --  & --  & --  & \cr

35769 & 8.7 & 0.689 & 0.755 & 0.412 & 0.241 & 4.28  & 5789 &
$+$0.24 & 4.98 & 0.19 & --  & --  & --  &  \cr

36152 & 8.3 & 0.657 & 0.752 & 0.408 & 0.224 & 0.37  & 5780 &
$+$0.08 & 4.99 & 0.13 & --  & --  & --  & \cr

41708 & 8.1 & 0.626 & 0.712 & 0.393 & 0.208 & 3.84  & 5862 &
$+$0.04 & 4.85 & 0.13 & --  & --  & --  & \cr

43180 & 8.4 & 0.658 & 0.734 & 0.390 & 0.212 & 4.16  & 5848 &
$+$0.12 & 4.87 & 0.10 & --  & --  & --  & $\surd$ \cr

45346 & 8.7 & 0.661 & 0.738 & 0.399 & 0.219 & 0.95  & 5820 &
$+$0.12 & 4.78 & 0.12 & --  & --  & --  & \cr

75288 & 8.6 & 0.673 & 0.754 & 0.417 & 0.243 & 5.17  & 5779 &
$+$0.19 & 4.75 & 0.13 & --  & --  & --  & \cr

76332 & 8.6 & 0.620 & 0.735 & 0.405 & 0.213 & 0.99  & 5802 &
$-$0.03 & 4.99 & 0.17 & --  & --  & --  & \cr

78130 & 8.8 & 0.674 & 0.755 & 0.413 & 0.236 & 2.75  & 5777 &
$+$0.16 & 4.85 & 0.16 & --  & --  & --  & \cr

78660 & 8.4 & 0.665 & 0.724 & 0.406 & 0.227 & 0.62  & 5816 &
$+$0.14 & 4.91 & 0.14 & --  & --  & --  & \cr

81700 & 8.6 & 0.650 & 0.733 & 0.402 & 0.211 & 0.53  & 5787 &
$-$0.01 & 4.89 & 0.11 & --  & --  & --  & \cr

90322 & 8.8 & 0.641 & 0.711 & 0.404 & 0.207 & 0.90  & 5785 &
$-$0.09 & 4.98 & 0.18 & --  & --  & --  & \cr

90333 & 8.4 & 0.676 & 0.738 & 0.404 & 0.216 & 0.49  & 5768 &
$+$0.02 & 4.91 & 0.12 & --  & --  & --  & \cr

93215 & 8.1 & 0.670 & 0.758 & 0.413 & 0.235 & 2.50  & 5774 &
$+$0.15 & 4.75 & 0.12 & --  & --  & --  & \cr

98649 & 8.1 & 0.658 & 0.741 & 0.405 & 0.227 & 0.58  & 5816 &
$+$0.15 & 4.91 & 0.11 & 120 & 140 & 420 & \cr

105901 & 8.3 & 0.626 & 0.728 & 0.395 & 0.197 & 4.58  & 5800 &
$-$0.11 & 4.78 & 0.13 & 110 & 250 & 240 & \cr

110668 & 8.3 & 0.672 & 0.752 & 0.408 & 0.223 & 0.54  & 5763 &
$+$0.06 & 4.74 & 0.15 & --  & --  & --  & \cr

110869 & 8.1 & 0.662 & 0.748 & 0.414 & 0.223 & 0.97  & 5737 &
$-$0.02 & 4.81 & 0.10 & --  & --  & --  & \cr

110979 & 8.1 & 0.654 & 0.751 & 0.408 & 0.214 & 0.16  & 5746 &
$-$0.05 & 4.77 & 0.13 & --  & --  & --  & \cr

111069 & 8.7 & 0.630 & 0.720 & 0.399 & 0.236 & 3.17  & 5921 &
$+$0.33 & 4.87 & 0.19 & --  & --  & --  & \cr

111938 & 8.5 & 0.632 & 0.713 & 0.407 & 0.207 & 0.98  & 5774 &
$-$0.12 & 4.91 & 0.13 & --  & --  & --  & \cr

115231 & 8.5 & 0.667 & 0.756 & 0.419 & 0.217 & 2.29  & 5671 &
$-$0.16 & 4.86 & 0.15 & --  & --  & --  & \cr

115382 & 8.5 & 0.630 & 0.731 & 0.408 & 0.205 & 1.16  & 5750 &
$-$0.16 & 4.80 & 0.17 & 140 & 110 & 330 & \cr

118598 & 8.3 & 0.652 & 0.721 & 0.407 & 0.213 & 0.12  & 5772 &
$-$0.05 & 4.84 & 0.13 & 130 & 130 & --  & \cr

121205 & 9.0 & 0.675 & 0.756 & 0.395 & 0.212 & 2.61  & 5791 &
$+$0.07 & 4.87 & 0.19 & --  & --  & --  & \cr

123682 & 8.4 & 0.690 & 0.756 & 0.410 & 0.209 & 1.56  & 5679 &
$-$0.14 & 4.93 & 0.14 & --  & --  & --  & \cr

126267 & 8.9 & 0.680 & 0.743 & 0.421 & 0.229 & 3.97  & 5700 &
$-$0.04 & 4.83 & 0.16 & --  & --  & --  & \cr

129920 & 8.3 & 0.659 & 0.711 & 0.409 & 0.211 & 0.40  & 5754 &
$-$0.10 & 4.90 & 0.09 & --  & --  & --  & \cr

133430 & 8.6 & 0.669 & 0.729 & 0.406 & 0.196 & 2.39  & 5695 &
$-$0.25 & 4.99 & 0.11 & --  & --  & --  & \cr

134702 & 8.4 & 0.645 & 0.716 & 0.405 & 0.213 & 0.28  & 5794 &
$-$0.03 & 4.91 & 0.16 & --  & --  & --  & \cr

140690 & 8.6 & 0.659 & 0.729 & 0.415 & 0.210 & 1.19  & 5699 &
$-$0.19 & 4.74 & 0.12 & 110 & 160 & --  & $\surd$ \cr

143337 & 8.1 & 0.639 & 0.729 & 0.406 & 0.180 & 7.03  & 5664 &
$-$0.45 & 4.80 & 0.15 & 80  & 180 & --  & $\surd$ \cr

153458 & 8.1 & 0.652 & 0.723 & 0.405 & 0.216 & 0.09  & 5794 &
$+$0.01 & 4.85 & 0.12 & 140 & 100 & --  & \cr

154221 & 8.7 & 0.640 & 0.710 & 0.402 & 0.232 & 1.78  & 5888 &
$+$0.25 & 4.77 & 0.21 & --  & --  & --  & \cr

155968 & 8.5 & 0.687 & 0.752 & 0.416 & 0.245 & 5.96  & 5782 &
$+$0.23 & 4.86 & 0.15 & --  & --  & --  & \cr

157750 & 8.1 & 0.670 & 0.721 & 0.405 & 0.231 & 1.26  & 5834 &
$+$0.20 & 4.88 & 0.16 & 110 & 120 & --  & $\surd$ \cr

158415 & 8.4 & 0.681 & 0.744 & 0.406 & 0.222 & 0.71  & 5769 &
$+$0.08 & 4.85 & 0.09 & --  & --  & --  & \cr

163441 & 8.5 & 0.685 & 0.750 & 0.402 & 0.217 & 1.14  & 5767 &
$+$0.06 & 4.76 & 0.15 & --  & --  & --  & \cr

163859 & 8.6 & 0.660 & 0.753 & 0.411 & 0.213 & 0.45  & 5717 &
$-$0.10 & 4.92 & 0.10 & --  & --  & --  & \cr

BD+15\,3364 & 8.7 & 0.647 & 0.733 & 0.411 & 0.213 & 1.38  & 5744 &
$-$0.10 & 4.83 & 0.17 & 110 & 150 & --  & \cr

171226 & 8.9 & 0.648 & 0.720 & 0.420 & 0.233 & 3.70  & 5767 &
$+$0.02 & 4.97 & 0.19 & --  & --  & --  & \cr

181199 & 8.2 & 0.656 & 0.752 & 0.407 & 0.199 & 1.73  & 5694 &
$-$0.22 & 4.78 & 0.26 & --  & --  & --  & \cr

183579 & 8.8 & 0.653 & 0.727 & 0.399 & 0.216 & 0.90  & 5824 &
$+$0.08 & 4.91 & 0.17 & --  & --  & --  & \cr

188298 & 8.5 & 0.657 & 0.718 & 0.408 & 0.229 & 0.81  & 5822 &
$+$0.14 & 4.92 & 0.17 & --  & --  & --  & \cr

191487 & 8.6 & 0.654 & 0.729 & 0.390 & 0.217 & 4.01  & 5873 &
$+$0.19 & 4.79 & 0.16 & 100 & 130 & --  & $\surd$ \cr

200633 & 8.4 & 0.639 & 0.728 & 0.406 & 0.201 & 1.46  & 5741 &
$-$0.19 & 4.76 & 0.16 & --  & --  & --  &  \cr

202072 & 8.2 & 0.665 & 0.725 & 0.399 & 0.198 & 2.80  & 5749 &
$-$0.14 & 4.78 & 0.13 & 90  & 100 & --  & $\surd$ \cr

204627 & 8.7 & 0.610 & 0.727 & 0.391 & 0.196 & 7.20  & 5833 &
$-$0.09 & 4.98 & 0.16 & --  & --  & --  & \cr

206772 & 8.4 & 0.653 & 0.709 & 0.394 & 0.210 & 2.77  & 5842 &
$+$0.06 & 4.79 & 0.10 & --  & --  & --  & \cr

209262 & 8.1 & 0.687 & 0.752 & 0.411 & 0.225 & 1.46  & 5739 &
$+$0.05 & 4.77 & 0.13 & --  & --  & --  & \cr

211786 & 8.1 & 0.666 & 0.726 & 0.394 & 0.197 & 4.46  & 5771 &
$-$0.10 & 4.96 & 0.11 & 130 & 100 & --  & $\surd$ \cr

214635 & 8.7 & 0.672 & 0.733 & 0.400 & 0.211 & 1.10  & 5779 &
$+$0.01 & 4.91 & 0.16 & --  & --  & --  & \cr

215942 & 8.1 & 0.664 & 0.723 & 0.404 & 0.213 & 0.31  & 5778 &
$-$0.01 & 4.80 & 0.11 & --  & --  & --  & \cr

216436 & 8.7 & 0.676 & 0.740 & 0.415 & 0.222 & 1.80  & 5720 &
$-$0.04 & 4.74 & 0.11 & 70  & 100 & --  & $\surd$ \cr

221343 & 8.4 & 0.657 & 0.733 & 0.404 & 0.235 & 1.75  & 5858 &
$+$0.26 & 4.84 & 0.14 & 50  & 40 & --  & $\surd$ \cr

BD+40\,5199 & 8.2 & 0.652 & 0.731 & 0.397 & 0.191 & 4.77  & 5742 &
$-$0.21 & 5.01 & 0.10 & --  & --  & --  & \cr

\noalign{\smallskip} \hline\hline
\end{tabular}
}
\end{table*}

\begin{table*}
\centering \caption[]{The same as Table~\ref{parphot1} for the
Galilean satellites and Vesta (taken as proxies of the solar flux
spectrum) plus stars selected in the Bright Star Catalogue
\citep{hoffleitjaschek1991, hoffleit1991} to have both their
($B-V$) and ($U-B$) colors similar to the solar ones.}
\label{parphot3}

{\tiny
\begin{tabular}{ l @{\hspace{1em}} c @{\hspace{1em}} c @{\hspace{1em}} c @{\hspace{1em}} c @{\hspace{1em}}
c @{\hspace{1em}} c @{\hspace{1em}} c @{\hspace{1em}} c
@{\hspace{1em}} c @{\hspace{1em}} c @{\hspace{1em}} c
@{\hspace{1em}} c @{\hspace{1em}} c @{\hspace{1em}} c
@{\hspace{1em}} c @{\hspace{1em}} c }

\noalign{\smallskip} \hline \hline \noalign{\smallskip}

HD  & $V^{\rm Tycho}$ & $(B-V)^{\rm Johnson}$ & $(B-V)^{\rm
Tycho}$ & $(b-y)$ & $m_{\rm 1}$ & $S_{\rm C}$ & \Teff$^{\rm phot}$
& [Fe/H]$_{\rm phot}$ & $M_{\rm V}^{\rm Tycho}$ & $\sigma$
($M_{\rm V}^{\rm Tycho}$) & $\lambda$6145 & $\lambda$6563 & FEROS
& UV \cr

\noalign{\smallskip} \hline \noalign{\smallskip}

Ganymede & -- & -- & -- & -- & -- & -- & -- & -- & -- & -- & 250 &
330 & 510 & $\surd$ \cr

Callisto & -- & -- & -- & -- & -- & -- & -- & -- & -- & -- & 440 &
230 & -- & $\surd$ \cr

Europa & -- & -- & -- & -- & -- & -- & -- & -- & -- & -- & 390 &
290 & -- & -- \cr

Io & -- & -- & -- & -- & -- & -- & -- & -- & -- & -- & -- & 250 &
-- & $\surd$ \cr

Vesta & -- & -- & -- & -- & -- & -- & -- & -- & -- & -- & -- & --
& -- & $\surd$ \cr

9562 & 5.8 & 0.639 & 0.709 & 0.389 & 0.221 & 5.43 & 5919 & +0.24 &
3.47 & 0.05 & --  & --  & --  & $\surd$ \cr

16141 & 6.9 & 0.670 & 0.751 & 0.424 & 0.211 & 4.87 & 5614 &
$-$0.31 & 4.12 & 0.11 & --  & --  & -- & $\surd$ \cr

19467 & 7.0 & 0.645 & 0.734 & 0.409 & 0.200 & 1.63 & 5709 &
$-$0.23 & 4.56 & 0.06 & --  & --  & -- & $\surd$ \cr

28255 & 6.0 & 0.659 & 0.730 & 0.397 & 0.227 & 1.94 & 5868 &
$+$0.24 & 3.81 & 0.08 & -- & -- & -- & $\surd$ \cr

94340 & 7.1 & 0.645 & 0.702 & 0.398 & 0.204 & 2.93 & 5811 &
$-0$.05 & 4.03 & 0.10 & 210 & 250 & -- & \cr

101563 & 6.5 & 0.651 & 0.719 & 0.410 & 0.206 & 0.91 & 5731 &
$-0$.17 & 3.41 & 0.08 & -- & -- & 210 & \cr

105590 & 6.8 & 0.666 & 0.753 & 0.404 & 0.233 & 4.20 & 5884 &
$+$0.34 & 4.65 & 0.63 & 170 & 180 & -- & \cr

111398 & 7.2 & 0.660 & 0.724 & 0.420 & 0.223 & 2.65 & 5717 &
$-0$.10 & 4.37 & 0.09 & -- & -- & 440 & \cr

119550 & 7.0 & 0.631 & 0.698 & 0.411 & 0.202 & 3.23 & 5744 &
$-0$.23 & 3.09 & 0.14 & 170 & 160 & -- & \cr

159222 & 6.6 & 0.639 & 0.722 & 0.404 & 0.219 & 0.60 & 5823 &
$+$0.06 & 4.72 & 0.04 & 240 & 200 & -- & $\surd$ \cr

159656 & 7.2 & 0.641 & 0.711 & 0.405 & 0.210 & 0.98 & 5790 &
$-0$.06 & 4.61 & 0.09 & 130 & 170 & -- & $\surd$ \cr

186408 & 6.0 & 0.643 & 0.728 & 0.410 & 0.211 & 0.51 & 5750 &
$-0$.11 & 4.36 & 0.04 & 180 & -- & -- & \cr

221627 & 6.9 & 0.666 & 0.739 & 0.421 & 0.204 & 3.82 & 5622 &
$-0$.35 & 3.47 & 0.10 & 250 & 160 & -- & $\surd$ \cr

\noalign{\smallskip} \hline\hline
\end{tabular}
}
\end{table*}

\section{Observations and reductions}

\subsection{OPD optical spectra}

Spectroscopic observations were performed with the coud{\'e}
spectrograph, coupled to the 1.60m telescope of Observat{\'o}rio
do Pico dos Dias (OPD, Braz{\'o}polis, Brazil), operated by
Laborat{\'o}rio Nacional de Astrof{\i}sica (LNA/CNPq), in a series
of runs from 1998 to 2002. Spectra were obtained for 47 stars in 2
spectral ranges, $\sim$150~\AA ~wide, centered at $\lambda$6145
and $\lambda$6563 (H$\alpha$). The nominal resolution was R $=$
20\,000 per resolution element, and the S/N ranged from 50 to 440,
with an average value of 160. Four stars were only observed in the
$\lambda$6145 range, and one star only in the $\lambda$6563 range.
Three objects, \object{HD\, 73350}, \object{HD\,146233} (18 Sco),
and \object{HD\, 189625} were observed in separate epochs and
reduced and analyzed separately as a procedure control. As proxies
of the solar flux spectrum, the Galilean satellites Europa,
Ganymede, and Callisto were observed in both spectral ranges; Io
was only observed in the $\lambda$6563 range.

\begin{figure*}
\begin{center}
\resizebox{8.5cm}{!}{\includegraphics{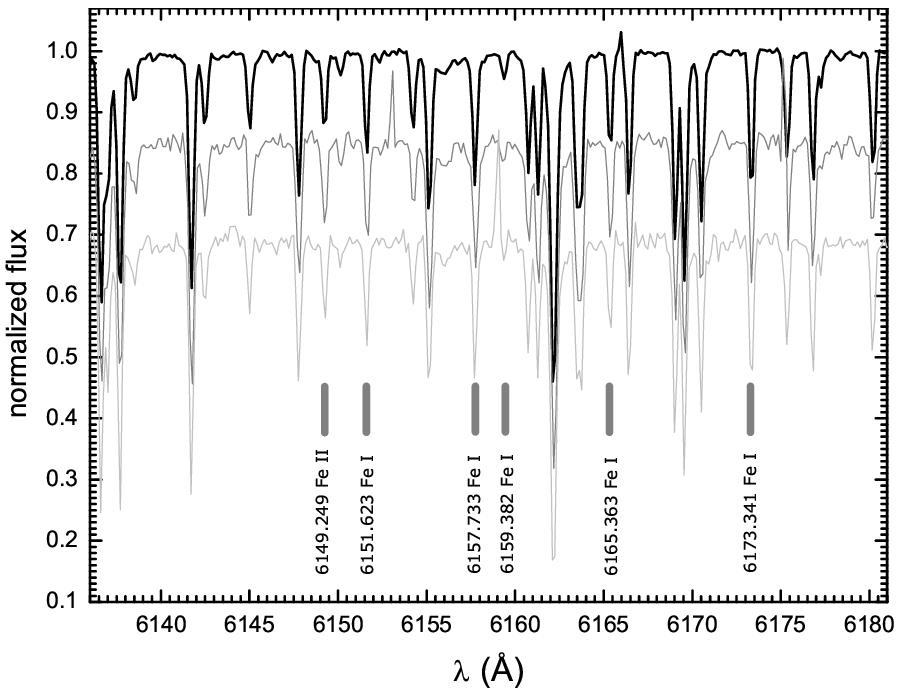}}
\resizebox{8.5cm}{!}{\includegraphics{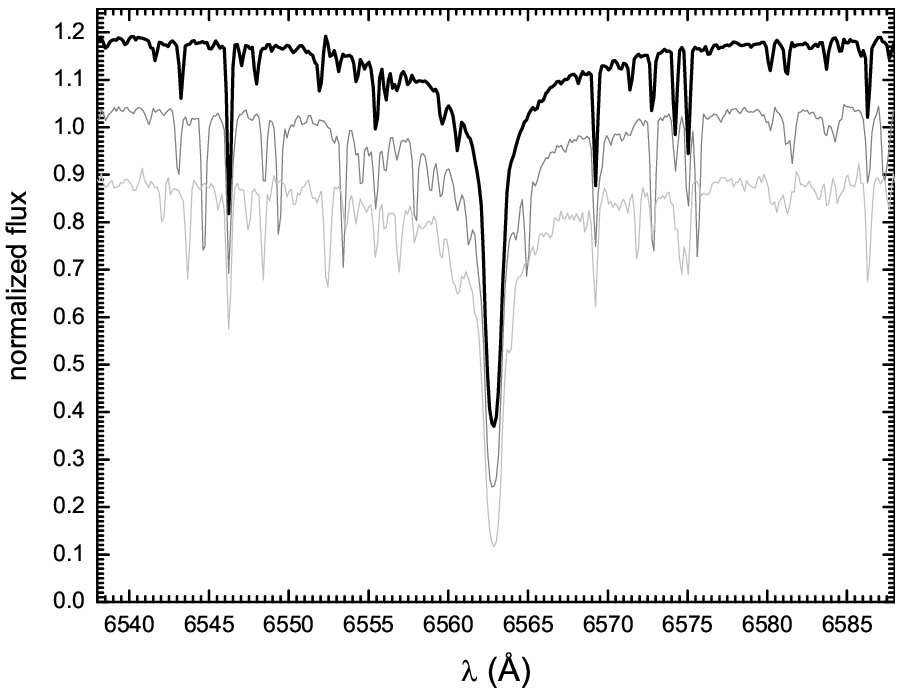}}
\end{center}
\caption[]{{\it Left}. Sample of normalized OPD spectra in the
$\lambda$6145 spectral range. The nominal resolution is R $=$
20\,000, and the spectra S/N, from top to bottom, are 340
(\object{HD\, 146233}), 120 (\object{HD\, 19518}) and 100
(\object{HD\, 191487}). Some of the \ion{Fe}{i} and \ion{Fe}{ii}
lines used in this spectral range for deriving atmospheric
parameters are marked by the vertical dashes. The spectra are
arbitrarily displaced on the vertical axis. {\it Right}. Sample of
normalized OPD spectra in the $\lambda$6563 spectral range. The
nominal resolution is R $=$ 20\,000, and the spectra S/N, from top
to bottom, are 230 (\object{HD\, 146233}), 200 (\object{HD\,
159222}), and 130 (\object{HD\, 191487}). The spectra are
arbitrarily displaced on the vertical axis.}
\label{sample-spectra}
\end{figure*}

Data reduction was carried out by the standard procedure using
IRAF{\footnote {{\it Image Reduction and Analysis Facility} (IRAF)
is distributed by the National Optical Astronomical Observatories
(NOAO), which is operated by the Association of Universities for
Research in Astronomy (AURA), Inc., under contract to the National
Science Foundation (NSF).}}. Bias and flat-field corrections were
performed, background and scattered light were subtracted, and the
one-dimensional spectra were extracted. The pixel-to-wavelength
calibration was obtained from the stellar spectra themselves, by
selecting isolated spectral lines in the object spectra and
checking for the absence of blends, the main screen for blends
being the Solar Flux Atlas \citep{kuruczatlas} and the Utrecht
spectral line compilation \citep{mooreetal1966}. There followed
the Doppler correction of all spectra to a rest reference frame.
Continuum normalization was performed by fitting low-order
polynomials to line-free regions, and a careful comparison to the
solar flux spectrum was carried out to ensure a high degree of
homogeneity in the procedure. The equivalent widths (hereafter
\Wlams) of a set of unblended, weak, or moderately strong lines of
\ion{Fe}{i} and \ion{Fe}{ii} were measured by Gaussian fitting in
all stars, totalling 15 and 2 lines, respectively, for the best
exposed spectra. Sample OPD spectra are shown in
Fig.~\ref{sample-spectra}.

\subsection{ESO/FEROS spectra}

The FEROS spectrograph \citep{kauferetal1999} was used to obtain
data for 17 stars, in three runs from 2000 to 2002. Six stars are
common between the FEROS and OPD data sets, as a homogeneity test:
\object{HD\, 19518}, \object{HD\, 73350}, \object{HD\, 98649},
\object{HD\, 105901}, \object{HD\, 115382}, and \object{HD\,
146233}. A FEROS spectrum of Ganymede was also secured. The
standard FEROS pipeline of spectral reduction was used, and the
spectra were degraded to the same resolution of the OPD spectra,
and clipped to the same wavelength range. The blaze curvature,
unfortunately, precluded our use of the H$\alpha$ profile of the
FEROS data. The S/N of the degraded FEROS spectra ranged from 100
to 700, the average value being 400. After this procedure, the
data reduction followed exactly the same steps as the OPD data.

\subsection{ESO UV spectra}

\begin{figure*}
\begin{center}
\resizebox{8.5cm}{!}{\includegraphics{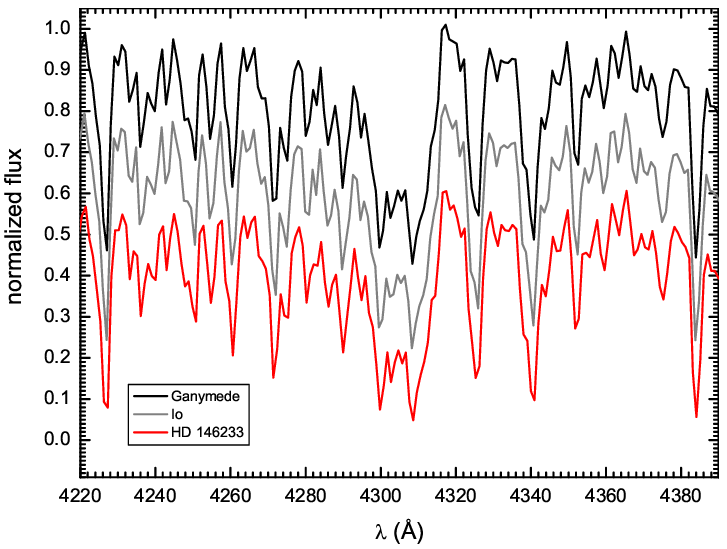}}
\resizebox{8.5cm}{!}{\includegraphics{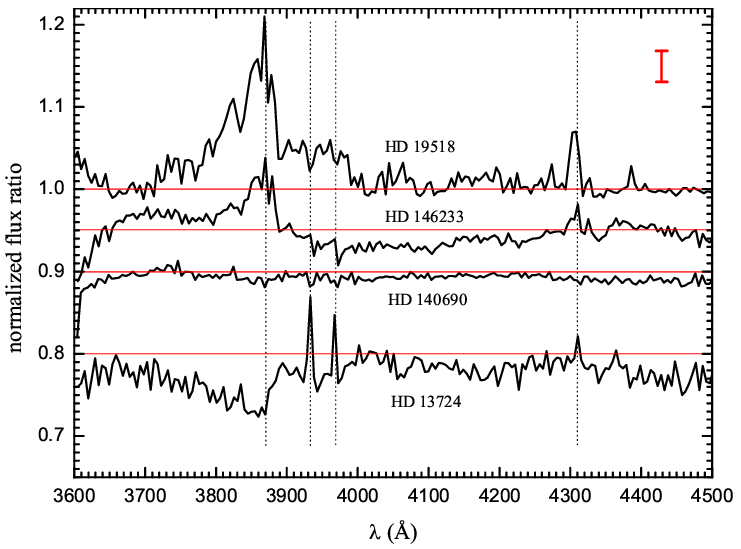}}
\end{center}
\caption[]{{\it Left}. Sample of normalized ESO/UV spectra in the
$\lambda$$\lambda$4220-4390 range, around the $\lambda$4310 CH
bandhead. The nominal resolution is R $=$ 800, and the spectra are
arbitrarily displaced on the vertical axis. {\it Right}. Sample
ratioed ESO/UV spectra, normalized to the solar (Ganymede)
spectra, in the $\lambda$$\lambda$3600-4500 range. The spectra are
arbitrarily displaced in the vertical axis, and the horizontal red
lines mark the unitary flux ratios for each object. The dotted
lines are, from left to right, respectively, the approximate
central wavelength of the $\lambda$3870 CN bandhead, the central
wavelengths of the K and H \ion{Ca}{ii} lines, and the approximate
central wavelength of the $\lambda$4310 CH bandhead. The 1$\sigma$
flux ratio error bar is shown in red.} \label{UV-spectra}
\end{figure*}

The cassegrain Boller \& Chivens spectrograph of the ESO 1.52m
telescope was used in two runs, in 1997 and 1998, to acquire
low-resolution spectra for 37 stars, plus four proxies of the
solar flux spectrum: Vesta, Ganymede, Callisto and Io. The useful
spectral range was $\lambda$$\lambda$3600-4600, and the nominal
spectral resolution was R = 2\,000. The exposure times were set to
obtain S/N around 100, and only a few cases fell short of this
goal. All spectra were reduced in a standard way and
wavelength-calibrated. Offsets in the wavelength calibration were
corrected by shifting all spectra to agree with the wavelength
scale of the Ganymede spectrum. Spectra were then degraded to a
resolution of R = 800, and the resulting S/N is better than 200 in
all cases. The spectra were normalized and ratioed to the Ganymede
spectrum. We estimated errors in the flux spectra remaining from
errors in the wavelength calibration as no larger than 1\%. Errors
due to the normalization procedure were estimated by comparing the
ratio spectra of Vesta, Callisto, and Io to that of Ganymede, and
attributing all fluctuations to random noise in the normalization:
an average value of 1$\sigma$ = 1.8\% results. The Ganymede
spectrum was chosen as the preferred reference solar proxy. A
similar exercise with the spectra of five stars very similar to
the Sun in their UV spectral features yielded 1$\sigma$ = 1.6\%.
We estimate that 1$\sigma$ $\sim$ 2\% is a conservative estimate
of the flux ratio uncertainty of the spectral features in the UV
spectra, between $\lambda$3700 and $\lambda$4500. Shortward of
$\lambda$3700 the lower S/N and very strong blending introduce a
larger uncertainty in the ratio spectra of strong-lined stars.
Samples of the unratioed and ratioed spectra are shown in
Fig.~\ref{UV-spectra}.

The main goal for the UV data is to verify to what extent, in the
spectra of solar analogs, the strength of the CN and CH bands,
both very sensitive to the stellar atmospheric parameters, remain
similar to the solar ones. Also, the flux ratio at the core of the
\ion{Ca}{ii} H and K lines, at $\lambda$$\lambda$ 3968 and 3934,
is very sensitive to the stellar chromospheric filling, and
therefore to age, at least within $\sim$2 Gyr of the ZAMS
\citep{pacepasquini2004}.\\

\begin{table}
\centering \caption[]{The \ion{Fe}{i} and \ion{Fe}{ii} transitions
used in the spectroscopic analysis. The first two columns are
self-explanatory; the third and fourth columns are, respectively,
the (raw) \Wlams measured off the OPD and FEROS Ganymede spectra
(both in m\AA); the fifth column is the lower excitation potential
(in $eV$); and the last column is the $\log{gf}$ derived from the
OPD spectra.} \label{lines}
\begin{tabular}{c l r@{.}l r@{.}l c c}
\hline\hline \noalign{\smallskip}
\parbox[c]{1.5cm}{\centering Wavelength {\tiny (\AA)}} &
Species & \multicolumn{2}{c}{\parbox[c]{0.8cm}{\centering \Wlam \\
{\tiny (OPD)}}} & \multicolumn{2}{c}{\parbox[c]{1.1cm}{\centering
\Wlam \\ {\tiny (FEROS)}}} &
\parbox[c]{0.5cm}{\centering $\chi$ {\tiny (eV)}} &
$\log{gf}$ \\
\noalign{\smallskip} \hline \noalign{\smallskip}
6078.499 & \ion{Fe}{I}  &  85&1 &  83&0 & 4.79 & $-$0.274 \\
6079.016 & \ion{Fe}{I}  &  51&8 &  51&0 & 4.65 & $-$0.942 \\
6084.105 & \ion{Fe}{II} &  25&7 &  25&1 & 3.20 & $-$3.717 \\
6089.574 & \ion{Fe}{I}  &  40&9 &  41&3 & 5.02 & $-$0.811 \\
6093.649 & \ion{Fe}{I}  &  36&8 &  37&6 & 4.61 & $-$1.263 \\
6096.671 & \ion{Fe}{I}  &  44&9 &  41&1 & 3.98 & $-$1.692 \\
6102.183 & \ion{Fe}{I}  &  92&2 &  86&9 & 4.83 & $-$0.144 \\
6149.249 & \ion{Fe}{II} &  38&4 &  37&5 & 3.89 & $-$2.752 \\
6151.623 & \ion{Fe}{I}  &  52&1 &  52&8 & 2.18 & $-$3.292 \\
6157.733 & \ion{Fe}{I}  &  69&4 &  66&1 & 4.07 & $-$1.143 \\
6159.382 & \ion{Fe}{I}  &  14&6 &  14&0 & 4.61 & $-$1.827 \\
6165.363 & \ion{Fe}{I}  &  47&6 &  48&1 & 4.14 & $-$1.492 \\
6173.341 & \ion{Fe}{I}  &  72&0 &  71&8 & 2.22 & $-$2.835 \\
6185.704 & \ion{Fe}{I}  &  20&7 &  17&2 & 5.65 & $-$0.695 \\
6187.995 & \ion{Fe}{I}  &  49&4 &  49&9 & 3.94 & $-$1.646 \\
6191.571 & \ion{Fe}{I}  & 134&1 & 136&4 & 2.43 & $-$1.600 \\
6200.321 & \ion{Fe}{I}  &  81&2 &  76&7 & 2.61 & $-$2.274 \\
\noalign{\smallskip} \hline\hline
\end{tabular}
\end{table}

There are 24 stars in our sample with V$^{\rm Tycho}$ $\leq$ 8.0,
color similarity index S$_{\rm C}$ $<$ 1.50 (corresponding to a
3$\sigma$ similarity) and accessible from the southern hemisphere.
All of them were observed at either the OPD or FEROS/ESO,
excepting \object{HD\, 28471} and \object{HD\, 88084}. Actually,
only four northern objects matching the above criteria,
\object{HD\, 70516}, \object{HD\, 98618}, \object{HD\, 139777},
and \object{HD\, 158222} are inaccessible from either the OPD or
FEROS locations. In Tables 1, 2, and 3, the available spectral
data for each star is indicated. Quality checks were performed on
the measured \Wlams following the same procedure as discussed in
detail by \cite{portodemelloetal2008}. Saturated lines were
eliminated by a 2$\sigma$ clipping on the relation of reduced
width \Wlam/$\lambda$ with line depth, and no lines were measured
beyond the linearity limit. Also, no trend is expected in the
relation of the line full-width-half-maximum and reduced width,
since the line widths are defined by the instrumental profile. The
measured \Wlams were corrected to bring them onto a system
compatible with the Voigt-fitted \Wlams of \cite{meylanetal1993}.
This correction is +5.0\% for the OPD and +6.0\% for the FEROS/ESO
\Wlams. In Table~\ref{lines} we list the \ion{Fe}{i} and
\ion{Fe}{ii} lines used with the corresponding excitation
potentials and $gf-$values derived. The \Wlams measurements of all
analyzed stars are available upon request.

\section{Results}

\subsection{Spectroscopic atmospheric parameters}

For the determination of the atmospheric parameters \Teff,
$\log{g}$ and [Fe/H], we employed a strictly differential analysis
with the Sun as the standard star. The expectation of this
approach is that systematic errors in the measurement of line
strengths, the representation of model atmospheres, and the
possible presence of non-local thermodynamic equilibrium (NLTE)
effects, will be greatly lessened, given the high similarity of
all program stars and the Sun. For each spectroscopic data set, at
least one point-source solar proxy was observed in a manner
identical to that of the stars.

Solar $gf$-values were determined for the \ion{Fe}{i} and
\ion{Fe}{ii} spectral lines, from solar \Wlams measured off the
Ganymede spectra (corrected to the Voigt scale), and an LTE, 1-D,
homogeneous and plane-parallel solar model atmosphere from the
grid described by \citet{edvardssonetal1993}. The adopted
parameters for the Sun were \Teff = 5780 K, surface gravity
$\log{g}$ = 4.44, $[$Fe/H$]$ = $+$0.00 and $\xi_t$ = 1.30
km\,s$^{\rm -1}$. The adopted solar absolute abundances (which are
inconsequential in a differential analysis) are those of
\citet{grevessenoels1993}, and $gf$ values were independently
generated for the OPD and FEROS data sets.

The atmospheric parameters of the program stars were determined by
iterating the photometric \Teff (calibrations for which are
described in the Appendix) determined from the $(B-V)^{\rm
Johnson}$, $(B-V)^{\rm Tycho}$, and $(b-y)$ color indices, coupled
to the spectroscopic metallicity derived from the \ion{Fe}{i}
lines. Model atmospheres were interpolated at each step, until the
spectroscopic [Fe/H] agreed with the model input. Once the
photometric \Teff are fixed, the $\log{g}$ was varied until
consistency was achieved between the \ion{Fe}{i} and \ion{Fe}{ii}
abundances, to a tolerance of 0.01 dex. The microturbulence
velocity in all steps was set by the relation of
\citet{edvardssonetal1993}, as a function of \Teff and $\log{g}$.
The photometric calibration and set of Fe \Wlams of each star
uniquely determines the atmospheric parameter solution.

It is noteworthy that excellent agreement was obtained for the
control stars between the atmospheric parameters from two
independent determinations based on OPD data, for three stars. The
average differences, respectively, for \Teff, $\log{g}$, and
$[$Fe/H$]$ are 13 K, 0.06 dex and 0.03 dex, well within the errors
of the analysis. Similarly, for seven stars in common between the
OPD and FEROS/ESO data sets, the mean difference in the sense OPD
minus FEROS, for \Teff, $\log{g}$ and $[$Fe/H$]$ is $+$19 K,
$+$0.02 dex and $+$0.06 dex, respectively, also well within the
errors of the analysis. We may thus regard the two data sets as
homogeneous, and we show the spectroscopic parameters for all
observed stars in Tables 5, 6, and 7. For the control objects, we
list the averaged values of all available determinations.

\subsection{Effective temperature from the H$\alpha$ profile}

Additional effective temperatures were determined for those stars
observed in the OPD in the $\lambda$6563 range by fitting the
H$\alpha$ line profiles by \citet{lyraportodemello2005}, so we
refer the reader to this paper for details. For the Galilean
satellites, no \Teff determination from H$\alpha$ was provided by
\cite{lyraportodemello2005}: for these objects and for some stars
also not analyzed by these authors, we determined the H$\alpha$
\Teff using exactly the same procedure. The H$\alpha$ profile
wings are very sensitive to \Teff but barely respond to the other
atmospheric parameters. They are particularly insensitive to
metallicity \citep{fuhrmannetal1993}, and therefore a robust
independent check on \Teff. The average uncertainty of the
\Teff($H\alpha$) determinations is a direct function of the
spectral S/N, given the very strong similarity in parameters of
all the program stars. This was estimated by the error analysis
provided by \cite{lyraportodemello2005}, and
$<\sigma>$(\Teff(H$\alpha$)) = 50 K resulted. These \Teffs are
very closely tied to the solar \Teff zero point since a perfectly
solar \Teff is retrieved for the spectra of all the solar proxies
(the Galilean satellites) (Table~\ref{paratmf-sim}). The
photometric and H$\alpha$ \Teffs scales therefore share the same
zero point, and any systematic offset still to be gauged only
remain in scale.

We note, however, that this good agreement should not be taken in
an absolute sense. More sophisticated modeling of the solar Balmer
profiles \citep{barklemetal2002} point to slight offsets between
observations and theory, possibly due to both inconsistencies in
the atmospheric models and the line broadening physics. Although
very successful in recovering many observational features of the
real Sun, even very recent 3D models \citep{pereiraetal2013} still
cannot reproduce the solar Balmer profiles perfectly. Our good
internal consistency between photometric and H$\alpha$ \Teff scale
should thus be regarded only in a relative sense for solar-type
stars in the context of classical 1D modelling.

\subsection{Uncertainties in the atmospheric parameters}

Formal errors are estimated as follows: for the metallicity
[Fe/H], we adopt the average standard deviation of the
distribution of abundances derived from the \ion{Fe}{i} lines.
This was $\sigma$([Fe/H]) = 0.08 dex, for the OPD spectra, and
only $\sigma$([Fe/H]) = 0.04 dex, for the FEROS spectra (owing to
the much better S/N of the latter). The error of the photometric
\Teff is affected by the metallicity error. For two stars with S/N
that are representative of the sample, we estimated the \Teff
uncertainty due to the internal \Teff standard deviation of the
color calibrations, adding the metallicity error. The values were
$\sigma$(\Teff) = 50 K for the OPD spectra and $\sigma$(\Teff) =
40 K for the FEROS spectra. The error in $\log{g}$ was estimated,
for the same three representative stars, by evaluating the
variation in this parameter which produces a disagreement of
1$\sigma$ between the abundances of \ion{Fe}{i} and \ion{Fe}{ii}.
The result was $\sigma(\log{g})$ = 0.20 dex for the OPD and 0.15
for the FEROS data.

\begin{table}
\centering \caption[]{Atmospheric parameters of the solar analogs
with V$_{\rm Tycho}$ $\leq$ 8.0. First col. is HD number; second
and third cols. are the photometric and H$\alpha$ \Teffs (the
former derived from the final adopted spectroscopic metallicity);
fourth col. is the ionization surface gravity; and last col. is
the spectroscopic metallicity.} \label{paratmf-bright}
\begin{tabular}{l c c c r@{}l}
\hline\hline \noalign{\smallskip} HD &
\parbox[c]{1.0cm}{\centering \Teff$^{\rm color}$ \\ ({\tiny K})} &
\parbox[c]{1.0cm}{\centering \Teff$^{{\rm H}\alpha}$ \\ ({\tiny K})} &
$\log{g}^{\rm \,ion}$ &
\multicolumn{2}{c}{[Fe/H]} \\
\noalign{\smallskip} \hline \noalign{\smallskip}
Sun         & 5777 & 5777 & 4.44 & $+$0&.00 \\
4308        & 5720 & 5695 & 4.44 & $-$0&.29 \\
9986        & 5820 & --   & 4.48 & $+$0&.09 \\
13724       & 5820 & 5790 & 4.16 & $+$0&.24 \\
19518       & 5780 & --   & 4.34 & $-$0&.11 \\
24293       & 5760 & 5690 & 4.10 & $-$0&.04 \\
25874       & 5770 & 5770 & 4.40 & $+$0&.04 \\
28821       & 5690 & 5680 & 4.58 & $-$0&.08 \\
30495       & 5840 & 5800 & 4.36 & $+$0&.09 \\
32963       & 5800 & --   & 4.44 & $+$0&.08 \\
35041       & 5810 & --   & 4.46 & $-$0&.05 \\
37773       & 5700 & --   & 4.32 & $+$0&.04 \\
66653       & 5840 & --   & 4.40 & $+$0&.15 \\
68168       & 5780 & --   & 4.42 & $+$0&.11 \\
71334       & 5770 & 5650 & 4.50 & $-$0&.06 \\
73350       & 5830 & 5790 & 4.45 & $+$0&.14 \\
88072       & 5800 & --   & 4.24 & $+$0&.05 \\
117939      & 5730 & 5800 & 4.44 & $-$0&.10 \\
134664      & 5810 & 5830 & 4.36 & $+$0&.13 \\
138573      & 5760 & 5740 & 4.42 & $+$0&.00 \\
142072      & 5790 & 5790 & 4.46 & $+$0&.20 \\
145825      & 5840 & 5830 & 4.52 & $+$0&.07 \\
146233      & 5790 & 5800 & 4.48 & $-$0&.03 \\
150248      & 5750 & 5750 & 4.38 & $-$0&.04 \\
155114      & 5830 & 5810 & 4.46 & $-$0&.02 \\
164595      & 5810 & 5770 & 4.67 & $-$0&.04 \\
187237      & 5850 & --   & 4.48 & $+$0&.16 \\
189625      & 5870 & 5810 & 4.45 & $+$0&.27 \\
190771      & 5840 & 5820 & 4.56 & $+$0&.19 \\
207043      & 5790 & 5760 & 4.55 & $+$0&.07 \\
214385      & 5730 & --   & 4.24 & $-$0&.26 \\
\noalign{\smallskip} \hline\hline
\end{tabular}
\end{table}

\begin{table}
\centering \caption[]{Same as Table~\ref{paratmf-bright} for the
8.0 $< V_{\rm T}$ $\leq$ 9.0 stars.} \label{paratmf-faint}
\begin{tabular}{l c c c r@{}l}
\hline\hline \noalign{\smallskip} HD &
\parbox[c]{1.0cm}{\centering \Teff$^{\rm color}$ \\ ({\tiny K})} &
\parbox[c]{1.0cm}{\centering \Teff$^{{\rm H}\alpha}$ \\ ({\tiny K})} &
$\log{g}^{\rm \,ion}$ &
\multicolumn{2}{c}{[FeH]} \\
\noalign{\smallskip} \hline \noalign{\smallskip}
8291        & 5810 & 5860 & 4.30 & $+$0&.03 \\
12264       & 5810 & 5810 & 4.54 & $+$0&.06 \\
98649       & 5770 & 5780 & 4.63 & $-$0&.02 \\
105901      & 5840 & 5850 & 4.50 & $-$0&.01 \\
115382      & 5780 & 5790 & 4.40 & $-$0&.08 \\
118598      & 5800 & 5730 & 4.52 & $+$0&.02 \\
140690      & 5780 & 5790 & 4.40 & $+$0&.06 \\
143337      & 5750 & 5760 & 4.36 & $-$0&.19 \\
153458      & 5850 & 5810 & 4.44 & $+$0&.20 \\
157750      & 5840 & 5850 & 4.54 & $+$0&.21 \\
BD+15\,3364 & 5800 & 5770 & 4.40 & $+$0&.07 \\
191487      & 5820 & 5820 & 4.24 & $-$0&.01 \\
202072      & 5750 & 5740 & 4.48 & $-$0&.17 \\
211786      & 5780 & 5800 & 4.42 & $-$0&.09 \\
216436      & 5750 & 5760 & 3.94 & $+$0&.04 \\
221343      & 5800 & 5710 & 4.05 & $+$0&.04 \\
\noalign{\smallskip} \hline\hline
\end{tabular}
\end{table}

\begin{table}
\centering \caption[]{Same as Table~\ref{paratmf-bright} for the
solar proxies and stars with UBV colors similar to the solar
ones.} \label{paratmf-sim}
\begin{tabular}{l c c c r@{}l}
\hline\hline \noalign{\smallskip} HD &
\parbox[c]{1.0cm}{\centering \Teff$^{\rm color}$ \\ ({\tiny K})} &
\parbox[c]{1.0cm}{\centering \Teff$^{{\rm H}\alpha}$ \\ ({\tiny K})} &
$\log{g}^{\rm \,ion}$ &
\multicolumn{2}{c}{[FeH]} \\
\noalign{\smallskip} \hline \noalign{\smallskip}

Ganymede    & --   & 5780 & --   & --       \\
Callisto    & 5770 & 5760 & 4.52 & $-$0&.04 \\
Europa      & 5780 & 5770 & 4.48 & $-$0&.02 \\
Sky         & 5770 & --   & 4.56 & $-$0&.03 \\
94340       & 5870 & 5840 & 3.99 & $+$0&.14 \\
101563      & 5750 & --   & 3.70 & $-$0&.12 \\
105590      & 5790 & 5760 & 4.58 & $+$0&.02 \\
111398      & 5780 & --   & 4.28 & $+$0&.08 \\
119550      & 5830 & 5780 & 3.98 & $+$0&.02 \\
159222      & 5850 & 5830 & 4.34 & $+$0&.14 \\
159656      & 5840 & 5850 & 4.32 & $+$0&.09 \\
186408      & 5820 & --   & 4.40 & $+$0&.11 \\
221627      & 5790 & 5810 & 4.14 & $+$0&.17 \\
\noalign{\smallskip} \hline\hline
\end{tabular}
\end{table}

\subsection{Masses and ages}

\begin{figure}
\centering
\includegraphics[width=8.5cm]{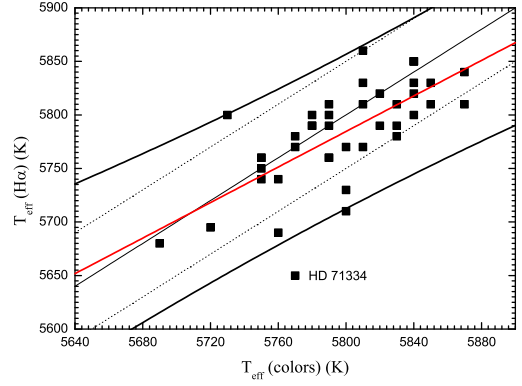}
\caption{Photometric and H$\alpha$ \Teffs compared. The thin black
line is the identity relation, the red line is a linear
least-squares fit, the two dotted lines bound the identity
relation by 50 K, and the two thick black lines are the 95\%
confidence limits of the fit. The only star to deviate
significantly from the identity is \object{HD\, 71334}.}
\label{compteffs}
\end{figure}

For all stars with both a photometric and H$\alpha$ \Teff
determination, we obtained a straight average to produce an
internally more precise value of \Teff, where the errors of the
determinations are very similar. A comparison of the two \Teff
scales (Fig.~\ref{compteffs}) reveals excellent internal
consistency: nearly all stars are contained within 1$\sigma$, and
only one object, \object{HD\, 71334}, deviates by more than
2$\sigma$ of the expected identity relation. The internal
compounded error of the average \Teff for stars with both
determinations is $\sigma \sim$ 30 K and $\sigma \sim$ 35 K, for
FEROS and OPD stars, respectively. We adopt, conservatively,
$\sigma$($<$\Teff$>$) = 40 K in the following discussion. For the
stars with both H$\alpha$ and photometric \Teffs, these averaged
values were used to plot them in a grid of theoretical HR diagrams
by the Geneva group \cite[][and references
therein]{schalleretal92,schaereretal92}. Only the photometric
\Teffs were used for the other stars. Bolometric corrections were
obtained from the tables of \cite{flower1996}, and masses and ages
were interpolated in the diagrams. Astrometric surface gravities
were derived from the well-known equation:

\begin{equation} \label{loggcalc}
\log\Bigg( \frac{g}{g_{\odot}}\Bigg) = \log\Bigg(
\frac{M}{M_{\odot}}\Bigg) + 4 \log\Bigg(
\frac{T_{\mbox{eff}}}{T_{\mbox{eff}\odot}}\Bigg) - \log\Bigg(
\frac{L}{L_{\odot}}\Bigg)~.
\end{equation}

In Tables 8, 9, and 10 we list astrometric surface gravities and
their uncertainties (compounding errors in mass, \Teff, and
luminosity in the equation above), along with bolometric
magnitudes (and uncertainties), masses, and ages. It is seen that
the internal errors of the astrometric surface gravities are much
smaller than those of the ionization ones, and should be preferred
in deciding the similarity of a given star to the Sun.

\subsection{UV spectra}

All stars observed in the UV at low resolution had their
atmospheric parameters obtained from spectroscopic \fehs but
\object{HD\, 6512} and \object{HD\, 28068}. We performed a
qualitative analysis of the UV stellar spectra divided by that of
the solar proxy Ganymede. Our analysis was limited to verifying
any significant structure in the ratio spectra differing from
unity above the photon noise and normalization uncertainties
(section 4.3), and we defer a more complete quantitative
investigation to a forthcoming paper. In Table~\ref{UV-results} we
list all results of this analysis, focused on the behavior of the
CN and CH bandheads, respectively centered roughly at
$\lambda$3870 and $\lambda$4310, and the \ion{Ca}{ii} H and K
lines. It is apparent that nearly all analyzed stars have
detectable or strong differences with respect to the solar
spectrum. Particularly significant are the differences in the CH
and CN bands, which are very sensitive to the stellar atmospheric
parameters. Differences in the H and K line cores can be ascribed
to different levels of chromospheric filling-in at the epoch of
the observations and is not a direct flag of atmospheric
parameters differing from the Sun's.

These data are particularly useful for selecting good UV analogs
of the Sun. Such stars are desirable as solar proxies for
observing comets, which generally have strong emission in the CH
and CN transitions \citep[e.g.,][]{feldmanetal2004,
grudzinskabarbon1968}. In the case of cometary observations
concentrated in the UV, the real issue is not whether the solar
proxy has a strong color similarity to the Sun in the visible, but
rather if its CH and CN features reproduce the solar ones well.
Cometary emission usually has weak continua, and a good
representation of the solar UV flux around the key molecular
emission wavelengths is a necessity. Our list contains four stars
that reproduce the solar CH and CN strengths very well, but not
the solar fill-in in the H and K lines; and two additional stars
that are indistinguishable from the Sun, within the errors, in the
CN/CH and H and K wavelengths, a fact of some importance since the
latter transitions lie in the wings of the $\lambda$ CN and
$\lambda$4056 C$_{\rm 3}$ cometary emission lines for
low-resolution observations. We discuss these objects in the next
section, along with the solar analog and solar twin candidates.

\section{Discussion}

\subsection{Different ways of masquerading as the Sun}

Our spectroscopic analysis revealed a number of stars that not
only possess a strong photometric similarity with the Sun
pertaining the Paschen colors and the m$_{\rm 1}$ index, but not
necessarily in the UV, as we discuss below. Many of these also
have atmospheric parameters, \Teff, $\log{g}$, and [Fe/H] which
are very similar to the solar ones within the errors. These
objects can be considered as excellent solar analogs, are expected
to have a spectral flux distribution very similar to the Sun's in
the blue and red spectral range, and can be used for any
observational procedure that requires removal of the solar
spectrophotometric signature. The inferred masses cluster tightly
around the solar value: indeed, in Tables 8 and 9, corresponding
to our Hipparcos sample, only two stars have masses differing from
solar by more than 0.1 M$_\odot$, and 74\% of the objects have
masses within $\pm$0.05 M$_\odot$ of the solar value. In
Table~\ref{massage-sim}, however, corresponding to stars selected
solely by UBV similarity and presence in the lists of
\cite{hardorp1982}, various objects differ in mass from the Sun by
more than this amount, illustrating the drawbacks of purely
photometric criteria in identifying solar twins. On the other
hand, the ages of all analyzed stars range very widely, from the
zero-age main sequence (ZAMS) to values twice as old as the Sun.
It is clear that stars with atmospheric parameters similar to the
solar ones span a wide range of evolutionary states, a fact
further stressing the differences between solar analogs and solar
twins, and the importance of accurately determining atmospheric
parameters and luminosities in order to successfully identify the
latter.

Cross-checking the color similarity indices of Tables 1 and 2 with
the spectroscopic parameters from Tables 5, 6 and 7, one gleams
that a number of stars that are photometrically similar to the Sun
appear so due to a combination of atmospheric parameters: they are
either hotter/metal-richer or cooler/metal-poorer than the Sun.
The most noteworthy of such objects among the brighter (V$_{\rm
Tycho}$ $\leq$ 8.0) sample stars are \object{HD\,9986},
\object{HD\,66653}, \object{HD\,73350}, \object{HD\,117939},
\object{HD\,134664}, \object{HD\,187237}, \object{HD\,189625},
\object{HD\,190771}. Examples among the fainter (8.0 $<$ V$_{\rm
Tycho}$ $\leq$ 9.0) sample stars are \object{HD\,153458} and
\object{HD\,157750}. They can be successfully employed as
photometric solar matches in a broad sense, for low-resolution
spectroscopy of Solar System objects, but any analysis that can be
influenced by subtle differences in the strength of metal lines
should avoid these stars as solar proxies. They are well spaced in
right ascension and span declinations from $-$64 to $+$30.

We divide our discussion of specific solar analogs and twins as
follows. Purely photometric matches of the Sun, for which we could
secure no spectroscopic data; solar analogs for which
spectroscopic data are available, some of them qualifying as solar
twin candidates; stars selected solely from UBV colors and
presence in the Hardorp lists; and, lastly, stars matching the low
resolution UV spectrum of the Sun. We close this section by
presenting a new list of solar twin candidates having a high
degree of photometric and spectroscopic resemblance to the Sun.

\subsection{Purely photometric matches to the Sun}

Stars for which no spectroscopic observations are available can be
judged as good photometric analogs to the Sun solely by the
S$_{\rm C}$ color similarity index and photometric atmospheric
parameters, besides the Tycho absolute magnitude. In this way
stars matching the solar colors can be revealed but not true solar
analogs and twins. Considering only stars with S$_{\rm C}$ $\leq$
1.50, there are seven stars in the V$_{\rm Tycho}$ $\leq$ 8.0
sample and 30 in the V$_{\rm Tycho}$ $>$ 8.0 sample. In the
brighter sample, one such object is the well-known solar twin
\object{HD\,98618} \citep{melendezetal2006}, which is reliably
recovered in our procedure. The other six stars are
\object{HD\,28471}, \object{HD\,70516}, \object{HD\,88084},
\object{HD\,139777}, \object{HD\,158222}, and \object{HD\,222143}.
\cite{hardorp1982} considered \object{HD\,70516} and
\object{HD\,139777} as poor UV matches to the Sun. Further data
may decide if they are good solar analog or twin candidates. All
are good photometric matches to the Sun, and are also probable
solar analogs except for \object{HD\,139777}, which is probably
less luminous and cooler than the Sun, as well as poorer in
metals. Particularly, \object{HD\,70516}, \object{HD\,88084},
\object{HD\,158222}, and \object{HD\,222143} should be further
investigated since their S$_{\rm C}$ plus the photometric \Teff
and \feh suggest a strong likeness to the Sun.

In the fainter sample, errors in the M$_{\rm V}^{\rm Tycho}$
absolute magnitudes are greater and the objects correspondingly
less interesting. Thirty stars are eligible as good solar
photometric matches by having S$_{\rm C}$ $\leq$ 1.50:
\object{HD\,6512}, \object{HD\,7678}, \object{HD\,15632},
\object{HD\,26736}, \object{HD\,27857}, \object{HD\,28068},
\object{HD\,31130}, \object{HD\,34599}, \object{HD\,36152},
\object{HD\,45346}, \object{HD\,76332}, \object{HD\,78660},
\object{HD\,81700}, \object{HD\,90322}, \object{HD\,90333},
\object{HD\,110668}, \object{HD\,110869}, \object{HD\,110979},
\object{HD\,111938}, \object{HD\,129920}, \object{HD\,134702},
\object{HD\,158415}, \object{HD\,163441}, \object{HD\,163859},
\object{HD\,183579}, \object{HD\,188298}, \object{HD\,200633},
\object{HD\,209262}, \object{HD\,214635}, and \object{HD\,215942}.
Most of the listed objects, additionally, have Tycho absolute
magnitude in agreement with the solar one under a 2$\sigma$
criterion, but for only three objects: \object{HD\,15632},
\object{HD\,34599}, and \object{HD\,36152}. Two stars in this list
have UV data, \object{HD\,6512} and \object{HD\,28068}: they are
very unlike the Sun in this wavelength range. This sample of 37
photometric matches to the Sun is widely scattered across the sky,
has conveniently faint magnitudes, except perhaps for 10m-class
telescopes, and may advantageously substitute Hardorp's lists in
many useful contexts.

\subsection{Solar analogs spectroscopically analyzed}

Solar twins are automatically solar analogs, but not the other way
round: bona fide stars successfully reproducing not only the solar
spectrophotometric properties but also its atmospheric parameters
and state of evolution must be gauged through spectroscopic
analyses, to which we now turn. In the following discussion, it is
important to keep in mind that: the \feh uncertainty of the FEROS
data is 0.04 dex, in contrast to 0.08 dex for the OPD data; that
the 1$\sigma$ internal uncertainty in \Teff is, approximately, 40K
for stars with both photometric and H$\alpha$ determinations, but
50K if only one \Teff determination is available; and that the
spectroscopic (ionization) $\log{g}$ (Tables 5, 6, and 7) is not a
good discriminator between unevolved and evolved stars, but the
astrometric $\log{g}$ (Tables 5, 6, and 7) is.

\subsubsection{Brighter Hipparcos sample}

The metric we adopt to judge a star as a good photometric match to
the Sun is S$_{\rm C}$ $\leq$ 1.50, a 3$\sigma$ match. When
available, we also consider activity data from the UV spectra
(Table~\ref{UV-results}), as well as H$\alpha$ radiative losses
from \cite{lyraportodemello2005}: on their scale, the value of the
solar flux is F'$_{\rm H\alpha}$ = 3.44 $\pm$ 0.45 (1$\sigma$), in
10$^{5}$~ergs$^{-1}$.~cm$^{-2}$.~sec$^{-1}$. We discuss first
those stars with only one \Teff determination, for which
conclusions carry less weight. In the brighter Hipparcos sample,
there are four stars in this situation, all of them analyzed with
FEROS data and all without a H$\alpha$ \Teff determination:

\object{HD\,9986} matches the Sun splendidly in its S$_{\rm C}$
index, but probably has higher metallicity than the Sun, \feh =
$+$0.09 $\pm$ 0.04 dex. Its absolute bolometric magnitude agrees
with the solar one only very narrowly in a 2$\sigma$ sense. It is
a fair solar analog candidate, although not a clear solar twin
candidate, yet it retains some possibility of a solar twin
candidacy and should be further investigated.

\object{HD\,66653} is very probably metal-richer at \feh = $+$0.15
$\pm$ 0.04 dex, but probably also hotter, which could explain its
very good photometric similarity to the Sun. Its UV spectrum (see
section 6.5) supports just such a \Teff and \feh match, balancing
the spectroscopic and photometric properties to resemble the
Sun's. Its chromospheric activity, judged by the H and K fill-in
(Table~\ref{UV-results}) is equal to the solar one, and its
absolute magnitude agrees with the solar one to nearly 1$\sigma$.
It is thus a good photometric {\it and} UV match to the Sun, but
not a real solar analog.

\object{HD\,88072} excellently matches the Sun in the photometric
sense, but its absolute bolometric magnitude points to a more
luminous and evolved star. Since its atmospheric parameters match
the solar ones very closely, it appears to be a good solar analog
candidate but not a solar twin case, but it is a close enough
match to warrant further study.

\object{HD\,187237} is a very good solar photometric match, but it
is very likely richer in metals than the Sun, at \feh = $+$0.16
$\pm$ 0.04 dex. Its bolometric magnitude agrees well with the
Sun's; yet its \Teff is probably hotter, explaining its position
in the HR diagram very close to the ZAMS. A good photometric
match, but neither a solar analog nor a solar twin candidate.

Concerning those stars in the brighter Hipparcos sample now with
two \Teff determinations and therefore more reliable data, still
adopting S$_{\rm C}$ $\leq$ 1.50, as a metric and considering
H$\alpha$ radiative losses from \cite{lyraportodemello2005}, we
have the following for each.

\object{HD\,24293} photometrically matches the Sun well, but it is
possibly cooler than the Sun with its lower average \Teff = 5735K;
its absolute bolometric magnitude is marginally solar at
2$\sigma$. Its UV data show the same chromospheric activity level
as the Sun, a similar CN feature but a weaker CH feature{, and its
activity as judged by the H$\alpha$ line agrees with the solar
level}. We conclude it is a possible solar analog but an unlikely
solar twin candidate.

\object{HD\,25874} has atmospheric parameters that are
indistinguishable from solar besides a good S$_{\rm C}$ index, so
it is an excellent solar analog. Its luminosity is higher than
solar, reliably established with $\sigma$(Tycho M$_{\rm bol}$) =
0.03 dex. Activity as judged by the H$\alpha$ line agrees with the
solar level. Our conclusion is that it is not a solar twin
candidate, but a prime solar analog.

\object{HD\,71334} is an excellent photometric match to the Sun
and also has atmospheric parameters very close to solar. Its FEROS
data and additional H$\alpha$ \Teff determination establish it as
a good solar twin candidate since its luminosity matches the Sun's
and it appears more inactive than the Sun in its H and K fill-in,
while its activity level as judged by the H$\alpha$ line agrees
with the Sun's. Its CH feature matches the Sun's, but the CN band
is much weaker. This object warrants closer study.

\object{HD\,73350} photometrically matches the Sun very closely,
but is probably richer in metals. It has both FEROS and OPD data,
so this is probably a robust result. \Teff is solar but luminosity
is lower with 2$\sigma$ reliability. It lies close to the ZAMS and
is a well known very active star \citep{lyraportodemello2005} with
much higher activity than the Sun, so it is a good photometric
match but not a solar analog or twin candidate. We note that
\cite{hardorp1982} did not consider it as a good UV match to the
Sun.

\object{HD\,117939} is an excellent photometric match for the Sun
and also has atmospheric parameters very close to solar. Since its
luminosity also matches the Sun, it is a good solar twin candidate
and deserves additional analysis, but its activity level from the
H$\alpha$ line is higher than solar, F'$_{\rm H\alpha}$ = 5.14
$\pm$ 0.45 (1$\sigma$).

\object{HD\,134664} is a close solar photometric match, but its
metallicity \feh = $+$0.13 $\pm$ 0.08 dex is not as good a  match.
\Teff is solar within the errors and the luminosity also matches
the Sun's within 1$\sigma$, yet this star has the largest parallax
error in the brighter Hipparcos sample. Its activity level from
the H$\alpha$ line is lower than solar, F'$_{\rm H\alpha}$ = 2.33
$\pm$ 0.45 (1$\sigma$), and we conclude it is a probable solar
analog but only a marginal solar twin candidate.

\object{HD\,138573} matches the Sun both photometrically and in
its atmospheric parameters. Its luminosity also agrees with the
solar one, and we conclude it is a good solar twin candidate, but
for a much enhanced radiative loss in the H$\alpha$ line, F'$_{\rm
H\alpha}$ = 6.22 $\pm$ 0.45 (1$\sigma$), a value compatible with
the Hyades cluster \citep{lyraportodemello2005}.

\object{HD\,146233} is the well known solar twin 18 Sco,
\object{HR\,6060} and the only star in our Hipparcos sample
brighter than V$^{\rm Tycho}$ = 6.0 (besides \object{HD\, 30495}).
It is an excellent photometric match in the S$_{\rm C}$ index, its
atmospheric parameters are solar within 1$\sigma$, and luminosity,
mass, and age are all very close to solar. Our method establishes
it firmly as a good solar twin candidate, lending confidence that
additional candidates can be thus revealed. Nevertheless its UV
features are not exactly solar: the CN band is weaker, and it
appears slightly less active than the Sun in the H and K
chromospheric fill-in. Its activity level from the H$\alpha$ line
is slightly lower than solar, F'$_{\rm H\alpha}$ = 2.71 $\pm$ 0.45
(1$\sigma$). We note that \cite{hardorp1982} did not consider
\object{HD\,146233} as a close UV match to the Sun.

\object{HD\,150248} is an excellent photometric match to the Sun,
and its atmospheric parameters and luminosity are sunlike within
1$\sigma$. Mass and age also agree very well, the H$\alpha$
radiative loss is solar within the errors, and thus it is still
another very good solar twin candidate.

\object{HD\,164595} is also a good photometric match and has
atmospheric parameters and luminosity within 1$\sigma$ of the
solar ones, the H$\alpha$ radiative loss is solar within the
errors, and this star is another excellent solar twin candidate.
Interestingly, \cite{fesenko1994} mentions this star as the one
most resembling the Sun, photometrically, in his survey of 10 700
stars with WBVR magnitudes in the Moscow Photometric Catalog.

\object{HD\,189625} is a good photometric match but is probably
metal-richer than the Sun, with \feh = $+$0.27 $\pm$ 0.08 dex. It
is also possibly hotter with $<$\Teff$>$ = 5840K, a likely
explanation for its good photometric similarity. Absolute
magnitude is marginally solar within 2$\sigma$, and mass is
probably higher than solar. The H$\alpha$ radiative loss is
slightly enhanced relative to solar, F'$_{\rm H\alpha}$ = 4.55
$\pm$ 0.45 (1$\sigma$). It is neither a solar analog nor a solar
twin case.

\object{HD\,190771} is probably hotter and more metal-rich than
the Sun, but a good photometric match. Its UV data point to
different CN and CH features and a much stronger chromospheric
fill-in in the H and K lines, which agrees with its evolutionary
position close to the ZAMS, despite absolute magnitude agreeing
with the solar one. This is confirmed by a much higher
chromospheric flux in the H$\alpha$ line, F'$_{\rm H\alpha}$ =
9.09 $\pm$ 0.45 (1$\sigma$), a value compatible with the Hyades
cluster or the very young Ursa Major moving group
\citep{lyraportodemello2005}. Neither a solar analog nor a solar
twin.

\object{HD\,207043} matches the Sun well photometrically and has
atmospheric parameters within the Sun's at the 1$\sigma$ level,
and yet its absolute magnitude and evolutionary position point to
a younger star that is much less evolved than the Sun, which is
confirmed by its higher H$\alpha$ radiative loss, F'$_{\rm
H\alpha}$ = 4.76 $\pm$ 0.45 (1$\sigma$). Thus, a very good solar
analog but no solar twin candidate.

The analysis of the atmospheric parameters, photometric
similarities, and the evolutionary state of the brighter stars in
the Hipparcos sample yields, therefore, three possible solar
analogs, \object{HD\,9986} and \object{HD\,88072} (with only one
\Teff determination) and \object{HD\,24293} (two \Teff
determinations), as well as two definite very good solar analogs,
\object{HD\,25874} (with higher luminosity than the Sun) and
\object{HD\,207043} (with lower luminosity). The real bounty,
however, is three possible solar twin candidates,
\object{HD\,71334}, \object{HD\,117939}, and \object{HD\,134664},
and three new excellent solar twin candidates: \object{HD\,138573}
(more active than the Sun, though), \object{HD\,150248}, and
\object{HD\,164595}, besides a confirmation by our method of the
known twin \object{HD\,146233}, \object{18 Sco}. Existing data for
H$\alpha$ radiative losses, however, point to \object{HD\,117939}
and \object{HD\,138573} having more enhanced chromospheric
activity than the Sun.

\subsubsection{Fainter Hipparcos sample}

The stars of the fainter Hipparcos sample with S$_{\rm C}$ $\leq$
1.50 all have two \Teff determinations, and there are eight cases
to discuss. Errors in absolute bolometric magnitude in this sample
range from 0.10 to 0.15, and conclusions concerning solar twin
candidacy carry correspondingly less weight than in the brighter
Hipparcos sample.

\object{HD\,12264} is a good photometric match and has atmospheric
parameters that all match the Sun's within 1$\sigma$. The absolute
bolometric magnitude also matches the solar one, but its UV
spectrum is very unlike the solar one, also presenting much
stronger fill-in in the H and K lines, confirmed by a much higher
chromospheric flux in the h$\alpha$ line, F'$_{\rm H\alpha}$ =
6.12 $\pm$ 0.45 (1$\sigma$). Thus, it is a good solar analog but
an unlikely solar twin candidate: the higher errors in luminosity
could have masked an evolutionary state close to the solar one,
but existing data on activity points towards a much younger star.

\object{HD\,98649} closely matches the Sun in photometry,
atmospheric parameters, and absolute magnitude. It has no UV data,
but the chromospheric flux gauged by H$\alpha$ is solar within
1$\sigma$. It is an excellent solar analog and a clear case for
solar twin candidacy, besides one of the lowest errors in absolute
magnitude in the fainter Hipparcos sample.

\object{HD\,115382} is a good photometric match and has
atmospheric parameters very close to solar. Its H$\alpha$
chromospheric flux also matches the Sun's, and the absolute
magnitude agrees with the solar value though only within a large
error of 0.16. Thus it is a good solar analog and still a solar
twin candidate.

\object{HD\,118598} is a nearly perfect photometric match to the
Sun, besides having atmospheric parameters and absolute magnitude
closely solar. Its H$\alpha$ radiative loss also agrees with the
Sun's, and its absolute magnitude error of 0.12 dex is at the
lower end in the fainter Hipparcos sample. We conclude it is a
good solar analog and a solar twin candidate.

\object{HD\,140690} is an interesting case in that it was selected
only in an incipient version of our Hipparcos color-absolute
magnitude boxes. It is more luminous than the Sun, a good
photometric match to it, and has atmospheric parameters that are
indistinguishable from solar. It is clearly a very good solar
analog but not a solar twin. In the UV, however, it is the closest
match to the Sun in our sample, indistinguishable in the CN and CH
features and also in the chromospheric fill-in in the H and K
lines. Its H$\alpha$ radiative loss also matches the Sun's, and it
is therefore a very good solar analog {\it and} a perfect UV
analog. It is a very interesting spectrophotometric proxy of the
Sun in a very wide wavelength range.

\object{HD\,153458} is an excellent photometric match to the Sun
but is without doubt hotter and richer in metals, with $<$\Teff$>$
= 5830K and \feh = $+$0.20 $\pm$ 0.08 dex. Its absolute magnitude
agrees very well with the Sun's, and the H$\alpha$ radiative loss
is much higher than solar, F'$_{\rm H\alpha}$ = 6.15 $\pm$ 0.45
(1$\sigma$). A good photometric match, but neither a solar analog
nor a twin.

\object{HD\,157750} is a fair photometric match but is another
case of a star hotter and richer in metals, with $<$\Teff$>$ =
5845K and \feh = $+$0.21 $\pm$ 0.04 dex. Its absolute magnitude
agrees very well with the Sun's, and the H$\alpha$ radiative loss
is much higher than solar, F'$_{\rm H\alpha}$ = 6.01 $\pm$ 0.45
(1$\sigma$). In the UV it has a solar CN feature but a weaker CH
one, besides much stronger fill-in in the H and K lines, in good
agreement with the H$\alpha$ data and an inferred evolutionary
position close to the ZAMS. We deem it a reasonable photometric
match, but neither a solar analog nor a twin.

Lastly, \object{HD\,BD$+$15 3364} is a fair photometric match with
atmospheric parameters, and absolute magnitude (with an error of
0.16) in very good agreement with the solar ones, and also a solar
level of radiative losses in H$\alpha$. This object is clearly a
good solar analog, and also a solar twin candidate:
\cite{hardorp1982} mentioned it as a close match to the solar UV
spectrum.

\subsection{UBV similarity and Hardorp list stars}

These stars were only selected on the basis of UBV similarity to
the Sun and presence in Hardorp's lists, and are therefore not
expected to have much resemblance to the solar atmospheric
parameters and state of evolution. Five stars from Hardorp's lists
were spectroscopically analyzed. \object{HD\,105590} has
atmospheric parameters close to solar (but only one \Teff
determination) but a very unsolar-like S$_{\rm C}$ index:
\cite{hardorp1982} regarded it, though, as a solar analog.
\object{HD\,186408} was not considered by Hardorp as a close case
as solar analogs go, but it is photometrically very similar to the
Sun in the S$_{\rm C}$ index: its \Teff is close to solar (but
this is judged from a single determination) and its \feh appears
higher than solar within the uncertainty. Previous analyses
\citep{frieletal93, portodemellodasilvatwin1997} confirm this star
as a good analog but not a solar twin. Two other Hardorp stars
with S$_{\rm C}$ indices close to solar turn out not to be solar
analogs at all: on the one hand, \object{HD\,159222}, considered a
close solar UV match by Hardorp, is hotter, richer in metals, and
more evolved than the Sun; on the other hand, \object{HD\,101563},
considered by Hardorp as a bad UV match to the Sun, is much more
massive, more evolved, and poorer in metals. One more Hardorp
star, but only judged by its UV spectroscopy, is
\object{HD\,28255}: it does not resemble the Sun in either its CN
and CH features or its S$_{\rm C}$ index.

Eight stars were selected by having solar UBV colors within the
adopted errors. Four of these have only UV spectroscopy, and only
one, \object{HD\,16141} turns out to have CN and CH features
resembling the Sun's (but a weaker chromospheric fill in the
\ion{Ca}{ii} H and K lines; we also note that \citet{hardorp1982}
found it as a bad UV match to the Sun, at variance with us. Its
S$_{\rm C}$ index is, however, very non-solar. The remaining four
stars with solar UBV colors were spectroscopically analyzed, but
none are photometrically similar to the Sun or has a strong
resemblance in atmospheric and/or evolutionary parameters.

The results for these stars further illustrate the danger of
choosing solar analogs from scarce data.

\subsection{Ultraviolet matches to the solar spectrum}

There are 37 stars that could be compared to the Sun in the UV
range at low resolution (Table~\ref{UV-results}). Still
undiscussed are \object{HD\,26767} and \object{HD\,43180} from the
fainter Hipparcos sample, and for which no additional data is
available but for the UV spectroscopy: none of them closely
resemble the Sun in their UV features or have a sunlike S$_{\rm
C}$ index. We found four stars with both a strong S$_{\rm C}$
similarity to the Sun and solar-like CN/CH features. Two of these
have already been discussed above: \object{HD\,66653} is
photometrically similar to the Sun, but this is probably owing, as
seen above, to a combination of higher \Teff and \feh. Its UV
features are very similar to the Sun, including the chromospheric
fill-in in the HK lines, and it is proposed as a good solar proxy
in the UV. The other one, \object{HD\,140690}, is a very good
solar analog strongly resembling the Sun in its S$_{\rm C}$ index,
besides having the UV spectrum indistinguishable from the solar
one. It is a rare case in which the atmospheric parameters, the
Paschen colors, and the UV spectral features are very solar-like,
and it can therefore be proposed as an excellent photometric
analog of the Sun in a wide wavelength range, a very interesting
object indeed. Nevertheless it is not a solar twin candidate,
since it is more luminous and probably more evolved than the Sun.

There are two other stars with S$_{\rm C}$ $\leq$ 1.50 and very
solar-like UV features. The first is \object{HD\,159656}. It was
selected from Hardorp's lists, is a good photometric match to the
Sun, but is definitely hotter than the Sun, as well as probably
being richer in metals. \cite{hardorp1982} did not mention it as a
good solar match in the UV. Its chromospheric H and K fill-in is
also much stronger than the Sun's. Secondly we have
\object{HD\,221343}: it is not a good photometric match, its
atmospheric parameters are determined from poor S/N data, and its
H and K emission is much stronger than solar. Our results suggest
it is probably hotter and richer in metals than the Sun. Our
analysis thus presents only \object{HD\,140690} as a truly good
photometric analog to the Sun also with a UV spectrum that is
strongly solar-like.

Two other objects merit comment: \object{HD\,16141}, already
discussed above, is a very poor photometric match to the Sun, and
it only has UV spectroscopic data, but its CN and CH features are
indistinguishable from solar. Its H and K emission is weaker than
solar. Its purely photometric \Teff and \feh point to its being
cooler and metal-poorer than the Sun. Finally, \object{HD\,68168}
is not a good photometric solar match, but again it has very
sun-like CN and CH features, yet weaker H and K fill in. Our
spectroscopic analysis suggests it is richer in metals than the
Sun but it possesses solar \Teff within errors, based on only one
\Teff determination.

The UV wavelength range is thus a very fine discriminator of solar
analogs and it is apparent that the CN and CH features, even in
low resolution, can bring out differences in \Teff and \feh
between stars and the Sun which are, at best, very hard to reveal
by spectroscopic analyses. The UV approach clearly warrants deeper
analysis with more data, which we plan to present in a forthcoming
paper.

\begin{figure*}
\begin{center}
\resizebox{9.1cm}{!}{\includegraphics{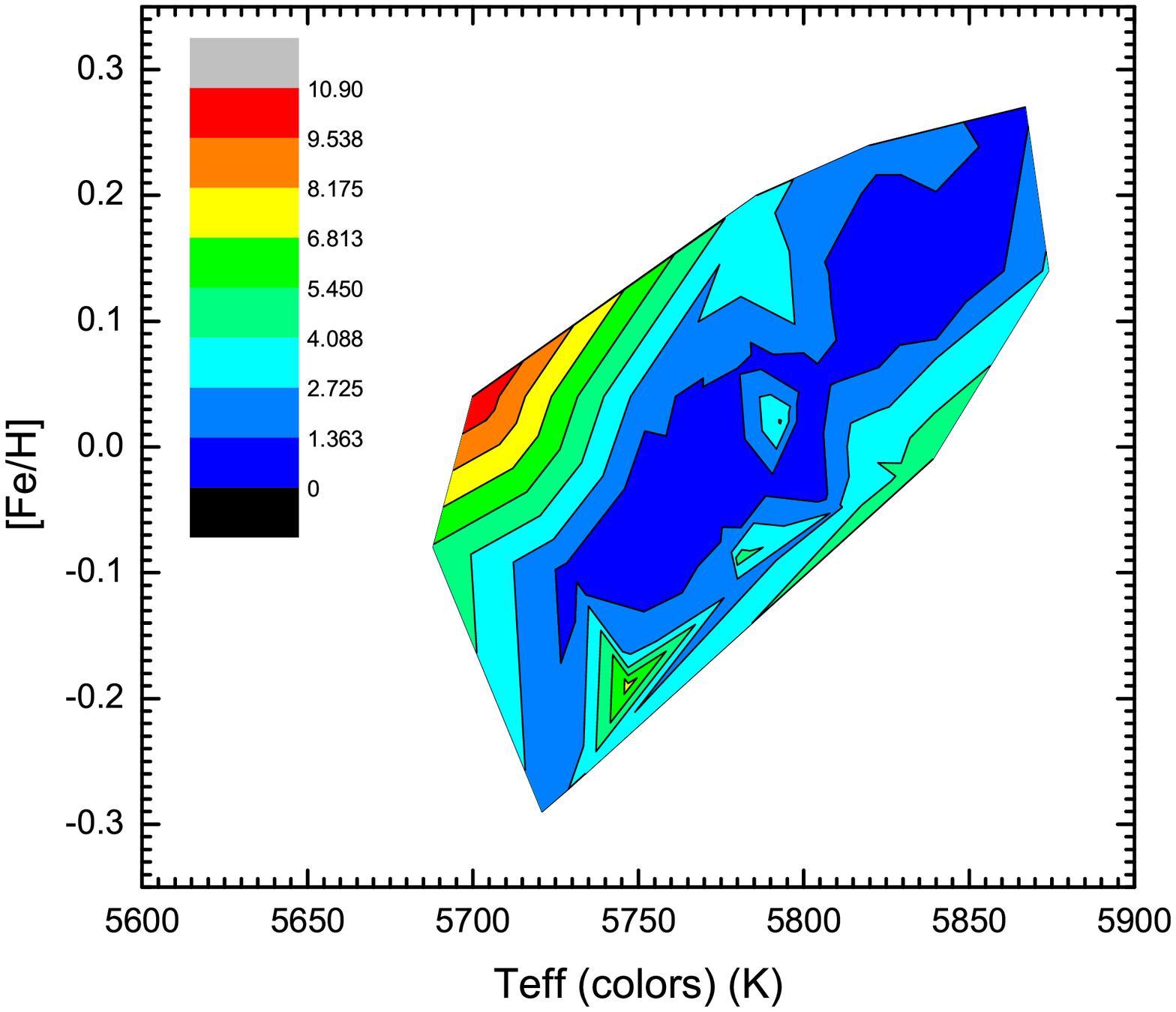}}
\resizebox{9.1cm}{!}{\includegraphics{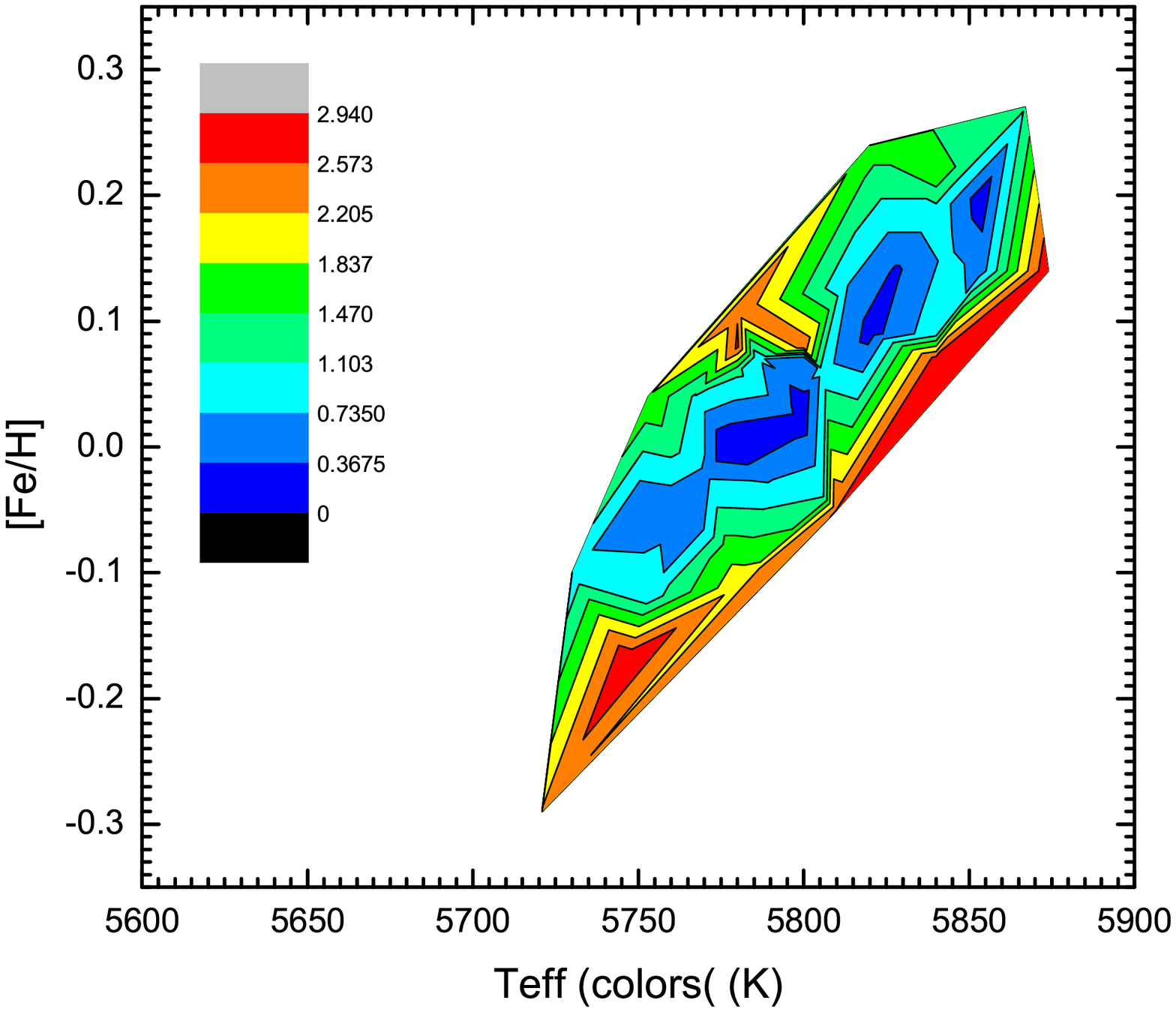}}
\end{center}
\caption[]{{\it Left}. The color similarity index S$_{\rm C}$
plotted {\it versus} the final photometric \Teffs (obtained from
the spectroscopic $[$Fe/H$]$s) and the spectroscopic $[$Fe/H$]$s
for all spectroscopically analyzed stars. The vertical colored bar
is coded by the S$_{\rm C}$ values: note that the scale of the
coding is different between the two plots. {\it Right}. The same
as the left panel, but for the stars with S$_{\rm C}$ $\leq$ 3.0.
It is clear that a combination of hot/metal rich and cool/metal
poor parameters defines an area of good photometric similarity to
the Sun, and also that the stars with the highest color similarity
to the Sun have atmospheric parameters more tightly clustered
around the solar values.} \label{scdetail}
\end{figure*}

Good solar twins appear unequivocally linked to a fair photometric
similarity to the Sun as inferred from our S$_{C}$ index. In
Figure~\ref{scdetail} we plot our \Teffs {\it versus} the
spectroscopic \fehs, shown as S$_{\rm C}$ contours, for two
different regimes of color similarity to the Sun: all analyzed
stars and only those with S$_{\rm C}$ $\leq$ 3.0. This plot
illustrates in greater detail what has already been gleamed from
Figure~\ref{teff-feh-phot}: stars with atmospheric parameters very
similar to the Sun's automatically produce high similarity in
S$_{\rm C}$ as well, but there is a locus in which stars with
\Teff and \feh values quite different from solar may mimic a high
degree of similarity to the Sun. Roughly, for every $+$0.1 dex
increase in \feh, a parallel $+$36 K increase in \Teff leaves
S$_{\rm C}$ unchanged. A spectroscopic analysis is then needed to
remove the degeneracy and separate true solar analogs from stars
merely mimicking a strong spectrophotometric similarity to the
Sun.

\subsection{New solar twin candidates}

In Table~\ref{new-solar-twin-list} we list ten stars pointed out
by our survey as interesting new solar twin candidates, along with
the Sun and the known twin \object{HD\,146233} (a.k.a. \object{18
Sco}) for comparison. At the top of the list there are six stars
from the brighter Hipparcos sample, three of them ``probable''
twins, and three only ``possible'' twins, with weaker claims. The
atmospheric and evolutionary parameters are shown with their
errors (excepting mass, for which errors are usually 0.02-0.03
solar masses, and age, for which errors are generally so large as
to preclude definite conclusions), and we also provide comments on
the chromospheric activity level. The last four entries in the
table correspond to stars from the fainter Hipparcos sample, two
of these being ``probable'' solar twins, and two more classified
as ``possible'' twins. The very best candidates are
\object{HD\,150248} and \object{HD\,164595}, which match the Sun
well in every parameter plus the level of chromospheric activity
and which belong to the brighter Hipparcos sample, so have reduced
errors in parallax and luminosity. In the fainter Hipparcos
sample, \object{HD\,98649} and \object{HD\,118598}, which also
match the Sun perfectly, have reasonable uncertainties in
luminosity, and are also chromospherically inactive. These stars
will be subjected to a more detailed scrutiny, including the
lithium abundance and additional chromospheric activity
indicators, in a forthcoming paper.

\object{HD\,146233}, \object{18 Sco}, remains the only one bright
(V$^{\rm Tycho}$ $\leq$ 6.0) solar twin candidate or confirmed
solar twin known so far. We may ask, given the completeness of the
data used as input in our survey, what the probability is of
finding still other solar twin candidates among the V$^{\rm
Tycho}$ $\leq$ 8.0 stars. This can be roughly estimated as
follows. Among the G-type, V$^{\rm Tycho}$ $\leq$ 8.0 stars in the
Hipparcos catalog, completeness in the $(B-V)^{\rm Tycho}$ color
is 95\%. We selected 52 stars for our survey within this magnitude
limit, and supposing that 5\% are missing, there are 2.6 stars we
failed to select. We spectroscopically analyzed 30 stars among the
52 star sample, and found four ``probable'' solar twin candidates,
which, plus the previously know solar twins \object{HD\,146233}
and \object{HD\,98618}, gives a total of six solid twin
candidates. Thus, among 30 analyzed stars, we have six twin
candidates, a rate of 20\%. There are, as a consequence, 2.6/5
$\sim$ 0.5 stars missing from our survey, owing to incompleteness,
which are probable twin candidates. Now, because there are seven
stars in our brighter sample for which we could not secure
spectroscopic data (leaving aside \object{HD\,98618}, accepted as
a known twin) and which have S$_{\rm C}$ $\leq$ 1.50, there
remains a possibility that $\sim$1.4 of these are probable twins
that we have so far failed to identify. Given that among 52 stars
selected for the brighter Hipparcos sample, only two, or 3.8\%,
are brighter than V$^{\rm Tycho}$ = 6.0, there is at best
(1.4+0.5) times 0.038 $\sim$ 0.076 stars with V$^{\rm Tycho}$
$\leq$ 6.0, which are probable solar twin candidates, and we
failed either to select in the first place or to analyze
specotroscopically. This figure is probably overestimated since
completeness in the Hipparcos catalog falls off between V$^{\rm
Tycho}$ $\leq$ 6.0 stars and our magnitude limit V$^{\rm Tycho}$ =
8.5, meaning that it is very unlikely that solar twins any
brighter than \object{HD\,146233} remain undetected.

\begin{table}
\centering \caption[]{Parameters derived from the HR diagram
analysis for the V$^{\rm Tycho}$ $\leq$ 8.0 stars. First column
shows the HD number, second column the astrometric surface gravity
and uncertainty, third column the absolute bolometric magnitude in
the V$^{\rm Tycho}$ band and uncertainty, and fourth and fifth
columns are the mass (in solar units) and age (in Gyr),
respectively. The Sun is shown in the first row for comparison.}
\label{massage-bright}
\begin{tabular}{l r@{\,}l r@{\,}l c c}
\hline\hline \noalign{\smallskip} HD &
\multicolumn{2}{c}{$\log{g}^{\rm \,astr}$} &
\multicolumn{2}{c}{$M_{\rm bol}$} &
\parbox[c]{1.0cm}{\centering Mass \\ {\tiny ($M_\odot$)}} &
\parbox[c]{1.0cm}{\centering Age \\ {\tiny (Gyr)}} \\
\noalign{\smallskip} \hline \noalign{\smallskip}
Sun         & 4.44 &          & 4.81 &$\pm$ 0.03 & 1.00 & 4.6  \\
4308        & 4.37 &$\pm$0.05 & 4.82 &$\pm$ 0.03 & 0.88 & 9.2  \\
9986        & 4.43 &$\pm$0.05 & 4.73 &$\pm$ 0.05 & 1.02 & 3.4  \\
13724       & 4.44 &$\pm$0.07 & 4.72 &$\pm$ 0.09 & 1.06 & 2.2  \\
19518       & 4.40 &$\pm$0.07 & 4.76 &$\pm$ 0.10 & 0.96 & 6.1  \\
24293       & 4.39 &$\pm$0.07 & 4.70 &$\pm$ 0.09 & 0.97 & 6.3  \\
25874       & 4.39 &$\pm$0.05 & 4.69 &$\pm$ 0.03 & 0.99 & 1.5  \\
28821       & 4.35 &$\pm$0.07 & 4.73 &$\pm$ 0.09 & 0.93 & 8.0  \\
30495       & 4.50 &$\pm$0.05 & 4.83 &$\pm$ 0.04 & 1.05 & 2.6  \\
32963       & 4.48 &$\pm$0.07 & 4.85 &$\pm$ 0.09 & 1.03 & 2.5  \\
35041       & 4.45 &$\pm$0.07 & 4.83 &$\pm$ 0.09 & 0.99 & 4.0  \\
37773       & 4.39 &$\pm$0.08 & 4.78 &$\pm$ 0.10 & 0.97 & 6.7  \\
66653       & 4.46 &$\pm$0.06 & 4.75 &$\pm$ 0.06 & 1.05 & 2.2  \\
68168       & 4.41 &$\pm$0.07 & 4.72 &$\pm$ 0.09 & 1.02 & 4.4  \\
71334       & 4.44 &$\pm$0.07 & 4.84 &$\pm$ 0.08 & 0.97 & 5.1  \\
73350       & 4.51 &$\pm$0.06 & 4.90 &$\pm$ 0.06 & 1.05 & ZAMS \\
88072       & 4.40 &$\pm$0.07 & 4.67 &$\pm$ 0.09 & 1.01 & 5.0  \\
117939      & 4.42 &$\pm$0.06 & 4.87 &$\pm$ 0.06 & 0.94 & 6.1  \\
134664      & 4.46 &$\pm$0.08 & 4.78 &$\pm$ 0.11 & 1.04 & 2.6  \\
138573      & 4.41 &$\pm$0.06 & 4.77 &$\pm$ 0.06 & 0.98 & 5.6  \\
142072      & 4.44 &$\pm$0.07 & 4.77 &$\pm$ 0.10 & 1.03 & 2.8  \\
145825      & 4.51 &$\pm$0.06 & 4.89 &$\pm$ 0.05 & 1.05 & ZAMS \\
146233      & 4.42 &$\pm$0.05 & 4.77 &$\pm$ 0.03 & 0.98 & 5.0  \\
150248      & 4.39 &$\pm$0.06 & 4.75 &$\pm$ 0.07 & 0.96 & 6.2  \\
155114      & 4.46 &$\pm$0.07 & 4.81 &$\pm$ 0.08 & 1.00 & 2.7  \\
164595      & 4.44 &$\pm$0.05 & 4.79 &$\pm$ 0.05 & 0.99 & 4.5  \\
187237      & 4.50 &$\pm$0.05 & 4.85 &$\pm$ 0.04 & 1.06 & ZAMS \\
189625      & 4.47 &$\pm$0.07 & 4.72 &$\pm$ 0.08 & 1.09 & 1.0  \\
190771      & 4.50 &$\pm$0.05 & 4.85 &$\pm$ 0.03 & 1.06 & ZAMS \\
207043      & 4.52 &$\pm$0.06 & 4.95 &$\pm$ 0.07 & 1.04 & ZAMS \\
214385      & 4.40 &$\pm$0.07 & 4.88 &$\pm$ 0.10 & 0.87 & 7.8  \\
\noalign{\smallskip} \hline\hline
\end{tabular}
\end{table}

\begin{table}
\centering \caption[]{Same as Table~\ref{massage-bright} for the
8.0 $< V^{\rm Tycho}$ $\leq$ 9.0 stars.} \label{massage-faint}
\begin{tabular}{l r@{\,}l r@{\,}l c c}
\hline\hline \noalign{\smallskip} HD &
\multicolumn{2}{c}{$\log{g}^{\rm \,astr}$} &
\multicolumn{2}{c}{$M_{\rm bol}$} &
\parbox[c]{1.0cm}{\centering Mass \\ {\tiny ($M_\odot$)}} &
\parbox[c]{1.0cm}{\centering Age \\ {\tiny (Gyr)}} \\
\noalign{\smallskip} \hline \noalign{\smallskip}
8291        & 4.45 &$\pm$0.10 & 4.78 &$\pm$ 0.16 & 1.02 & 3.9  \\
12264       & 4.45 &$\pm$0.08 & 4.80 &$\pm$ 0.12 & 1.02 & 3.4  \\
98649       & 4.44 &$\pm$0.08 & 4.84 &$\pm$ 0.10 & 0.98 & 4.7  \\
105901      & 4.43 &$\pm$0.08 & 4.74 &$\pm$ 0.12 & 0.99 & 3.8  \\
115382      & 4.39 &$\pm$0.10 & 4.73 &$\pm$ 0.16 & 0.96 & 6.1  \\
118598      & 4.44 &$\pm$0.08 & 4.78 &$\pm$ 0.12 & 1.01 & 4.3  \\
140690      & 4.39 &$\pm$0.08 & 4.66 &$\pm$ 0.11 & 1.01 & 5.3  \\
143337      & 4.35 &$\pm$0.09 & 4.72 &$\pm$ 0.14 & 0.91 & 7.9  \\
153458      & 4.50 &$\pm$0.08 & 4.82 &$\pm$ 0.11 & 1.07 & ZAMS \\
157750      & 4.50 &$\pm$0.09 & 4.84 &$\pm$ 0.15 & 1.06 & ZAMS \\
191487      & 4.43 &$\pm$0.10 & 4.74 &$\pm$ 0.15 & 1.00 & 4.5  \\
202072      & 4.35 &$\pm$0.08 & 4.70 &$\pm$ 0.12 & 0.92 & 7.9  \\
211786      & 4.46 &$\pm$0.08 & 4.89 &$\pm$ 0.10 & 0.97 & 4.3  \\
216436      & 4.37 &$\pm$0.09 & 4.65 &$\pm$ 0.11 & 0.99 & 6.1  \\
221343      & 4.44 &$\pm$0.09 & 4.78 &$\pm$ 0.13 & 1.01 & 4.0  \\
BD+15\,3364 & 4.44 &$\pm$0.10 & 4.77 &$\pm$ 0.16 & 1.02 & 3.8  \\
\noalign{\smallskip} \hline\hline
\end{tabular}
\end{table}

\begin{table}
\centering \caption[]{Same as Table~\ref{massage-bright} for the
stars with UBV colors similar to the solar ones. \object{HD\,
105590} has a very large parallax error, and no age was derived
for it.} \label{massage-sim}
\begin{tabular}{l r@{\,}l r@{\,}l c c}
\hline\hline \noalign{\smallskip} HD &
\multicolumn{2}{c}{$\log{g}^{\rm \,astr}$} &
\multicolumn{2}{c}{$M_{\rm bol}$} &
\parbox[c]{1.0cm}{\centering Mass \\ {\tiny ($M_\odot$)}} &
\parbox[c]{1.0cm}{\centering Age \\ {\tiny (Gyr)}} \\
\noalign{\smallskip} \hline \noalign{\smallskip}
94340       & 4.19 &$\pm$0.08 & 3.98 &$\pm$ 0.10 & 1.14 & 4.9  \\
101563      & 3.89 &$\pm$0.10 & 3.24 &$\pm$ 0.08 & 1.22 & 3.6  \\
105590      & 4.34 &$\pm$0.28 & 4.54 &$\pm$ 0.60 & 1.00 & --   \\
111398      & 4.46 &$\pm$0.08 & 4.84 &$\pm$ 0.12 & 1.02 & 3.3  \\
119550      & 3.85 &$\pm$0.09 & 3.01 &$\pm$ 0.13 & 1.31 & 2.9  \\
159222      & 4.43 &$\pm$0.05 & 4.66 &$\pm$ 0.04 & 1.05 & 2.8  \\
159656      & 4.36 &$\pm$0.07 & 4.54 &$\pm$ 0.09 & 1.03 & 4.4  \\
186408      & 4.25 &$\pm$0.05 & 4.29 &$\pm$ 0.04 & 1.02 & 2.4  \\
221627      & 3.95 &$\pm$0.08 & 3.36 &$\pm$ 0.10 & 1.23 & 3.6  \\
\noalign{\smallskip} \hline\hline
\end{tabular}
\end{table}

\begin{table}
\centering \caption[]{Qualitative assessment of the stellar
spectral feature deviations from the solar spectrum, expressed as
measured in the ratio spectra between the stars and Ganymede. In
the second and third columns we classify the strength of ratio
features around the $\lambda$3870 CN bandheads and the
$\lambda$4310 CH bandhead, respectively. In the fourth column, the
chromospheric filling in the \ion{Ca}{ii} H and K lines is given.}
\label{UV-results}
\begin{tabular}{l c c c r@{}}
\hline\hline \noalign{\smallskip} HD &
 CN & CH & H \& K \\
\noalign{\smallskip} \hline \noalign{\smallskip}
4308        & much weaker   & solar         & weaker \\
6512        & stronger      & weaker        & solar \\
8291        & weaker        & weaker        & stronger \\
9562        & much weaker   & much weaker   & much weaker \\
12264       & weaker        & weaker        & much stronger \\
13724       & stronger      & solar         & much stronger \\
16141       & solar         & solar         & weaker \\
19467       & much weaker   & solar         & much weaker \\
19518       & much weaker   & weaker        & solar \\
24293       & solar         & weaker        & solar \\
26767       & much weaker   & much weaker   & much stronger \\
28068       & much weaker   & much weaker   & much stronger \\
28255       & much weaker   & weaker        & solar \\
28821       & much weaker   & solar         & solar \\
32963       & solar         & weaker        & solar \\
35041       & much weaker   & much weaker   & much stronger \\
37773       & much stronger & weaker        & solar \\
43180       & much weaker   & much weaker   & stronger \\
66653       & solar         & solar         & solar \\
68618       & solar         & solar         & much weaker \\
71334       & much weaker   & solar         & weaker \\
140690      & solar         & solar         & solar \\
143337      & much weaker   & solar         & solar \\
146233      & weaker        & solar         & weaker \\
155114      & much weaker   & much weaker   & much stronger \\
157750      & solar         & weaker        & much stronger \\
159222      & much weaker   & solar         & stronger \\
159656      & solar         & solar         & much stronger \\
189625      & solar         & stronger      & weaker \\
190771      & much weaker   & weaker        & much stronger \\
191487      & much weaker   & much weaker   & much stronger \\
202072      & much weaker   & solar         & solar \\
211786      & much weaker   & weaker        & solar \\
214385      & much weaker   & weaker        & weaker \\
216436      & weaker        & solar         & weaker \\
221343      & solar         & solar         & much stronger \\
221627      & much weaker   & weaker        & much weaker \\
\noalign{\smallskip} \hline\hline
\end{tabular}
\end{table}

\begin{table*}
\centering \caption{New solar twin candidates identified in this
work. Columns 2, 3, and 4 give the final atmospheric parameters
\Teff~(K), \feh and $\log{g}$, respectively. Column 5 presents the
apparent magnitude in the Tycho band, and column 6 the color
similarity index S$_{\rm C}$. Columns 7, 8, and 9 provide the
absolute bolometric magnitude in the Tycho band, mass, and age
respectively. The tenth column comments on the chromospheric
activity level as compared to the Sun, when available, and gives
relevant remarks.} \label{new-solar-twin-list}
\begin{tabular}{l c c c c c c c c c c c}
\hline\hline \noalign{\smallskip} HD &
\parbox[c]{1.0cm}{\centering $T_{\rm eff} \pm\ \sigma$ \\ {\tiny [K]}} &
$\log{g}^{\rm \,astr} \pm\ \sigma$ & [Fe/H] $\pm\ \sigma$ &
V$^{\rm Tycho}$ & $S_{\rm C}$ & $M_{\rm bol}^{\rm Tycho} \pm\
\sigma$ &
\parbox[c]{0.8cm}{\centering Mass \\ {\tiny [$M_\odot$]}} &
\parbox[c]{0.8cm}{\centering Age \\ {\tiny [Gyr]}} &
remarks \\
\noalign{\smallskip} \hline \noalign{\smallskip}
Sun    & 5777          & 4.44            &   +0.00            & --  & 0.00 & 4.81 $\pm$ 0.03 & 1.00 & 4.6 &  -- \\
\noalign{\smallskip} \hline \noalign{\smallskip}
71334  & 5710 $\pm$ 30 & 4.44 $\pm$ 0.07 & $-$0.06 $\pm$ 0.04 & 7.9 & 0.22 & 4.84 $\pm$ 0.08 & 0.97 & 5.1 & inactive; possible twin \\
\noalign{\smallskip} \hline \noalign{\smallskip}
117939 & 5765 $\pm$ 40 & 4.42 $\pm$ 0.06 & $-$0.10 $\pm$ 0.08 & 7.4 & 0.83 & 4.87 $\pm$ 0.06 & 0.94 & 6.1 & active; possible twin \\
\noalign{\smallskip} \hline \noalign{\smallskip}
134664 & 5820 $\pm$ 40 & 4.46 $\pm$ 0.08 &   +0.13 $\pm$ 0.08 & 7.8 & 0.76 & 4.78 $\pm$ 0.11 & 1.04 & 2.6 & inactive; possible twin \\
\noalign{\smallskip} \hline \noalign{\smallskip}
138573 & 5750 $\pm$ 40 & 4.41 $\pm$ 0.06 &   +0.00 $\pm$ 0.08 & 7.3 & 1.24 & 4.77 $\pm$ 0.06 & 0.98 & 5.6 & very active; probable twin \\
\noalign{\smallskip} \hline \noalign{\smallskip}
146233 & 5795 $\pm$ 30 & 4.42 $\pm$ 0.05 & $-$0.03 $\pm$ 0.04 & 5.6 & 0.78 & 4.77 $\pm$ 0.03 & 0.98 & 5.0 & known solar twin \\
\noalign{\smallskip} \hline \noalign{\smallskip}
150248 & 5750 $\pm$ 40 & 4.39 $\pm$ 0.06 & $-$0.04 $\pm$ 0.08 & 7.1 & 0.52 & 4.75 $\pm$ 0.07 & 0.96 & 6.2 & inactive; very probable twin \\
\noalign{\smallskip} \hline \noalign{\smallskip}
164595 & 5790 $\pm$ 40 & 4.44 $\pm$ 0.05 & $-$0.04 $\pm$ 0.08 & 7.1 & 1.11 & 4.79 $\pm$ 0.05 & 0.99 & 4.5 & inactive; very probable twin \\
\noalign{\smallskip} \hline \noalign{\smallskip}
98649  & 5775 $\pm$ 30 & 4.44 $\pm$ 0.08 & $-$0.02 $\pm$ 0.04 & 8.1 & 0.58 & 4.84 $\pm$ 0.10 & 0.98 & 4.7 & inactive; probable twin \\
\noalign{\smallskip} \hline \noalign{\smallskip}
115382 & 5775 $\pm$ 30 & 4.39 $\pm$ 0.10 & $-$0.08 $\pm$ 0.04 & 8.5 & 1.16 & 4.73 $\pm$ 0.16 & 0.96 & 6.1 & inactive; possible twin \\
\noalign{\smallskip} \hline \noalign{\smallskip}
118598 & 5755 $\pm$ 40 & 4.44 $\pm$ 0.08 &   +0.02 $\pm$ 0.08 & 8.3 & 0.12 & 4.78 $\pm$ 0.12 & 1.01 & 4.3 & inactive; probable twin \\
\noalign{\smallskip} \hline \noalign{\smallskip}
BD+15 3364 & 5785 $\pm$ 30 & 4.44 $\pm$ 0.10 &    +0.07 $\pm$ 0.08 & 8.2 & 1.38 & 4.77 $\pm$ 0.16 & 1.02 & 3.8 & inactive; possible twin \\
\noalign{\smallskip} \hline\hline
\end{tabular}
\end{table*}

\section{Conclusions}

We have reported a photometric and spectroscopic survey of solar
twin stars that is photometrically all-sky, complete out to 40 pc,
and partially complete out to 50 pc, and involving 136 solar-type
stars. We derived photometric \Teffs and photometric metallicities
\feh for the whole sample and ranked these stars relative to the
Sun by means of a photometric similarity index. Spectroscopic
parameters based on moderate-resolution, high-S/N spectra were
also derived for a subsample of 55 stars, and for these we derived
spectroscopic metallicities, photometric \Teffs based on the
spectroscopic metallicities, and \Teffs derived from the fitting
of H$\alpha$ profiles. Masses and ages were also provided for the
spectroscopically analyzed stars. Low-resolution UV spectra are
available for a subsample of 37 stars, allowing the evaluation of
their relative similarity with respect to the Sun in the CH and CN
molecular features, as well as the chromospheric fill-in in the
\ion{Ca}{ii} H and K lines. Our conclusions are as follows.\\

1) The color-similarity index is very successful in selecting
stars having colors $\it and$ atmospheric parameters that are very
similar to the solar ones. A large number of new solar analogs
were identified, and these objects proposed as useful
spectrophotometric proxies of the Sun, satisfying various degrees
of accuracy and covering essentially all of the sky with a
magnitude limit V$^{\rm Tycho}$ $\leq$ 8.5. They should be
particularly useful as solar proxies for photometry and/or
low-resolution spectroscopy of Solar System bodies.\\

2) Two stars were also shown to have all the near UV spectral
features indistinguishable from solar and were suggested as solar
UV templates. Only one, however, \object{HD\,140690}, possesses
atmospheric parameters equal to the Sun's, making it a solar
analog, and it also photometrically matches the Sun well in the
Paschen continuum colors and the Str\"omgren m$_{\rm 1}$ index. It
was therefore proposed as a prime solar analog from the UV out to
the visible wavelength range. Other stars were shown to resemble
the Sun in the UV owing to a fortuitous composition of atmospheric
parameters, and care should be exercised in selecting stars to
represent the Sun {\it both} in the UV and visible ranges. Good UV
solar proxies may be particularly important for the
observation of UV emission lines in comets.\\

3) The spectroscopic and evolutionary analysis revealed five new
``probable'' solar-twin candidates, plus five new ``possible''
twin candidates, besides successfully identifying two previously
known solar twins, \object{HD\,146233} and \object{HD\,98618}. The
four probable new solar twin candidates, \object{HD\,98649},
\object{HD\,118598}, \object{HD\,150248}, and \object{HD\,164595},
have atmospheric and evolutionary parameters indistinguishable
from the solar ones within the uncertainties, besides a low level
of chromospheric activity, so they clearly warrant closer scrutiny.\\

In a forthcoming paper, we will discuss these objects in more
detail, including a multi-element abundance analysis, additional
criteria to determine \Teff, a deeper study of their spectroscopic
chromospheric indicators, the determination of their kinematics,
and a more detailed evolutionary analysis.\\

{\bf
\appendix{APPENDIX A: A Metallicity-dependent IRFM \Teff calibration
for solar-type stars}}\\

Theoretical calculations in stellar modeling predict relations
between structural quantities that see little change during
stellar evolution, such as mass and metallicity, and others that
vary extensively, such as effective temperature, radius, and
luminosity. These quantities are not straightforward to determine,
and their match to accessible observational data such as colors
lies at the heart of stellar astrophysics. The effective
temperature is the most basic stellar parameter that affects the
model atmosphere abundance analysis of stars. Moreover, at least
for nearby stars for which very precise parallaxes are presently
available \citep{hipparcos1997}, the \Teff is now the single most
important source of error in placing stars in theoretical HR
diagrams.

The aim of the \Teff calibrations presented here is not to emulate
or be an alternative to the many excellent resources available
nowadays \citep[e.g.,][]{casagrandeetal2010, casagrandeetal2006,
masanaetal2006, ramirezmelendez2005b}, but to provide, in the
context of a solar analog search, a solid base for comparing
photometric \Teffs to those inferred from spectroscopy and Balmer
line profiles with the specific aim of better distinguishing small
\Teff differences between the Sun and candidate solar analogs and
twins. To apply differential philosophy to the greatest possible
extent and to ensure maximum homogeneity, it is desirable that all
\Teff scales employed in the present study be tied to a similar
suite of model atmospheres, in our case the MARCS system of model
atmospheres, as described by Edvardsson et al. (1993, see
http://marcs.astro.uu.se; Gustafsson et al. 2008).

An ``ideal'' direct \Teff calibration should be based on an
extensive set of precise measurements of bolometric fluxes and
angular diameters and be independent of any grid of model
atmospheres. Among the various ``indirect'' methods employed so
far alternatively, one of the most advantageous is the infrared
flux method (IRFM) \citep[originally described
by][]{blackwellshallis1977, blackwelletal1986} since it relies
only weakly on theoretical representations of stellar atmospheres.
Details on the method are given by Blackwell et a. (1986, 1990).

Some recent determinations of the relation between \Teff(IRFM) and
stellar colors \citep[e.g.,][]{masanaetal2006, casagrandeetal2010,
ramirezmelendez2005b} report that the \Teff scale of FGK stars is
established to better than $\sim$1$\%$ or $\sim$60 K.
\citet{casagrandeetal2006} find good agreement between empirical
and synthetic colors both for the ATLAS and MARCS families of
models in the visible, but less so in the infrared, and also
report, concurrently with \citet{masanaetal2006} and
\citet{dasilvaetal2012}, that good agreement is realized between
the spectroscopic and photometric \Teff scales for solar-type
stars, although disagreements of a few percent are found between
different authors. However, much equally recent work
\citep[e.g.,][]{ramirezetal2007, ramirezmelendez2005a,
yongetal2004} state that the spectroscopic \Teff scale of
solar-type stars is hotter than the photometric one by $\sim$100
K. This is in line with \citet{portodemelloetal2008}, who find, in
a detailed analysis of the very well-studied double system
$\alpha$ Cen AB, a discrepancy between the spectroscopic \Teff
scale and those from photometry and the fitting of Balmer line
profiles. Nonetheless \citet{ramirezmelendez2005a} and
\citet{dasilvaetal2012} obtain good consistency between
photometric \Teffs and those derived from the fitting of Balmer
line profiles. Despite recent efforts \citep{casagrandeetal2010}
to clarify these offsets, a somewhat confusing picture still
emerges from the literature concerning the overall agreement of
the photometric, Balmer line, and spectroscopic \Teff scales over
a wide parameter range; fortunately, much better consistency can
be found near the solar parameters \citep{dasilvaetal2012}, thus
the broader issues of the \Teff scale of FGK stars need not concern us here.\\

\begin{table*}
\centering \caption{Table A.1. Objects selected for the \Teff {\it
versus} color calibrations. Sources for ($b-y$) and $\beta$ are
given in Table 14. The $(R-I)$, $(V-R)$, $(V-I)$, and $(V-K)$
colors are all in the Johnson system \citep{glass1974,johnson1964,
johnsonetal1966, johnsonetal1968}. When only narrow-band
($V-K_{\rm n}$) from \cite{selbyetal1988} is available, they have
been transformed to Johnson ($V-K$) using the relations provided
by these same authors. The MK spectral type and Johnson
($B-V$)$^{\rm J}$ colors were taken from the Bright Star Catalogue
\citep{hoffleitjaschek1991}. The Tycho ($B-V$)$^{\rm T}$ colors
come from \cite{hipparcos1997}. The third column gives \Teff as
obtained by \cite{blackwelletal1991} (1) or
\cite{saxnerhammarback1985} (2). When both authors provide a
T$_{\rm eff}$ value, the straight average is tabulated.}
\label{Table A1}
\begin{tabular}{l l l l c c c c c c c c c}
\hline\hline \noalign{\smallskip} HR & HD &
\parbox[c]{1.02cm}{\centering Spectral type} &
\parbox[c]{1.0cm}{\centering $T_{\rm eff}$ (IRFM) \\ {\tiny (K)}} &
\parbox[c]{1.0cm}{\centering [Fe/H]} &
\parbox[c]{1.0cm}{\centering $(B-V)^{\rm J}$} &
\parbox[c]{1.0cm}{\centering $(B-V)^{\rm T}$} &
\parbox[c]{1.0cm}{\centering $(R-I)$} &
\parbox[c]{1.0cm}{\centering $(V-R)$} &
\parbox[c]{1.0cm}{\centering $(V-I)$} &
\parbox[c]{1.0cm}{\centering $(V-K)$} &
\parbox[c]{1.0cm}{\centering ($b-y$)} &
\parbox[c]{1.0cm}{\centering $\beta$} \\
\noalign{\smallskip} \hline \noalign{\smallskip}
  33 &    693 & F7V    & 6148   (2) & $-$0.55 & 0.487 & 0.519 & --   & --   & --   & 1.23 & 0.326 & 2.617 \\
  98 &   2151 & G2IV   & 5860   (2) & $-$0.17 & 0.618 & --    & 0.34 & 0.50 & 0.84 & 1.48 & 0.394 & 2.597 \\
 244 &   5015 & F8V    & 6042   (2) & $-$0.08 & 0.540 & 0.610 & 0.30 & 0.48 & 0.78 & 1.28 & 0.346 & 2.613 \\
 417 &   8799 & F5IV   & 6547   (1) &   --    & 0.421 & 0.463 & 0.23 & 0.41 & 0.64 & 1.07 & 0.288 & 2.672 \\
 458 &   9826 & F8V    & 6177 (1/2) &   +0.12 & 0.536 & 0.589 & 0.29 & 0.46 & 0.75 & 1.25 & 0.344 & 2.629 \\
 483 &  10307 & G1.5V  & 5856   (2) & $-$0.02 & 0.618 & 0.686 & 0.33 & 0.53 & 0.86 & 1.39 & 0.389 & 2.604 \\
 509 &  10700 & G8V    & 5341   (1) & $-$0.50 & 0.727 & --    & 0.42 & 0.62 & 1.09 & 1.82 & 0.449 & 2.555 \\
 544 &  11443 & F6IV   & 6350   (1) &   +0.06 & 0.488 & 0.526 & 0.28 & 0.42 & 0.70 & 1.18 & 0.316 & 2.637 \\
 740 &  15798 & F4IV   & 6409   (2) & $-$0.19 & 0.454 & 0.478 & 0.27 & 0.41 & 0.68 & 1.12 & 0.297 & 2.640 \\
 799 &  16895 & F8V    & 6373   (2) &   +0.02 & 0.514 & 0.541 & 0.30 & 0.46 & 0.76 & 1.15 & 0.326 & 2.625 \\
 818 &  17206 & F5/F6V & 6350   (2) &   +0.05 & 0.481 & 0.515 & 0.27 & 0.43 & 0.70 & 1.14 & 0.328 & 2.646 \\
 937 &  19373 & G0V    & 5997 (1/2) &   +0.08 & 0.595 & 0.668 & 0.29 & 0.53 & 0.82 & 1.36 & 0.201 & 2.605 \\
 996 &  20630 & G5Vv   & 5692   (1) &   +0.06 & 0.681 & 0.756 & 0.36 & 0.57 & 0.93 & 1.52 & 0.420 & 2.585 \\
1101 &  22484 & F9V    & 5953 (1/2) & $-$0.13 & 0.575 & 0.626 & 0.32 & 0.49 & 0.81 & 1.36 & 0.370 & 2.608 \\
1325 &  26965 & K1V    & 5125   (1) & $-$0.32 & 0.820 & --    & 0.45 & 0.69 & 1.14 & 2.02 & 0.487 & 2.543 \\
1543 &  30652 & F6V    & 6373   (2) &   +0.09 & 0.484 & 0.504 & 0.26 & 0.42 & 0.68 & 1.12 & 0.299 & 2.651 \\
1729 &  34411 & G2IV-V & 5866 (1/2) &   +0.11 & 0.630 & 0.696 & 0.32 & 0.53 & 0.85 & 1.43 & 0.389 & 2.598 \\
1983 &  38393 & F6V    & 6259   (2) & $-$0.12 & 0.481 & 0.530 & 0.26 & 0.45 & 0.71 & 1.16 & 0.315 & 2.634 \\
2047 &  39587 & G0V    & 5839   (2) & $-$0.08 & 0.594 & 0.659 & 0.31 & 0.51 & 0.82 & 1.44 & 0.378 & 2.601 \\
2943 &  61421 & F5IV-V & 6601   (2) & $-$0.07 & 0.432 & --    & 0.23 & 0.42 & 0.65 & 1.02 & 0.272 & 2.671 \\
4054 &  89449 & F6IV   & 6374   (2) & $-$0.15 & 0.452 & 0.503 & 0.23 & 0.45 & 0.68 & 1.15 & 0.301 & 2.654 \\
4540 & 102870 & F9V    & 6147   (2) &   +0.11 & 0.518 & 0.613 & 0.28 & 0.48 & 0.76 & 1.27 & 0.354 & 2.629 \\
4785 & 109358 & G0V    & 5842   (2) & $-$0.20 & 0.588 & 0.655 & 0.31 & 0.54 & 0.85 & 1.43 & 0.385 & --    \\
5072 & 117176 & G2.5Va & 5480   (1) & $-$0.11 & 0.714 & 0.804 & 0.39 & 0.61 & 1.00 & 1.74 & 0.454 & 2.576 \\
5185 & 120136 & F6IV   & 6383   (2) & $-$0.04 & 0.508 & 0.534 & 0.24 & 0.41 & 0.65 & 1.11 & 0.313 & 2.656 \\
5235 & 121370 & G0IV   & 6044   (1) &   +0.21 & 0.580 & 0.670 & 0.29 & 0.44 & 0.73 & 1.31 & 0.370 & 2.625 \\
5304 & 123999 & F9IVw  & 6173   (1) & $-$0.07 & 0.541 & 0.582 & 0.29 & 0.44 & 0.73 & 1.30 & 0.347 & 2.631 \\
5868 & 141004 & G0Vv   & 5940   (2) &   +0.00 & 0.604 & 0.672 & 0.32 & 0.51 & 0.83 & 1.38 & 0.384 & 2.606   \\
5914 & 142373 & F8V    & 5861 (1/2) & $-$0.43 & 0.563 & 0.615 & 0.32 & 0.48 & 0.80 & 1.53 & 0.381 & 2.601 \\
5933 & 142860 & F6V    & 6246   (2) & $-$0.14 & 0.478 & --    & 0.24 & 0.49 & 0.73 & 1.20 & 0.321 & 2.632 \\
5986 & 144284 & F8IV   & 6147   (1) &   +0.20 & 0.528 & 0.590 & 0.25 & 0.45 & 0.70 & 1.24 & 0.354 & 2.639 \\
6623 & 161797 & G5IV   & 5496   (1) &   +0.03 & 0.758 & 0.856 & 0.38 & 0.53 & 0.91 & 1.65 & 0.464 & 2.614 \\
7061 & 173667 & F6V    & 6368   (1) & $-$0.09 & 0.483 & 0.502 & 0.26 & 0.39 & 0.65 & 1.10 & 0.314 & 2.654 \\
7602 & 188512 & G8IVv  & 5080   (1) & $-$0.22 & 0.855 & 0.984 & 0.49 & 0.66 & 1.15 & 2.01 & 0.523 & 2.554 \\
7957 & 198149 & K0IV   & 4997   (1) & $-$0.29 & 0.912 & 1.065 & 0.49 & 0.67 & 1.16 & 2.15 & 0.553 & --    \\
8181 & 203608 & F6V    & 6065   (2) & $-$0.84 & 0.494 & 0.522 & 0.30 & 0.47 & 0.77 & 1.31 & 0.335 & 2.618 \\
\noalign{\smallskip} \hline\hline
\end{tabular}
\end{table*}

\begin{table*}
\centering \caption{Table A.2. References for the uvby$\beta$
photometry and [Fe/H] of Table 13.}\label{Table A2}
\begin{tabular}{l c c c}
\hline\hline \noalign{\smallskip}
HR & $(b-y)$ & $\beta$ & [Fe/H] \\
\noalign{\smallskip} \hline \noalign{\smallskip}
33   & Gr\"onbech \& Olsen (1976)  & Gr\"onbech \& Olsen (1977) &  Balachandran (1990)        \\
98   & Crawford et al. (1970)      & Heck \& Manfroid (1980)    &  Abia et al. (1988)              \\
244  & Str\"omgren \& Perry (1965) & Crawford et al. (1966) &  Lambert et al. (1991)           \\
417  & Str\"omgren \& Perry (1965) & Crawford et al. (1966) &  --                      \\
458  & Reglero et al. (1987)       & Olsen (1983)       &  Boesgaard \& Lavery (1986)          \\
483  & Str\"omgren \& Perry (1965) & Crawford et al. (1966) &  Clegg et al. (1981)             \\
509  & Olsen (1983)        & Olsen (1983)       &  Arribas \& Crivellari (1989)        \\
544  & Str\"omgren \& Perry (1965) & Crawford et al. (1966) &  Balachandran (1990)             \\
740  & Gr\"onbech \& Olsen (1976)  & Olsen (1983)       &  Balachandran (1990)             \\
799  & Str\"omgren \& Perry (1965) & Crawford et al. (1966) &  Clegg et al. (1981)             \\
818  & Crawford et al. (1970)      & Crawford et al. (1970) &  Luck \& Heiter (2005)           \\
937  & Crawford \& Barnes (1970)   & Crawford et al. (1966) &  Chen et al. (2000)            \\
996  & Olsen (1983)        & Olsen (1983)       &  Cayrel de Strobel \&Bentolila(1989) \\
1101 & Olsen (1983)        & Gr\"onbech \& Olsen (1977) &  Nissen \& Edvardsson (1992)         \\
1325 & Olsen (1983)        & Schuster \& Nissen (1988)  &  Steenbock (1983)            \\
1543 & Olsen (1983)        & Olsen (1983)       &  Clegg et al. (1981)             \\
1729 & Crawford \& Barnes (1970)   & Crawford et al. (1966) &  Friel \& Boesgaard (1992)           \\
1983 & Olsen (1983)        & Olsen (1983)       &  Cayrel de Strobel et al. (1988)     \\
2047 & Warren \& Hesser (1977)     & Olsen (1983)       &  Boesgaard \& Friel (1990)           \\
2943 & Crawford \& Barnes (1970)   & Crawford et al. (1966) &  Steffen (1985)              \\
4054 & Olsen (1983)        & Crawford et al. (1966) &  Th\'evenin et al. (1986)        \\
4540 & Olsen (1983)        & Olsen (1983)       &  Nissen \& Edvardsson (1992)         \\
4785 & Warren \& Hesser (1977)     & Warren \& Hesser (1977)    &  Boesgaard \& Lavery (1986)          \\
5072 & Manfroid \& Sterken (1987)  & Crawford et al. (1966) &  da Silva et al. (2012)            \\
5185 & Manfroid \& Sterken (1987)  & Crawford et al. (1966) &  Th\'evenin et al. (1986)        \\
5235 & Warren \& Hesser (1977)     & Gr\"onbech \& Olsen (1977) &  Clegg et al. (1981)             \\
5304 & Olsen (1983)        & Crawford et al. (1966) &  Balachandran (1990)             \\
5868 & Olsen (1983)        & Olsen (1983)       &  Boesgaard \& Lavery (1986)          \\
5914 & Olsen (1983)        & Crawford et al. (1966) &  Boesgaard \& Lavery (1986)          \\
5933 & Reglero et al. (1987)       & Olsen (1983)       &  Boesgaard \& Lavery (1986)          \\
5986 & Warren \& Hesser (1977)     & Crawford et al. (1966) &  Boesgaard \& Lavery (1986)          \\
6623 & Bond (1970)         & Heck (1977)        &  McWilliam (1990)            \\
7061 & Str\"omgren \& Perry (1965) & Reglero et al. (1987)  &  Boesgaard \& Friel (1990)           \\
7602 & Olsen (1983)        & Stokes (1972)      &  Edvardsson (1988)               \\
7957 & Bond (1970)         & Bond (1970)        &  McWilliam (1990)            \\
8181 & Olsen (1983)        & Olsen (1983)       &  Zhao \& Magain (1991)           \\
\noalign{\smallskip} \hline\hline
\end{tabular}
\end{table*}

\begin{figure}
\centering
\includegraphics[width=9.0cm]{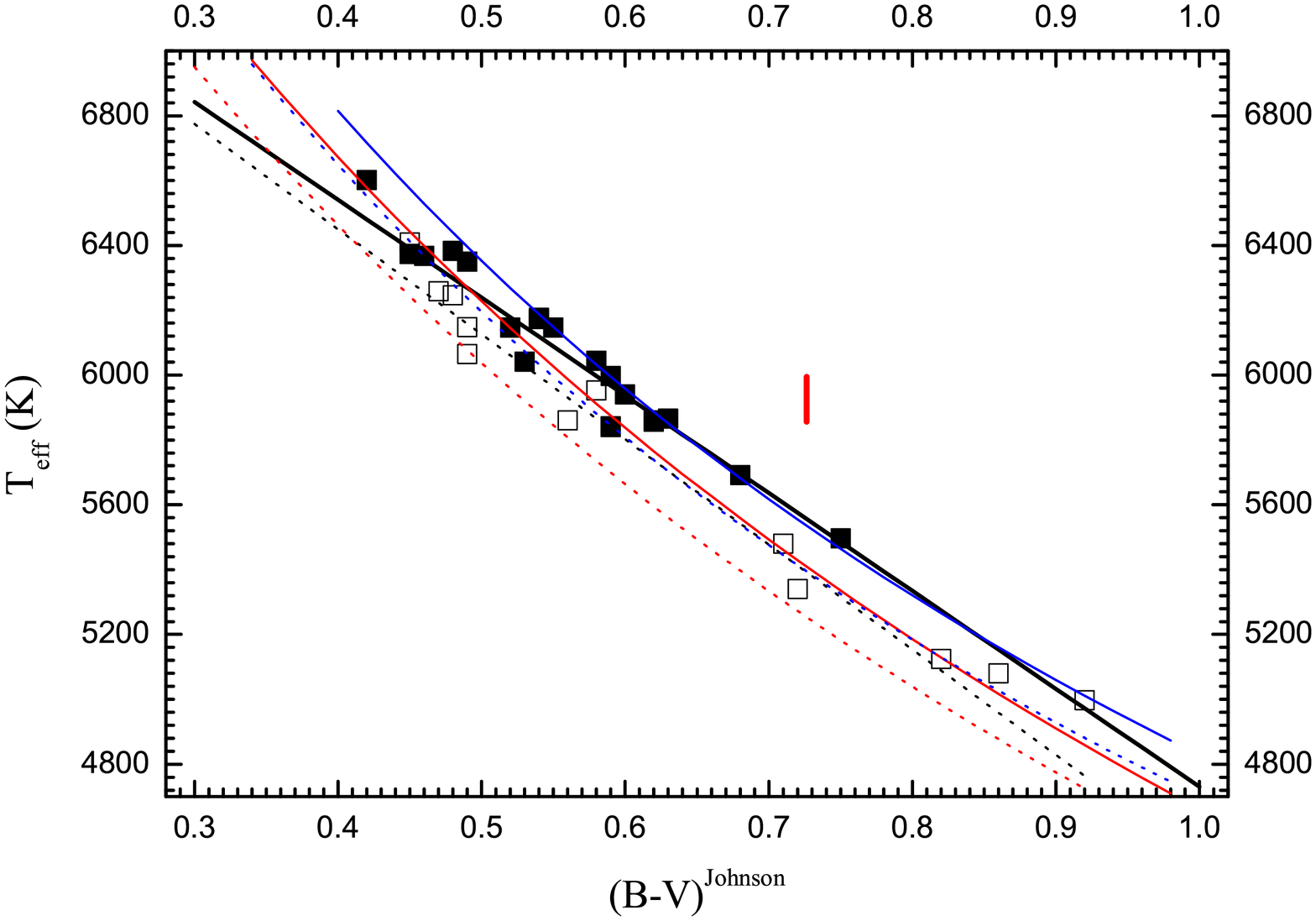}
\caption{Figure A.1. Our calibration for the ($B-V$)$^{\rm
Johnson}$ index compared to literature results. The adopted
1$\sigma$ \Teff error (see text) is shown as the thick red
vertical bar. Full squares are stars with $[$Fe/H$]$ $>$ $+$0.10
dex, open squares with $[$Fe/H$]$ $<$ $+$0.10 dex. The black full
and dotted lines refer, respectively, to our calibrations for
$[$Fe/H$]$ = $+$0.00 and $[$Fe/H$]$ = $-$0.50; the red and blue
lines follow the same \feh convention for the relations of
\cite{alonsoetal1996} and \cite{casagrandeetal2006},
respectively.} \label{Figure A.1}
\end{figure}

\begin{figure}
\centering
\includegraphics[width=9.0cm]{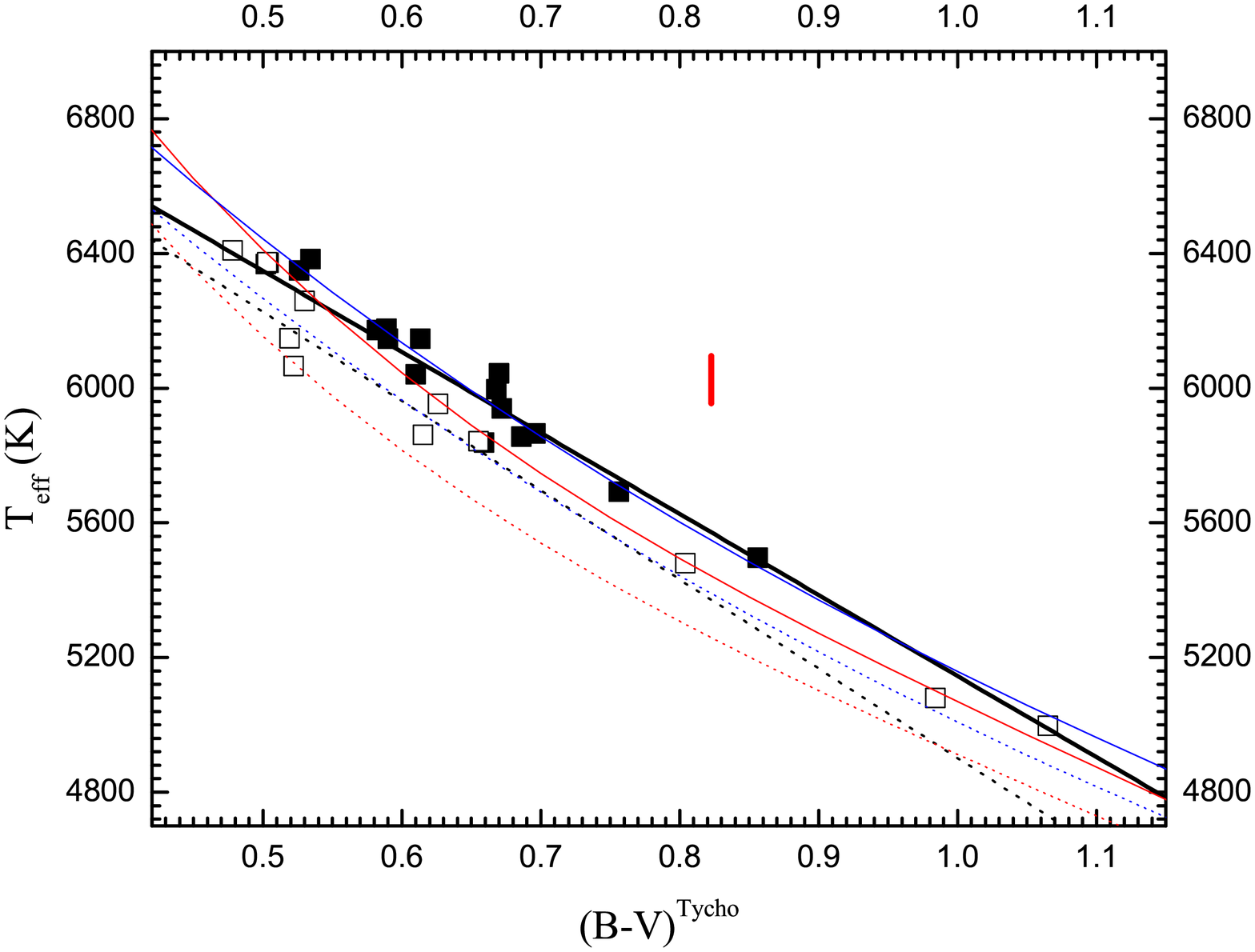}
\caption{Figure A.2. The same as Fig. A.1 for the ($B-V$)$^{\rm
Tycho}$ index: the red and blue lines are the relations of
\cite{ramirezmelendez2005b} and \cite{casagrandeetal2010},
respectively.} \label{Figure A.2}
\end{figure}

\begin{figure}
\centering
\includegraphics[width=9.0cm]{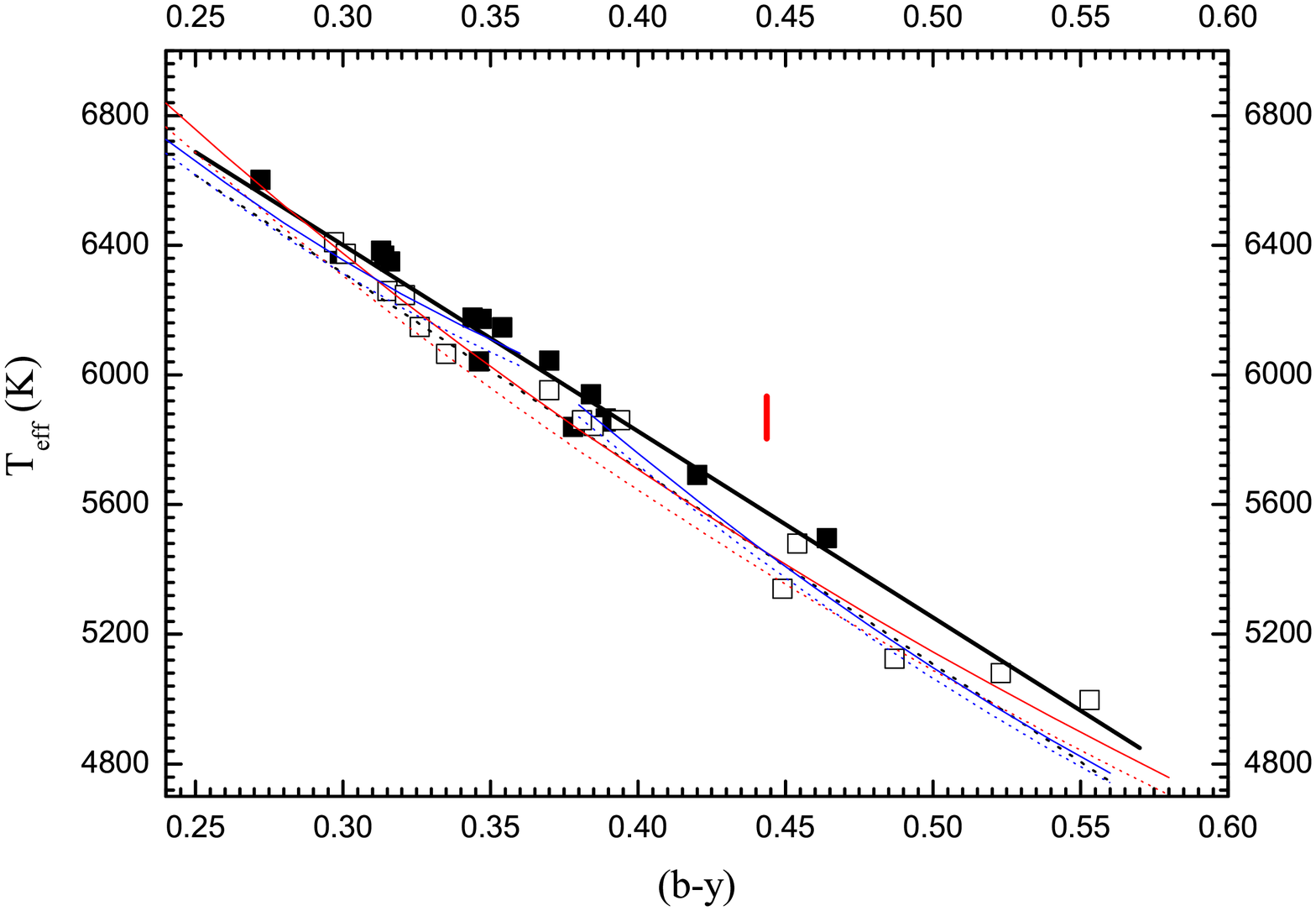}
\caption{Figure A.3. The same as Fig. A.1 for the ($b-y$) index:
the red and blue lines are the relations of \cite{alonsoetal1996}
and \cite{blackwelllynasgray1998}, respectively.} \label{Figure
A.3}
\end{figure}

\begin{figure}
\centering
\includegraphics[width=9.0cm]{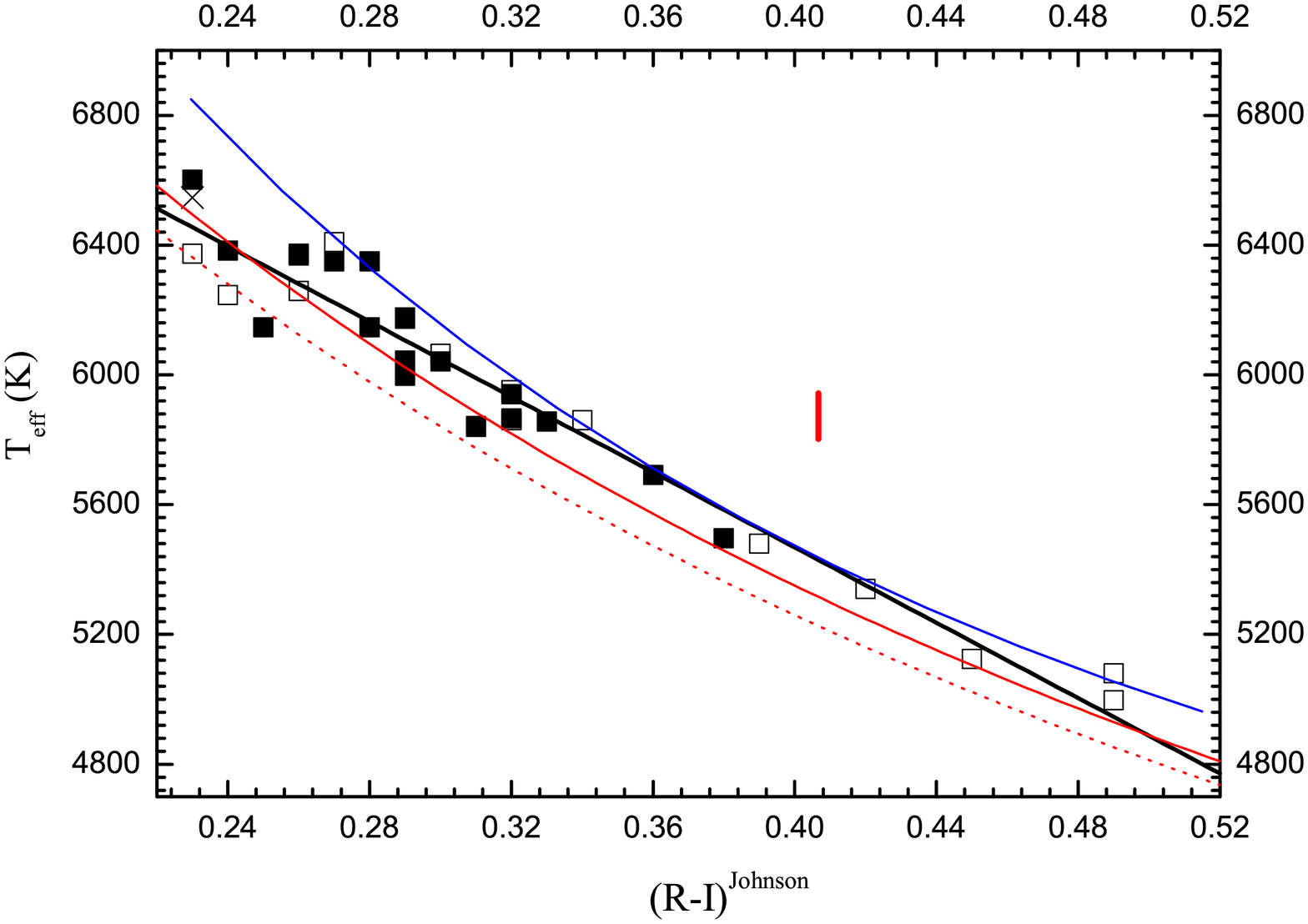}
\caption{Figure A.4. The same as Fig. A.1 for the ($R-I$)$^{\rm
Johnson}$ index. The cross stands for HR 417, without known
$[$Fe/H$]$. The black full line is our calibration, the blue line
that of \cite{casagrandeetal2006}, and the red full and dotted
lines follow the same $[$Fe/H$]$ convention as Fig. A.1 for the
calibration of \cite{alonsoetal1996}.} \label{Figure A.4}
\end{figure}

\begin{figure}
\centering
\includegraphics[width=9.0cm]{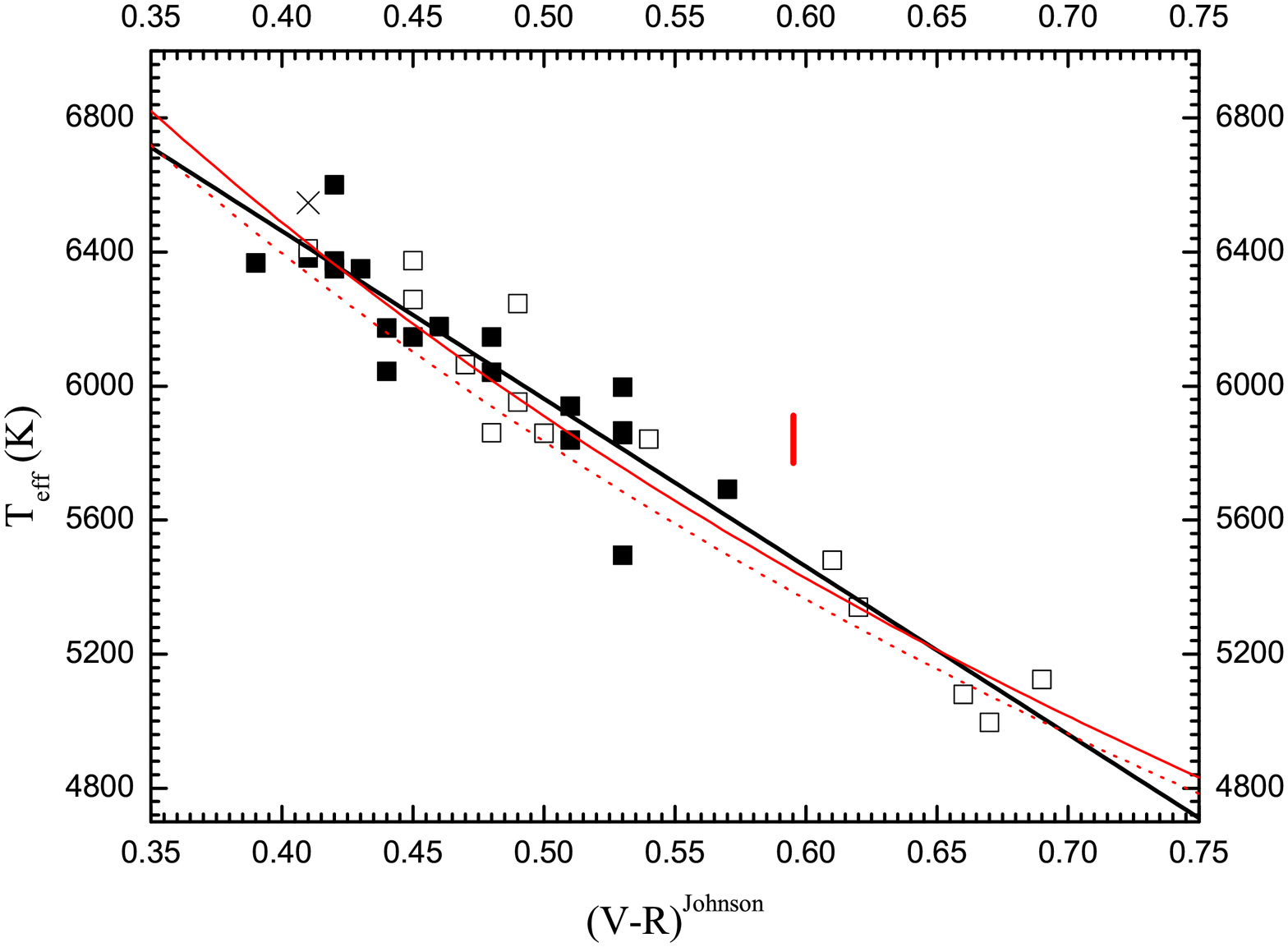}
\caption{Figure A.5. The same as Fig. A.4 for the ($V-R$)$^{\rm
Johnson}$ index: the red full and dotted lines follow the same
$[$Fe/H$]$ convention as Fig. A.1 for the calibration of
\cite{alonsoetal1996}.} \label{Figure A.5}
\end{figure}

\begin{figure}
\centering
\includegraphics[width=9.0cm]{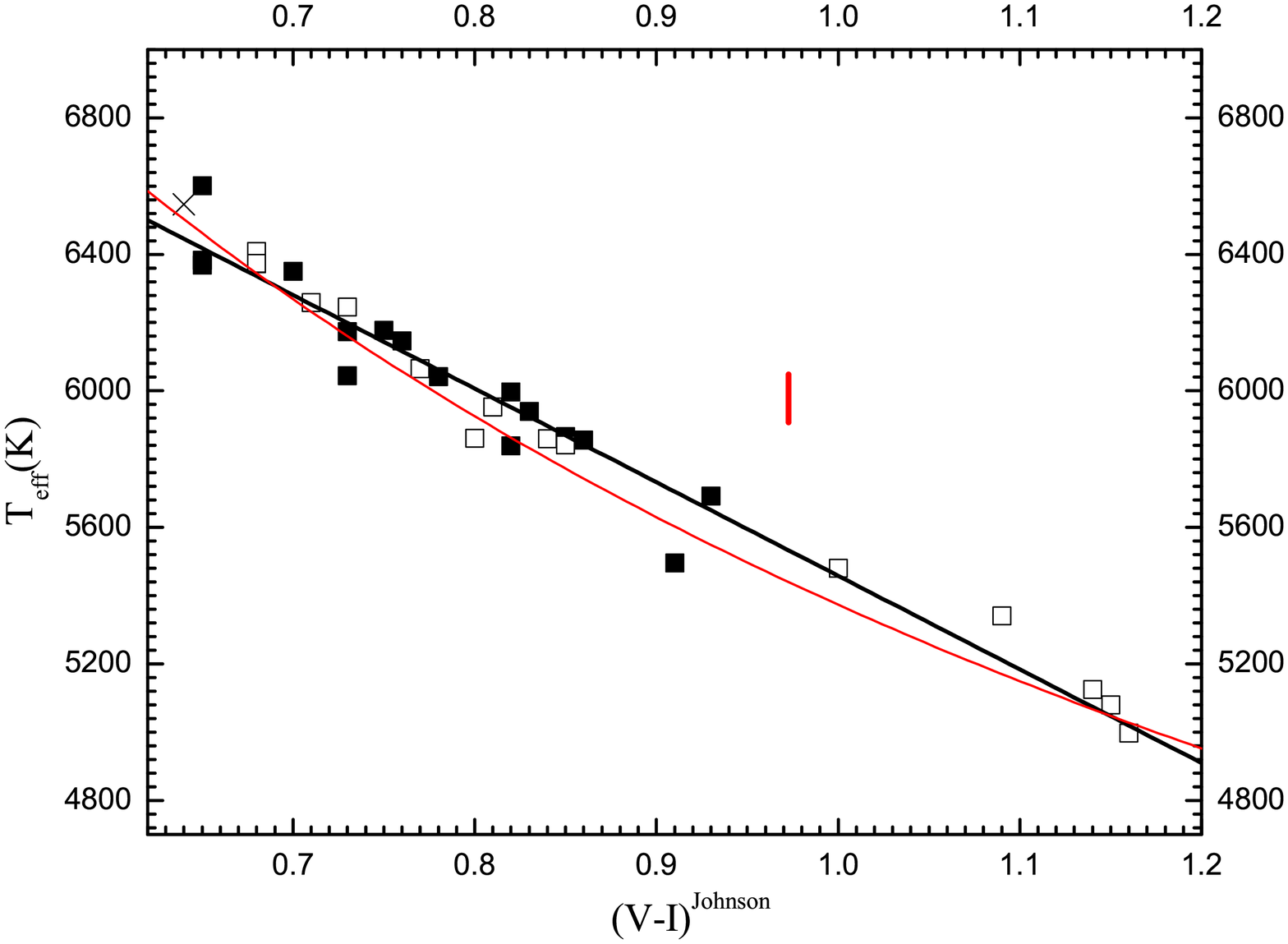}
\caption{Figure A.6. The same as Fig. A.4 for the ($V-I$)$^{\rm
Johnson}$ index. The red line is the calibration of
\cite{alonsoetal1996}.} \label{Figure A.6}
\end{figure}

\begin{figure}
\centering
\includegraphics[width=9.0cm]{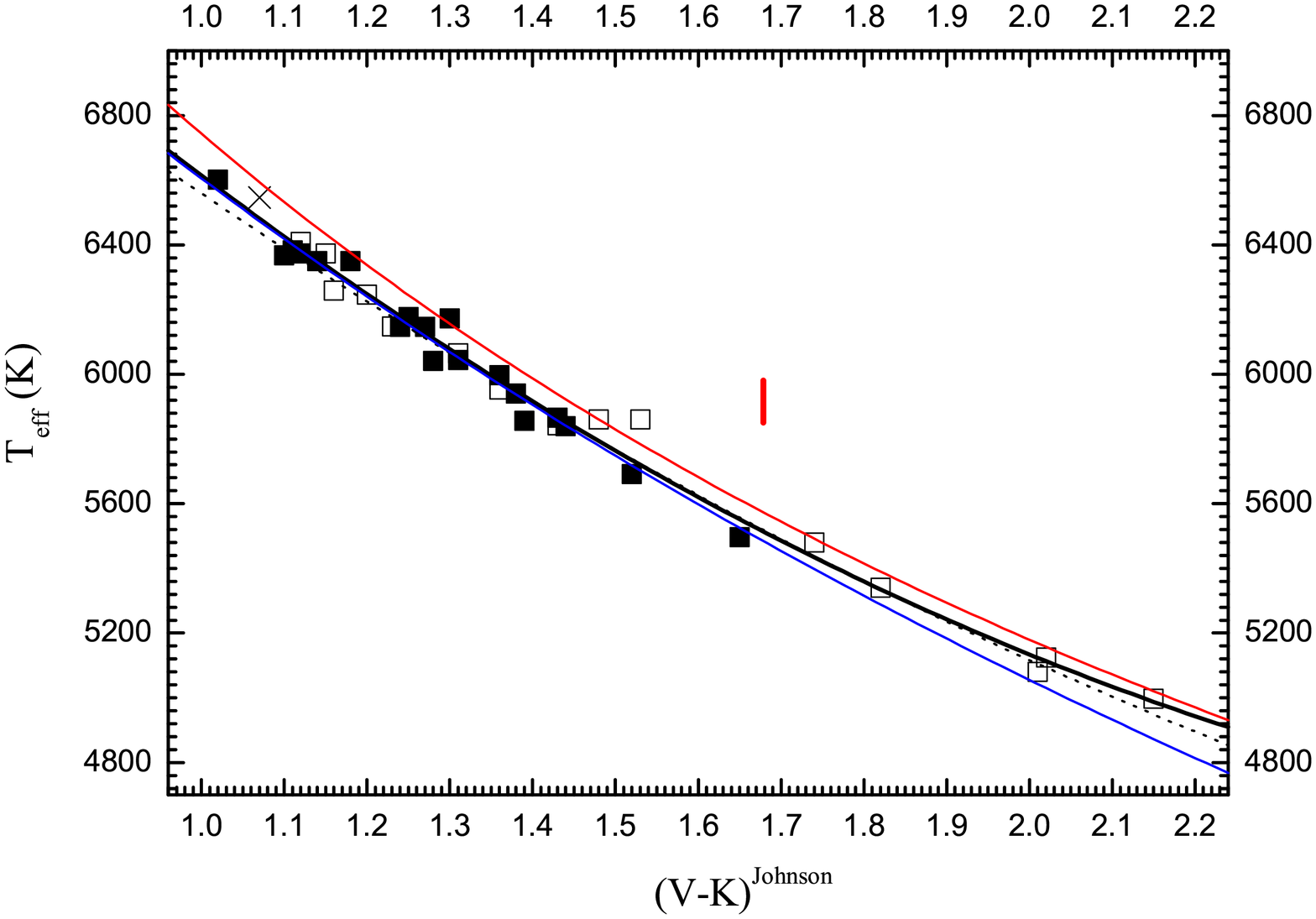}
\caption{Figure A.7. The same as Fig. A.4 for the ($V-K$)$^{\rm
Johnson}$ index. The black full, black dotted, red, and blue lines
correspond to the calibrations of this work,
\cite{blackwelllynasgray1998}, \cite{masanaetal2006}, and
\cite{alonsoetal1996}, respectively.} \label{Figure A.7}
\end{figure}

\begin{figure}
\centering
\includegraphics[width=9.0cm]{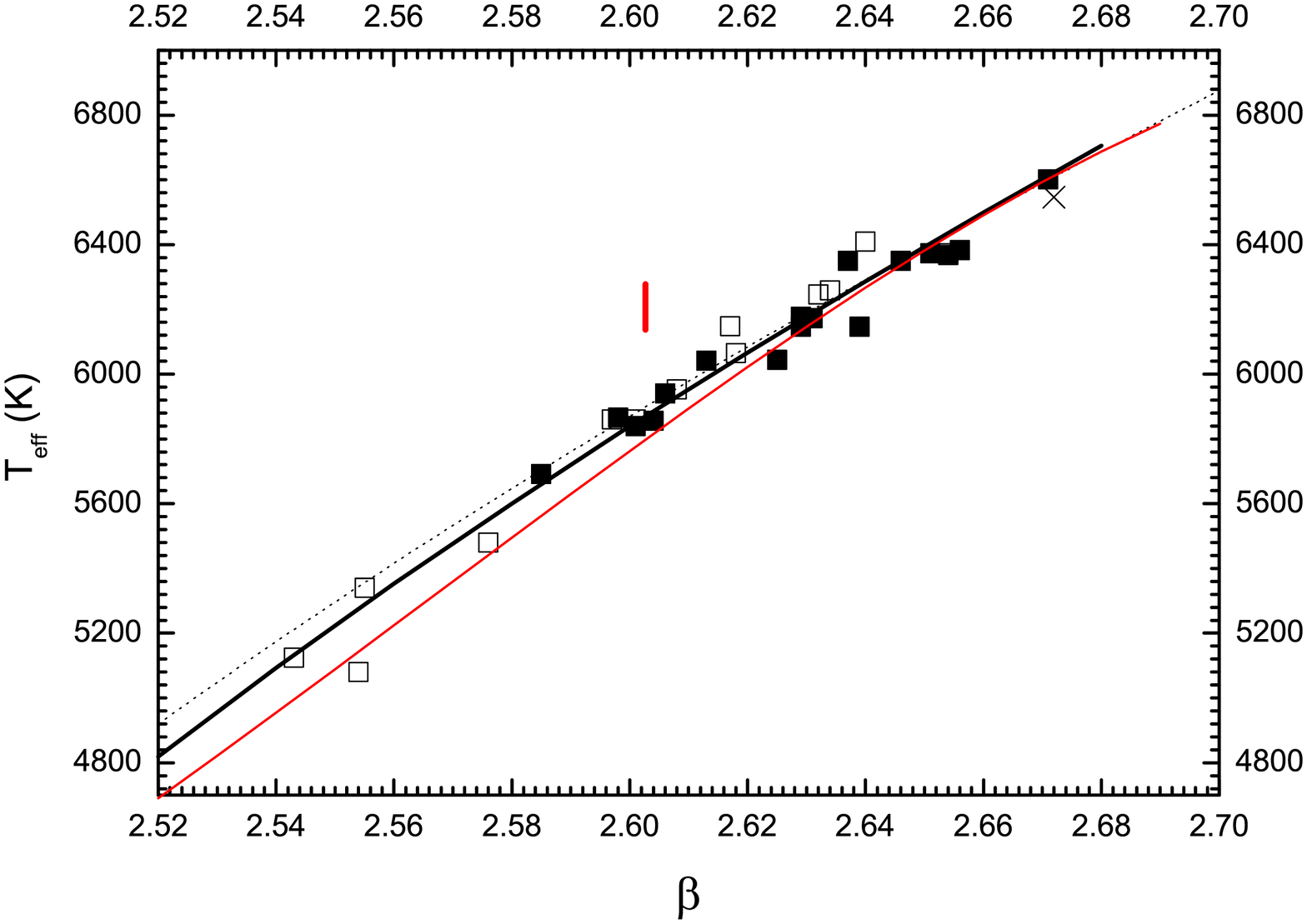}
\caption{Figure A.8. The same as Fig. A.4 for the $\beta$ index.
The black full, black dotted, and red lines stand for the
calibrations of this work, \cite{saxnerhammarback1985}, and
\cite{alonsoetal1996} respectively.} \label{Figure A.8}
\end{figure}

{\bf Selection of photometry, IRFM \Teff and $[$Fe/H$]$ data}\\

Our choice of older IRFM \Teff determinations is based mainly on
the goal of realizing strict consistency with the MARCS models
used in our $[$Fe/H$]$ (section 5.1) and Balmer line \Teff
(section 5.2) determinations, and therefore our emphasis was not
on the latest resources. The adopted \Teff values, given in Table
A.1, come from \cite{saxnerhammarback1985} and
\cite{blackwelletal1991}, who used MARCS models in their
derivation of IRFM \Teff values. In Table A.1 we also list the
$[$Fe/H$]$ measurements and photometry used in the calibrations.
As in section 2, strong preference was given to the series of
papers by Olsen and co-authors as sources of the ($b-y$) indices.
Sources of these data are given in detail in Table A.2. We have
selected 18 stars from \cite{saxnerhammarback1985} and 23 stars
from \cite{blackwelletal1991}, with 5 stars in common, totalling
36 stars: \Teffs for the common stars between these two sources
agree within a mean value of 0.8\%, and straight averages were
used in these cases. All objects have FGK-types and are classified
as dwarfs or subgiants (surface gravities confirm this status in
all cases); they span approximately a 5000K $\leq$ \Teff $\leq$
6500K range, the range in metallicity being $-$0.4 $\leq$
$[$Fe/H$]$ $\leq$ $+$0.3, with a modest extension to more
metal-poor stars; and are close enough that reddening corrections
are unnecessary. These atmospheric parameter ranges bracket those
of our solar analog and solar twin candidates perfectly.

\cite{blackwelllynasgray1994} and \cite{megessier1994} discuss
discrepancies between the use of older MARCS models and more
up-to-date ATLAS models in the deriving of IRFM \Teffs. These
offsets might amount to $\sim$1\% at the extremes of the main
sequence, but were found to be much smaller for solar-type stars.
Differences between the \Teff values of
\cite{blackwelllynasgray1994} and \cite{blackwelletal1991} average
$+$10 $\pm$ 30 K, which is inconsequential to our purposes.
Moreover, we show below that the calibrations derived here are in
very good agreement with recent ones widely cited in the
literature. \cite{casagrandeetal2006} have discussed systematic
differences between MARCS and ATLAS models and conclude that while
good agreement is found in the visible, in the IR offsets still
remain and may be traceable to lingering uncertainties in the
absolute flux calibration of Vega, which they regard as a factor
still influencing the accuracy of IRFM \Teff determinations.

Our calibrations aim derive empirical \Teff relations as a
function of the most commonly employed photometric colors and
additionally, at calibrating the metallicity dependency of the
blanketing-sensitive color indices precisely, by using exclusively
spectroscopically derived $[$Fe/H$]$ values. The literature was
searched for spectroscopic metallicity determinations based on
high-resolution spectra and employing a large number of
\ion{Fe}{i} lines. The $[$Fe/H$]$ values were corrected to full
consistency to the \Teff (IRFM) scale: for this we used the
$\Delta$$[$Fe/H$]$/$\Delta$\Teff provided by the authors
themselves, when available, or else the representative value of
$-$0.06 dex for 100K, to correct $[$Fe/H$]$ as a function of the
difference between the \Teff adopted in the spectroscopic analysis
and the IRFM \Teff for each star. These $[$Fe/H$]$ values have
considerable heterogeneity in what concerns observational data and
methods of analysis and part of this heterogeneity should be
removed by the correction to the common \Teff (IRFM) scale,
besides realizing full consistency with the metallicities we
derived in section 5.1 for our solar-analog and twin candidates,
which are strictly tied to the photometric \Teff scale.\\

{\bf The calibrations and comparison to other authors}\\

Our calibrations of \Teff against various color indices in
widespread use are given in figures A.1 to A.8, along with their
(non-exhaustive) comparisons with published results in wide use.
As expected, the data points define tight correlations with little
scatter. We tested for possible trends in the correlations with
$[$Fe/H$]$ and $\log{g}$, and in all cases surface gravity did not
affect the calibrations, as expected from the quite narrow range
of this parameter among our sample. Some colors are clearly
affected, however, by blanketing effects to different degrees,
even in our narrow \Teff and $[$Fe/H$]$ range. Below we give
details on the functional forms of the adopted calibrations and
their comparison with the literature. We performed two experiments
concerning the use of homogenized values from the literature as
compared to those selected from individual references. The first
one concerns the ($b-y$) and $\beta$ photometry, and we tested the
use in the calibrations of values taken from individually selected
papers compared to mean values given by Hauck \& Mermilliod (1998,
1990). The other test involved comparing the calibrations obtained
with $[$Fe/H$]$ taken from selected works as opposed to mean
values from $[$Fe/H$]$ catalogs \citep{cayreldestrobeletal2001}.
For the photometry, significantly tighter regressions and smaller
errors in the coefficients were attained when critically  selected
values from the literature were employed. A similar yet less
clear-cut result was obtained for \feh. We therefore conclude that
catalogs of homogenized values, while extremely helpful, should be
used with some care. Not all possible comparisons to recent
literature are shown in the plots in order not to clutter the
diagrams unnecessarily, but relevant remarks are given in the
text.

In figures A.1 to A.8 the sample is stratified in two \feh
intervals separated by \feh = $-$0.10. The necessity of a \feh
term is clear in the Paschen continuum colors, and for the
($B-V$)$^{\rm Johnson}$, ($B-V$)$^{\rm Tycho}$ and ($b-y$)
regressions we have adopted the same functional relationship as in
\cite{saxnerhammarback1985}. For the ($R-I$)$^{\rm Johnson}$,
($V-R$)$^{\rm Johnson}$ and ($V-I$)$^{\rm Johnson}$ regressions, a
linear form was found satisfactory, and no \feh term was
necessary, while for ($V-K$)$^{\rm Johnson}$, and $\beta$~
second-order functions improved the regressions significantly, and
also no \feh trends appear. As a formal estimate of the
uncertainties, in figures 8 to 15 we plot an error bar
quadratically composed of the standard deviations of our
regressions and a 2\% formal error on the \Teffs used for the
calibrations \citep[as estimated by][]{casagrandeetal2006}. While
probably underestimating the total error budget, since these two
error sources are not independent, this estimate should properly
account for differences between \Teff scales from different
authors, as well as the internal uncertainty in our regressions.

The corresponding regressions are given below, where we also
provide the uncertainties in the coefficients, the standard
deviation of the fit, and the number of stars employed in each
regression:

\begin{displaymath}
\begin{array}{rclclcl}
T_{\rm eff} & = & 7747       & - & 3016\,(B-V)^{\rm Johnson} (1 & - & 0.15\,{\rm [Fe/H]}) \\
            &   & \pm 58     &   & \pm 100                      &   & \pm 0.04 \\[0.15cm]
\sigma      & = & 65~{\rm K} &   & ({\rm 36~stars})             &
&
\end{array}
\end{displaymath}

\begin{displaymath}
\begin{array}{rclclcl}
T_{\rm eff} & = & 7551       & - & 2406\,(B-V)^{\rm Tycho} (1 & - & 0.20\,{\rm [Fe/H]}) \\
            &   & \pm 57     &   & \pm 88                     &   & \pm 0.05 \\[0.15cm]
\sigma      & = & 64~{\rm K} &   & ({\rm 31~stars})           &
&
\end{array}
\end{displaymath}

\begin{displaymath}
\begin{array}{rclclcl}
T_{\rm eff} & = & 8124       & - & 5743\,(b-y)^{\rm F} (1 & - & 0.10\,{\rm [Fe/H]}) \\
            &   & \pm 58     &   & \pm 156                &   & \pm 0.02 \\[0.15cm]
\sigma      & = & 55~{\rm K} &   & ({\rm 36~stars})       &   &
\end{array}
\end{displaymath}

\begin{displaymath}
\begin{array}{rclclcl}
T_{\rm eff} & = & 8481       & - & 6516\,(b-y)^{\rm G} (1 & - & 0.09\,{\rm [Fe/H]}) \\
            &   & \pm 67     &   & \pm 177                &   & \pm 0.02 \\[0.15cm]
\sigma      & = & 58~{\rm K} &   & ({\rm 36~stars})       &   &
\end{array}
\end{displaymath}

\begin{displaymath}
\begin{array}{rclcl}
T_{\rm eff} & = & 7790       & - & 5805\,(R-I)^{\rm Johnson} \\
            &   & \pm 80     &   & \pm 249 \\[0.15cm]
\sigma      & = & 100{\rm K} &   & ({\rm 35~stars})
\end{array}
\end{displaymath}

\begin{displaymath}
\begin{array}{rclcl}
T_{\rm eff} & = & 8465       & - & 5005\,(V-R)^{\rm Johnson} \\
            &   & \pm 139    &   & \pm 276 \\[0.15cm]
\sigma      & = & 125{\rm K} &   & ({\rm 35~stars})
\end{array}
\end{displaymath}

\begin{displaymath}
\begin{array}{rclcl}
T_{\rm eff} & = & 8234      & - & 3523\,(V-I)^{\rm Johnson} \\
            &   & \pm 75    &   & \pm 116 \\[0.15cm]
\sigma      & = & 77{\rm K} &   & ({\rm 35~stars})
\end{array}
\end{displaymath}

\begin{displaymath}
\begin{array}{rclclcl}
T_{\rm eff} & = & 8974       & - & 2880\,(V-K)^{\rm Johnson} & + & 440\,(V-K)_{\rm Johnson}^{\rm 2} \\
            &   & \pm 219    &   & \pm 292                   &   & \pm 94 \\[0.15cm]
\sigma      & = & 50~{\rm K} &   & ({\rm 36~stars})          &   &
\end{array}
\end{displaymath}

\begin{displaymath}
\begin{array}{rcll}
T_{\rm eff} & = & 11654      & \sqrt{\beta - 2.349} \\
            &   & \pm 182    & \hspace{0.88cm} \pm 0.008 \\[0.15cm]
\sigma      & = & 70~{\rm K} & ({\rm 34~stars}).
\end{array}
\end{displaymath}

Some of these calibrations have already been successfully employed
in determining photometric \Teffs for solar-type stars
\citep{delpelosoetal2005c, dasilvaetal2012}. We now comment on the
agreement of the calibrations shown in figs. 8 to 15 with various
published ones in regular use.

In figure A.1, for ($B-V$)$^{\rm Johnson}$, a comparison with
\cite{alonsoetal1996} shows good agreement within the 1$\sigma$
uncertainty we adopt for our target interval of 5000K $\leq$ \Teff
$\leq$ 6500K. The results of \cite{casagrandeetal2006} are in very
good agreement with ours in the full \Teff range but for a
2$\sigma$ offset at the hotter end, where their values are hotter.
The magnitude of the \feh sensitivity is similar in the three
calibrations, but the curvature in our relation is significantly
less than in the others, which may be explained by our shorter
\Teff interval. The \cite{ramirezmelendez2005b} calibration is
cooler than in Casagrande et al. by $\sim$100K, and more similar
to the Alonso et al. one. The latter relation implies a solar
($B-V$) bluer than ours by about 0.03 mag, while the value implied
in Casagrande et al. agrees very well with ours. The comparison of
our ($B-V$)$^{\rm Tycho}$ calibration in figure A.2 was done with
\cite{ramirezmelendez2005b} and \cite{casagrandeetal2010}. There
is good agreement with the former at the ends of our \Teff range,
but at the solar \Teff their relation is cooler by $\sim$130K.
Their sensitivity to \feh is similar to ours, but their regression
has a much more pronounced curvature. Very good agreement,
however, is found between our relation and that of
\cite{casagrandeetal2010}. This color should be more explored
more, since little use has been made of it in IRFM \Teff
calibrations. Its errors are similar to those of ($B-V$)$^{\rm
Johnson}$, and its \feh sensitivity is comparable. For ($b-y$), we
separately provide relations for the photometry of the F
\citep{olsen1983} and G \citep{olsen1993} catalogs. We compare in
figure A.3 our relation due to the F catalog with those of
\cite{blackwelllynasgray1998} and \cite{alonsoetal1996}, and the
regressions by \cite{blackwelllynasgray1998} are given separately
for two different ranges separated at $\sim$6000K according to
their prescription. These authors do not specify which of the
Olsen photometry systems is used in their calibrations. The
regression of \cite{alonsoetal1996} depends on the Str\"omgren
c$1$ index: for the comparison we fixed this index at the middle
of our interval, namely the value of 18 Sco, a warranted
approximation given our narrow $\log{g}$ interval. Agreement is
good particularly at the hotter end and still within 2$\sigma$
down to \Teff $\sim$ 5000K. The three calibrations have similar
curvatures in this range, but our \feh sensitivity is higher than
that of \cite{alonsoetal1996}, which in turn is higher than for
\cite{blackwelllynasgray1998}. The solar ($b-y$) color implied by
the \cite{alonsoetal1996} calibration is bluer than ours by 0.02
mag., while good agreement is found with
\cite{blackwelllynasgray1998}.

The infrared Johnson colors are found to be independent of \feh in
our relations but for ($V-K$). We compare our ($R-I$) calibration
in figure A.4 with those of \cite{alonsoetal1996} and
\cite{casagrandeetal2006}, where the latter relation is
independent of \feh, while the former is not. The relation of
\cite{casagrandeetal2006} has been transformed from the Cousins to
the Johnson system with the prescription given by
\cite{bessell1979}. Our results agree very well with
\cite{alonsoetal1996}, especially at the ends of our \Teff range:
near the solar \Teff the offset is still within 1$\sigma$. The
\cite{casagrandeetal2006} calibration is in good agreement with
ours near the solar \Teff but is hotter at both ends of our \Teff
range, particularly near \Teff = 6500K where the disagreement
reaches 2$\sigma$. For the Johnson ($V-R$) and ($V-I$) colors, we
compare respectively in figures A.5 and A.6, our relations to
\cite{alonsoetal1996}. For ($V-I$) we both found no sensitivity to
\feh, and agreement is good for the full \Teff range. For ($V-R$),
the \cite{alonsoetal1996} relation depends on \feh, unlike ours,
but again good agreement is found for the full range: however, a
larger scatter is seen, and the standard deviation of our fit for
this color is the highest. For the Johnson ($V-K$) color, we
compared in figure A.7 our calibrations with those of
\cite{alonsoetal1996}, \cite{blackwelllynasgray1998}, and
\cite{masanaetal2006}. Only the \cite{alonsoetal1996} regression
is very slightly \feh-sensitive and we set \feh to the solar value
for comparison. The agreement of the three calibrations is very
good for the full \Teff range, and we note that the relation given
by \cite{mcwilliam1990} is also compatible with the previously
mentioned calibrations in this \Teff range.

For the Str\"omgren $\beta$ index, we compare our regression with
those of \cite{saxnerhammarback1985} and \cite{alonsoetal1996} in
figure A.8. Not surprisingly, our results are in line with
\cite{saxnerhammarback1985}. The calibration of
\cite{alonsoetal1996} is only weakly \feh-dependent and we set
\feh = $+$0.00 for the comparison: this calibration also shows
good agreement in our \Teff range. {The $\beta$ color index},
though more difficult to obtain for faint stars due to the
narrowness of the filters, has good \Teff sensitivity in this
range, is reddening-free, and could be explored more for
solar-type stars.

Taking into account that IRFM \Teffs are accurate to no more than
a few percent \citep{ramirezmelendez2005a, casagrandeetal2006},
the agreement of our calibrations with many widely adopted recent
resources ranges from fair to very good, within a 2$\sigma$
assessment but for a few cases. In the range of \Teff we are
exploring, they provide good internal consistency and collectively
enable the derivation of photometric \Teffs for solar-type stars,
within a reasonable range around the solar atmospheric parameters,
with internal errors less than 1\%. The external errors of
the IRFM \Teff scale are, of course, larger by at least a factor of two.\\

\begin{acknowledgements}

This paper is dedicated, {\it in memorian}, to Giusa Cayrel de
Strobel, for her dedicated pioneering in the subject of solar
analogs and twins. G. F. P. M. acknowledges financial support by
CNPq grant n$^{\circ}$ 476909/2006-6, FAPERJ grant n$^{\circ}$
APQ1/26/170.687/2004, and a CAPES post-doctoral fellowship
n$^{\circ}$ BEX 4261/07-0. R. S. acknowledges a scholarship from
CNPq/PIBIC. L. S. thanks CNPq for the grant 30137/86-7. We thank
the staff of the OPD/LNA for considerable support in the observing
runs needed to complete this project. Use was made of the Simbad
database, operated at the CDS, Strasbourg, France, and of NASA's
Astrophysics Data System Bibliographic Services. We thank Giusa
Cayrel de Strobel, {\it in memorian}, Edward Guinan, Jos\'e Dias
do Nascimento Jr., Jos\'e Renan de Medeiros, and Jeffrey Hall for
interesting discussions. We also thank the referee, Dr. Martin
Asplund, for suggestions and criticism which considerably improved
the paper.

\end{acknowledgements}

\bibliographystyle{aa}
\bibliography{Porto-de-Mello-et-al-Solar-Twin-Survey_PREPRINT}

\end{document}